\documentclass[aps, pre, groupedaddress]{revtex4-2}
\usepackage{hyperref,lineno,natbib,amsmath,amssymb,graphicx,multirow,algorithmic,color,microtype}
\usepackage{gensymb}
\usepackage{dcolumn}
\usepackage{bm}
\usepackage{xcolor}
\usepackage{booktabs}

\usepackage{textcomp}

\definecolor{dblue}{rgb}{0.0,0.0,0.5}
\definecolor{dmag}{rgb}{0.831,0.165,1.0}
\definecolor{dred}{rgb}{1.0,0,0}
\definecolor{jade}{rgb}{0.1333,0.5647,0.4784}
\definecolor{lblue}{rgb}{0,0.6745,1.0}
\definecolor{pmag}{rgb}{0.580,0.129,0.572}
\definecolor{pgry}{rgb}{0.572,0.572,0.572}

\usepackage{hyperref}

\definecolor{webgreen}{rgb}{0,.35,0}
\definecolor{webbrown}{rgb}{.6,0,0}
\definecolor{RoyalBlue}{rgb}{0,0,0.9}
\definecolor{purp}{rgb}{0.4,0.2,0.8}
\definecolor{mywhite}{rgb}{1.0,1.0,1.0}

\hypersetup{
   colorlinks=true, linktocpage=true, pdfstartpage=3, pdfstartview=FitV,
   breaklinks=true, pdfpagemode=UseNone, pageanchor=true, pdfpagemode=UseOutlines,
   plainpages=false, bookmarksnumbered, bookmarksopen=true, bookmarksopenlevel=1,
   hypertexnames=true, pdfhighlight=/O,
   urlcolor=webbrown, linkcolor=RoyalBlue, citecolor=webgreen,
   pdfauthor={Nicholas M. Boffi and Chris H. Rycroft},
   pdfsubject={A coordinate transformation methodology for simulating quasi-static elastoplastic solids},
   pdfkeywords={fluid mechanics, projection method, plasticity, elastoplasticity},
   pdfcreator={pdfLaTeX},
   pdfproducer={LaTeX with hyperref}
}

\newcommand{\tK}{\text{~K}}
\newcommand{\bigO}{\mathcal{O}}

\newcommand{\half}{\frac{1}{2}}

\newcommand{\sbar}{\bar{s}}
\newcommand{\p}{\partial}
\renewcommand{\vec}[1]{\mathbf{#1}}
\newcommand{\ten}[1]{\mathbf{#1}}
\newcommand{\tC}{\ten{C}}

\newcommand{\vv}{\vec{v}}

\newcommand{\prX}[1]{\frac{\p #1}{\p X}}

\newcommand{\dt}{\Delta t}

\newcommand{\bx}{\mathbf{x}}
\newcommand{\by}{\mathbf{y}}
\newcommand{\bX}{\mathbf{X}}
\newcommand{\vV}{\mathbf{V}}
\newcommand{\tD}{\ten{D}}

\newcommand{\Dpl}{\tD^\text{pl}}

\newcommand{\dpl}{D^\text{pl}}

\newcommand{\Dt}{\Delta t}
\newcommand{\bsig}{\boldsymbol\sigma}
\newcommand{\bSig}{\boldsymbol\Sigma}

\newcommand{\Nab}{\nabla_\bX}
\newcommand{\nab}{\nabla_\bx}

\newcommand{\bL}{\mathbf{L}}
\newcommand{\Trans}{\mathsf{T}}
\newcommand{\bT}{\mathbf{T}}

\DeclareMathOperator{\tr}{tr}

\begin{document}

\title{A coordinate transformation methodology for simulating quasi-static elastoplastic solids}
\author{Nicholas M. Boffi}
\email{boffi@g.harvard.edu}
\affiliation{John A. Paulson School of Engineering and Applied Sciences, Harvard University, Cambridge, MA 02138}

\author{Chris H. Rycroft}
\email{chr@seas.harvard.edu}
\affiliation{John A. Paulson School of Engineering and Applied Sciences, Harvard University, Cambridge, MA 02138}
\affiliation{Computational Research Division, Lawrence Berkeley Laboratory, Berkeley, CA 94720}

\begin{abstract}
  Molecular dynamics simulations frequently employ periodic boundary
  conditions where the positions of the periodic images are manipulated in
  order to apply deformation to the material sample. For example,
  Lees--Edwards conditions use moving periodic images to apply simple
  shear. Here, we examine the problem of precisely comparing this type of
  simulation to continuum solid mechanics. We employ a hypo-elastoplastic
  mechanical model, and develop a projection method to enforce quasi-static
  equilibrium. We introduce a simulation framework that uses a fixed Cartesian
  computational grid on a reference domain, and which imposes deformation via a
  time-dependent coordinate transformation to the physical domain. As a test
  case for our method, we consider the evolution of shear bands in a bulk
  metallic glass using the shear transformation zone theory of amorphous
  plasticity. We examine the growth of shear bands in simple shear and pure
  shear conditions as a function of the initial preparation of the bulk
  metallic glass.
\end{abstract}

\date{\today}
\maketitle

\section{Introduction}
\newlength{\subpanelwid}
\setlength{\subpanelwid}{0.46\textwidth}
Molecular dynamics (MD) simulations, whereby atoms or molecules are individually simulated according to Newton's laws~\cite{allen}, are widely used across the physical sciences~\citep{Hossain2010, Guan2013, Puosi2014, regev15}. Open source sofware packages such as LAMMPS~\cite{lammps,plimpton95} and GROMACS~\cite{berendsen95} have enabled simulations to be performed with millions of particles on modern parallel computer hardware. MD simulations provide a detailed view of the material physics and are able to capture discrete particle-level effects~\cite{maloney04,maloney06}. Despite these advantages, MD simulations are computationally expensive, and it is usually only possible to simulate microscopic material samples. Furthermore, since the simulations must resolve rapid interaction timescales between particles, the applied deformation rates in MD are often orders of magnitude larger than deformation rates in laboratory tests~\cite{falk99,bailey04a,sepulveda-macias16}.

Because MD simulations simulate microscopic domains, it is difficult to apply deformation via moving walls, as simulation data may be affected by finite-size effects~\cite{landry03,rycroft12c}. Instead, the standard approach is to apply periodic boundary conditions, but manipulate the periodic images of the primary simulation domain to achieve different applied deformations. For example, in three-dimensional Lees--Edwards boundary conditions, the periodic images have a horizontal velocity proportional to their $z$ position in order to impose simple shear~\citep{LE_orig} (Fig.~\ref{fig:le_trans}(a)). The Kraynik--Reinelt boundary conditions~\cite{kraynik92,todd98,todd99,baranyai99}, plus a recent generalization by Dobson~\cite{dobson14}, use a combination of moving periodic images and domain remapping in order to simulate different extensional flows.

A complementary approach to MD is to use continuum modeling, which has the ability to simulate large system sizes and long, physically realistic timescales. However, continuum-scale theories involve a substantial theoretical hurdle, in that the transition from a particle-level theory to a continuum theory involves a coarse-graining procedure. The coarse-graining procedure defines a representative volume element (RVE)~\cite{RVE1, RVE2} throughout which local deviations of material field values from their average within the RVE are neglected. The fundamental assumption of every continuum theory is that such an RVE is well-defined, and that neglecting the discrepancy between the relevant system variables and their mean within an RVE is well-justified~\citep{Drugan1996, Chaboche2008}.

In effect, coarse-graining reduces the complex many-body system of interacting particulate constituents to a much lower degree-of-freedom system well-described by a set of nonlinear partial differential equations. This reduction in complexity is primarily responsible for the well-behaved scaling with system size in continuum simulations, in that all the classical techniques of numerical analysis become available for evolving the system over time. However, the process of coarse-graining to the continuum is difficult in general, and has primarily been successful when tailored to specific phenomena. The coarse-graining procedure introduces internal state variables that summarize the many particulate degrees of freedom, and accurate initial conditions for such internal variables can be difficult to construct. Some equilibrium systems are amenable to rigorous approaches by explicitly averaging over unwanted degrees of freedom in the system partition function~\citep{Dijkstra1999, Bolhuis2001}, but these approaches are intractable for many out-of-equilibrium systems.

\setlength{\unitlength}{0.01\textwidth}
\begin{figure*}
    \centering
    \begin{picture}(95,29)
        \put(0,4){\includegraphics[width=0.4\textwidth]{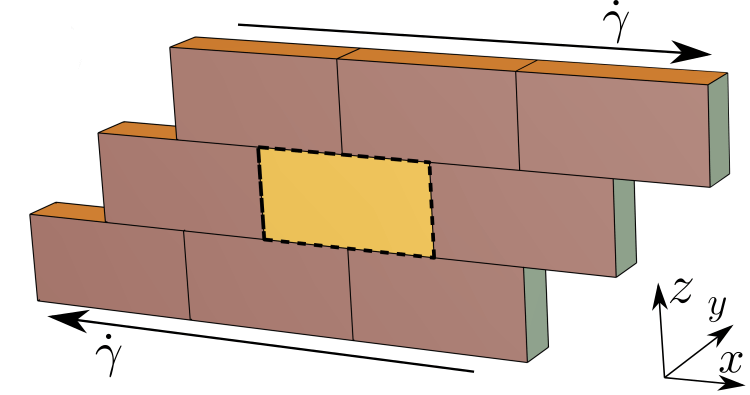}}
        \put(45,0){\includegraphics[width=0.5\textwidth]{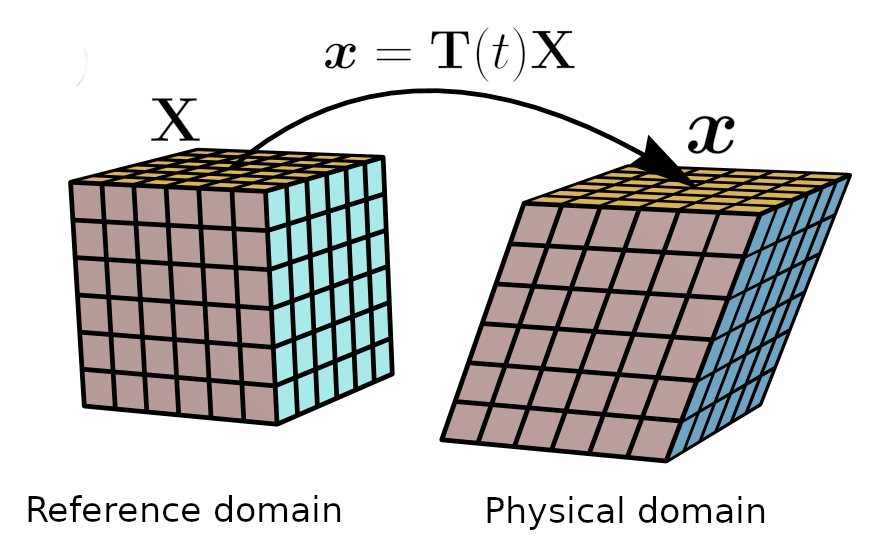}}
        \put(1,25){(a)}
        \put(46,25){(b)}
    \end{picture}
    \caption{(a) Lees--Edwards boundary conditions in three dimensions where
    the $z$ coordinate points upward. The system of interest is shown in yellow
    and outlined in dashed black lines. Periodic copies of the system above and
    below are set to move with a specific velocity, imposing a specific strain
    rate $\dot{\gamma}$ on the system. (b) A graphical depiction of a
    domain transformation $\bT(t)$ that maps a fixed reference domain $\bX$ to
    a sheared physical domain $\bx$.\label{fig:le_trans}}
\end{figure*}

To quantitatively explore the effect of coarse-graining MD to the continuum, it is therefore useful to perform the two types of simulation using the same geometry and conditions. However, precisely recreating the boundary conditions from MD for use in continuum simulations poses some numerical challenges. Consider the Lees--Edwards boundary conditions and suppose that the primary simulation domain is discretized on a Cartesian grid. Because the periodic images are moving, their grids will generally not align with the primary domain. This could be handled numerically via interpolation, but grid points near the boundary will incur different discretization errors. If the continuum model involves an elliptic problem, then the shifted grids will result in a complex connectivity structure in the associated linear system, which is less well-suited to some numerical linear algebra techniques.

In this work, we address this problem by developing a continuum solid mechanics simulation that permits MD boundary conditions to be recreated precisely. We use the hypo-elastoplasticity model~\cite{truesdell55} in which the deformation rate tensor $\tD$ is decomposed additively into a sum of elastic and plastic parts~\cite{hill58}. There are a number of different frameworks for simulating elastoplastic materials~\cite{xiao06}, but the hypo-elastoplastic model is well-suited for problems that involve large plastic deformation. This regime is appropriate for matching to typical MD simulations, where large total strain may be applied.

Combining the additive decomposition of $\tD$ with Newton's second law results in a closed hyperbolic system of partial differential equations (PDEs) for the velocity $\vv$ and stress $\bsig$, plus coupling to evolution equations for any internal state variables. Due to the small size of MD simulations, it is usually a good approximation to say that elastic waves are fast compared to the simulation timescale, allowing for Newton's second law to be replaced by the constraint that the stresses remain in quasi-static equilbrium, $\nabla \cdot \bsig =\vec{0}$.

The resulting constrained PDE system has a mathematical correspondence to the incompressible Navier--Stokes equations, where the fluid velocity must satisfy the constraint that $\nabla \cdot \vv=0$. For incompressible fluids a standard numerical technique is the projection method of Chorin~\cite{chorin67,chorin68}. By exploiting the mathematical correspondence, a new projection method for quasi-static hypo-elastoplasticity was recently introduced~\cite{rycroft15} and extended to three dimensions~\cite{boffi2019-1} (Sec.~\ref{sec:mp}).

To match the MD boundary conditions, we introduce a coordinate transformation framework for the quasi-static hypo-elastoplastic system. It is based on an abstract linear mapping $\bT(t)$ from a \textit{reference domain} to the \textit{physical domain} (Fig.~\ref{fig:le_trans}(b)). Lees--Edwards conditions can be implemented in the continuum setting with this methodology by imposing shear through a transformation, and additionally enforcing periodic boundary conditions in all directions. Effectively, our method decouples the application of material deformation from the application of a specific boundary condition.

In addition to Lees--Edwards boundary conditions, the transformation framework is flexible, and enables simple implementation of otherwise potentially difficult deformation, such as pure shear. Any applied deformation that can be written down as a linear transformation of a reference domain can be implemented just by implementing the matrix and its time derivatives. We show that the projection method for hypo-elastoplasticity can be generalized to simulate this case by working with transformed velocities and stresses in the reference domain. The projection step in the method requires solving an elliptic problem for the velocity, and the resulting linear system has a simple mathematical structure that is well-suited for solution via numerical linear algebra techniques such as the geometric multigrid method~\cite{demmel,Briggs2000}.

The new method is capable of simulating a wide range of elastoplastic materials, but here we consider the example of a bulk metallic glass (BMG), a new type of alloy where the atoms have a random and amorphous arrangement, in constrast to most metals~\cite{johnson99}. BMGs have attracted considerable research interest during the past two decades. They have many favorable properties, such as high strength and wear resistance, that make them attractive for a variety of applications~\cite{schroers13}. However, the amorphous arrangement of atoms makes the study of dynamic mechanical phenomena---such as deformation and failure---in these materials exceptionally challenging~\cite{falk_langer_rev}.

To date, a general theory of the microscopic origins of plastic deformation in amorphous solids has remained elusive. However, several prominent theories capable of making accurate qualitative and quantitative predictions have been developed, such as free-volume based theories~\cite{spaepen77, argon79, cohen79, boutreux97} and the shear transformation zone (STZ) theory~\cite{langer08, bouchbinder07, bouchbinder07b, bouchbinder09, bouchbinder09c}. Ultimately, free-volume theories and the STZ theory are flow-defect theories that attempt to connect microscopic rearrangements of groups of atoms with macroscopic plastic deformation, in rough analogy to the dislocation-mediated theory of plasticity in crystalline materials~\cite{disloc}.

We employ an elastoplastic model of a BMG based on the STZ theory. A key feature of the model is the effective temperature (Sec.~\ref{ssec:plast}), which characterizes the amorphous particle structure via a continuum field~\cite{bouch_eff_dyn, cugli1, cugli2, bouchbinder09b}. The effective temperature can be measured indirectly~\cite{song05}, but there is currently no complete method to connect it to the microscopic particle configuration. This was recently explored by Hinkle \textit{et al.}~\cite{hinkle15}, who directly compared continuum and MD simulations, and examined how measurable features of MD such as the coarse-grained atomic potential energy are connected to the effective temperature. A key limitation of this study is that the MD simulations use Lees--Edwards conditions, whereas the deformation was imposed in the continuum simulation using moving parallel plates, meaning that the two could not be exactly compared. The numerical techniques that we develop here remove this limitation.

The STZ theory has proven useful for examining the failure properties of BMGs. The elastoplastic model that we employ has been used to explain the large experimental variations in notched fracture toughness of BMGs~\cite{rycroft12}. This was subsequently extended to make predictions about BMG fracture toughness for a range of parameters~\cite{vasoya16}. Recent experimental work suggests that these predictions are broadly correct~\cite{schroers2018}. BMGs also exhibit shear bands, a strain-softening instability characterized by the localization of shear strains along a thin band~\cite{greer-2013}, which can be the precursor to failure~\cite{bing-2005,schuh-2007,maass15}. In our simulations, we examine how shear bands nucleate as a function of the initial inhomogeneities in the effective temperature field.

The paper is organized as follows. In Sec.~\ref{sec:mp}, we describe the equations of quasi-static hypo-elastoplasticity and provide an introduction to the physics of the STZ theory of amorphous plasticity. In Sec.~\ref{sec:coords}, we introduce the coordinate transformation methodology and develop the transformed projection method. In Sec.~\ref{sec:numerical_tests} we provide numerical experiments demonstrating convergence of the solution of the transformed method to the original quasi-static method in physically equivalent situations as the grid spacing is decreased. In Sec.~\ref{sec:examples}, we study shear banding in a bulk metallic glass subject to simple shear, Lees--Edwards, and pure shear boundary conditions. We highlight differences in results between Lees--Edwards and simple shear boundary conditions and examine how the shear band formation depends on the initial effective temperature.

\section{Mathematical preliminaries}
\label{sec:mp}
\subsection{Quasi-static hypo-elastoplasticity}
\label{ssec:qshep}
We consider an elastoplastic material with Cauchy stress tensor $\bsig(\bx, t)$ and velocity field $\vv(\bx, t)$. We denote by $\bL = \nabla \vv$ the velocity gradient tensor and $\tD = \frac{1}{2}\left(\bL + \bL^\Trans\right)$ the rate of deformation tensor. We adopt the framework of hypo-elastoplasticity, which assumes the rate of deformation tensor can be additively decomposed into a sum of elastic and plastic parts, $\tD = \tD^{\text{el}} + \Dpl$. Writing linear elasticity in rate form yields
\begin{equation}
    \frac{\mathcal{D} \bsig(\bx, t)}{\mathcal{D} t} = \tC : \left(\tD - \Dpl\right)
    \label{eqn:elas}
\end{equation}
where $\tC$ is the stiffness tensor. For simplicity, the material is taken to be isotropic and homogeneous, so that $C_{ijkl} = \lambda\delta_{ij}\delta_{kl} + \mu\left(\delta_{ik}\delta_{jl} + \delta_{il}\delta_{jk}\right)$ where $\lambda$ is Lam\'e's first parameter and $\mu$ is the shear modulus. The time derivative in Eq.~\ref{eqn:elas} is the Truesdell derivative~\footnote{This expression is typically presented with $\bL$ transposed with respect to the definition here. We adopt the convention that $\left(\nabla \mathbf{f}\right)_{ij} = \p_i f_j$ for a vector field $\bf$, i.e., partial derivatives go row-wise in gradients of vector fields. Typically, the symbol $\bL = \nabla \vv$ is used to denote the Jacobian or Fr\'echet derivative of $\vv$, which formally is the transpose of the gradient~\cite{gurtin10}. The transformation formalism developed in this work involves both Jacobians and vector field gradients, and for physically consistent answers it is necessary to make this distinction.},
\begin{equation}
    \frac{\mathcal{D} \bsig}{\mathcal{D} t} = \frac{d \bsig}{dt} - \bL^\Trans \bsig - \bsig \bL + \tr(\bL)\bsig,
    \label{eqn:orig_sig}
\end{equation}
with \smash{$\frac{d}{dt} = \frac{\p}{\p t} + \vv \cdot \nabla$} denoting the advective derivative. The velocity field satisfies a continuum version of Newton's second law,
\begin{equation}
    \rho \frac{d \vv}{dt} = \nabla \cdot \bsig,
    \label{eqn:nwtn}
\end{equation}
with $\rho$ the material density. Taken together, Eqs.~\ref{eqn:elas} \& \ref{eqn:nwtn} form a closed hyperbolic system that could form the basis of a numerical method. However, an explicit numerical method used to solve this system will resolve elastic waves. Stable resolution of elastic waves places a limit on the simulation timestep according to the well-known Courant--Friedrichs--Lewy (CFL) condition~\cite{courant67}. The CFL condition requires $\Delta t \leq \frac{h}{c_e}$ where $c_e$ is a typical elastic wave speed and $h$ is the grid spacing.

In metals and other materials of interest, the elastic wave speed $c_e$ can be large, and the grid spacing $h$ needed to resolve fine-scale features such as shear bands can be small. The CFL condition thus poses a prohibitive limit on the timestep for probing realistic timescales and system sizes, and the development of alternative simulation approaches which avoid resolving elastic waves is necessary. It is often appropriate to take the long-timescale and small-velocity limit, in which the material acceleration is negligible and Eq.~\ref{eqn:nwtn} can be replaced by the constraint
\begin{equation}
    \nabla \cdot \bsig = \mathbf{0},
    \label{eqn:qs_const}
\end{equation}
which states the stresses remain in quasi-static equilibrium, and conveniently avoids the description of elastic waves. In this quasi-static limit, Eq.~\ref{eqn:elas} depends on the material velocity field through $\tD$, but the evolution equation for the velocity field has been exchanged for the constraint in Eq.~\ref{eqn:qs_const}. It is thus unclear how to solve Eq.~\ref{eqn:elas} subject to the global constraint in Eq.~\ref{eqn:qs_const}.

\subsection{Projection method}
\label{ssec:orig_proj}
As noted by Rycroft \textit{et al.}~\cite{rycroft15}, Eqs.~\ref{eqn:elas} \& \ref{eqn:nwtn} have a close mathematical correspondence with the Navier--Stokes equations for incompressible fluid flow. The Navier--Stokes equations consist of an explicit partial differential equation for the fluid velocity along with a constraint that the velocity must be divergence-free. Much like Eqs.~\ref{eqn:elas} \& \ref{eqn:qs_const}, the constraint on the velocity field is obtained from a limiting procedure applied to an explicit partial differential equation for the pressure, and the equation for the velocity still depends on the pressure after this limit has been taken.

In this setting, a well-established numerical technique is the projection method of Chorin~\cite{chorin67,chorin68}. In Chorin's projection method, the update for the velocity field is split into two steps. In the first step, an intermediate velocity field is computed which does not obey the divergence-free constraint. In the second step, a linear system is solved for the pressure field which simultaneously projects the intermediate velocity field onto the manifold of divergence-free solutions.

By using the correspondence between quasi-static hypo-elastoplaticity and incompressible fluid flow, Rycroft \textit{et al.}~\cite{rycroft15} developed a new projection method for quasi-static elastoplasticity. Consider taking a timestep of size $\Delta t$, and use superscripts of $n$ and $n+1$ to denote the simulation fields before and after the step, respectively. To begin, an intermediate stress $\bsig^*$ is computed by dropping the $\tC:\tD$ term in Eq.~\ref{eqn:elas} to obtain
\begin{equation}
  \frac{\bsig^* - \bsig^n}{\Delta t} = (\bL^n)^\Trans \bsig^n + \bsig^n \bL^n - \tr(\bL^n)\bsig^n - (\vv^n \cdot \nabla) \bsig^n - \tC : (\Dpl)^n.
  \label{eqn:sproj1}
\end{equation}
If the velocity $\vv^{n+1}$ were known, and hence if the total deformation rate
$\tD^{n+1}$ could be calculated, then the final stress would be given by
\begin{equation}
  \frac{\bsig^{n+1} - \bsig^*}{\Dt} = \tC : \tD^{n+1}.
  \label{eqn:sproj2}
\end{equation}
Taking the divergence of this equation and enforcing that $\nabla \cdot
\bsig^{n+1}=\vec{0}$ yields
\begin{equation}
  \Dt\, \nabla \cdot (\tC : \tD^{n+1}) = - \nabla \cdot \bsig^*.
  \label{eqn:sproj3}
\end{equation}
After finite-difference expansion of the definition of $\tD^{n+1}$, Eq.~\ref{eqn:sproj3} forms a linear system for the velocity field $\vv^{n+1}$ with source term given by the known vector $-\nabla\cdot\bsig^*$, and it can be solved via standard techniques of numerical linear algebra. After solution of Eq.~\ref{eqn:sproj3}, $\bsig^{n+1}$ can be computed according to Eq.~\ref{eqn:sproj2}, which can be shown to orthogonally project $\bsig^*$ onto the manifold of quasi-static solutions. In this manner, the two-step projection method enables solving Eq.~\ref{eqn:elas} subject to the global constraint Eq.~\ref{eqn:qs_const} despite the dependence of Eq.~\ref{eqn:elas} on $\vv$. We refer the reader to papers by Rycroft \textit{et al.}~\cite{rycroft15}, and Rycroft and Boffi~\cite{boffi2019-1} for complete details on this method.

\subsection{Plasticity model}
\label{ssec:plast}
As our plasticity model for a bulk metallic glass, we use an athermal form of the shear transformation zone (STZ) theory of amorphous plasticity suitable for studying glassy materials below the glass transition temperature~\cite{stz_basic, bouchbinder09}. The STZ theory postulates that ephemeral and localized fluctuations of the configurational bath---STZs---occur sporadically throughout an otherwise elastic material. The STZs may be conceptualized as clusters of atoms susceptible to shear-induced configurational rearrangements when local stresses surpass the material yield stress $s_Y$. Each such rearrangement leads to a small increment of plastic strain, and many such rearrangements conspire to bring about macroscopic plastic deformation.

In the athermal theory used here, thermal fluctuations of the atomic configurations are neglected, and molecular rearrangements are assumed to be driven entirely by external mechanical forces. Thermal theories introduce an additional coupling between the configurational subsystem governing the rearrangements that occur at STZs, and a kinetic/vibrational subsystem governing the thermal vibrations of atoms in their cage of nearest neighbors~\cite{kamrin14a}. Such thermal theories, with an additional field tracking the thermodynamic temperature that evolves according to a diffusion equation, could in principle be incorporated into our framework.

Each rearrangement corresponds to a transition in the configurational energy landscape; these transitions are usually towards a lower-energy configuration, but there is a small probability for a reverse transition. Before the application of external shear, the material sample sits at a local minimum. External shear alters the shape of the energy landscape, and can make transitions to other states considerably more likely. The density of STZs in space follows a Boltzmann distribution in an effective disorder temperature denoted by $\chi$~\cite{bouchbinder09b, bouch_eff_dyn, cugli1, cugli2}.

$\chi$ governs the out-of-equilibrium configurational degrees of freedom of the material and has many properties of the usual temperature: it is measured in Kelvin, and it can be obtained as the derivative of a configurational energy with respect to a configurational entropy~\cite{falk_langer_rev}. $\chi$ is distinct from the thermodynamic temperature $T$, though it plays the same role for the configurational subsystem as $T$ does for the kinetic/vibrational subsystem.

We define the deviatoric stress tensor $\bsig_0 = \bsig - \frac{1}{3}\tr(\bsig)\mathbf{I}$. The total rate of plastic deformation tensor is proportional to the deviatoric stress, $\Dpl = D^{\text{pl}}\frac{\bsig_0}{\bar{s}}$, where $\bar{s}^2 = \frac{1}{2}\bsig_{0, ij}\bsig_{0, ij}$ is a local scalar measure of the total deviatoric stress. The STZ theory provides the magnitude of the plastic rate of deformation as
\begin{equation}
    \tau_0 D^{\text{pl}} = e^{-e_z/k_B \chi}e^{-\Delta/k_B T}\cosh\left(\frac{\Omega \epsilon_0 \bar{s}}{k_B T}\right) \left(1 - \frac{s_Y}{\bar{s}}\right).
    \label{eqn:stz_dpl}
\end{equation}
$\tau_0$ is a molecular vibration timescale, $e_z$ is a typical STZ formation energy, $k_B$ is the Boltzmann constant, $T$ is the thermodynamic temperature, $\Delta$ is a typical energetic barrier for a transition, $\Omega$ is a typical STZ volume, and $\epsilon_0$ is a typical local strain. The effective temperature satisfies a heat equation~\citep{stz_basic, manning07, langer07, manning09, bouchbinder09b}
\begin{equation}
    c_0\frac{d \chi}{dt} = \frac{\left(\Dpl : \bsig_0\right)}{s_Y}\left(\chi_\infty - \chi\right) + l^2 \nabla \cdot \left(D^{\text{pl}}\nabla\chi\right).
    \label{eqn:chi_evo}
\end{equation}
The interdependence of Eqs.~\ref{eqn:stz_dpl} \& \ref{eqn:chi_evo} enables the development of shear bands through a positive feedback mechanism, as increasing one of $\chi$ or $D^{\text{pl}}$ also leads to an increase in the other~\citep{manning07, manning09}.

\section{Coordinate transformation framework}
Let $\bT(t)$ denote a time-varying mapping from a reference domain $\bX$ to the physical domain of interest $\bx$ such that
\label{sec:coords}
\begin{equation}
    \bx = \bT \bX,
    \label{eqn:X}
\end{equation}
as shown in Fig.~\ref{fig:le_trans}(b). Here, $\bX \in [a_X, b_X]\times[a_Y, b_Y]\times[a_Z, b_Z]$. Capital letters denote quantities in the reference frame and lower case letters denote quantities in the physical frame. $\Nab$ and $\nab$ denote spatial differentiation in the reference and physical frame, respectively. We emphasize that $\bX$ exists in a fixed frame on which the quasi-static hypo-elastoplastic equations will be solved, and not in the Lagrangian frame of coordinates. To clarify this point, let $\mathcal{R} = (\mathcal{X}, \mathcal{Y}, \mathcal{Z})$ denote a set of fixed Lagrangian coordinates. For an Eulerian frame $(x, y, z)$, we define the Eulerian displacements,
\begin{equation}
    u_i = x_i - \mathcal{R}_i.
    \label{eqn:disp}
\end{equation}
We then define the Eulerian velocities \smash{$v_i = \frac{\p u_i}{\p t}|_\mathcal{R}$}. The same procedure can be performed in the reference frame. We first define the physical displacements,
\begin{align}
    \mathbf{u} = \bT \bX - \mathcal{R}.
    \label{eqn:disp_phys}
\end{align}
Taking a time derivative of both sides of Eq.~\ref{eqn:disp_phys} at fixed Lagrangian coordinates $\mathcal{R}$, we arrive at an expression for the physical velocity,
\begin{equation}
    \vv = \frac{\p \bT}{\p t} \bX + \bT \vV.
    \label{eqn:v}
\end{equation}
Above, we have identified the transformed velocity $\vV = \frac{\p X}{\p t}|_\mathcal{R}$. Equation \ref{eqn:v} can be used to compute the \textit{physical} velocity $\vv$ from the \textit{transformed} velocity $\vV$, if $\vV$ is known. By inversion, it can also be used as a definition of the transformed velocity,
\begin{equation}
    \vV = \bT^{-1}\left(\vv - \frac{\p \bT}{\p t}\bX\right).
    \label{eqn:V}
\end{equation}
Using the chain rule, spatial derivatives are transformed as
\begin{equation}
    \nab = \bT^{-\Trans}\Nab.
    \label{eqn:spat_deriv}
\end{equation}
Taking an advective time derivative of Eq.~\ref{eqn:V}, using Eq.~\ref{eqn:nwtn} for $\dot{\vv}$, and transforming physical spatial derivatives to transformed spatial derivatives, the transformed velocity evolves according to the transformed generalization of Newton's second law,
\begin{equation}
    \frac{\p \vV}{\p t} = -\left(\vV \cdot \Nab \right)\vV + \frac{\p \bT^{-1}}{\p t} \bT \vV + \bT^{-1}\left(\bT^{-\Trans} \Nab \cdot\left(\bT \bSig \bT^{\Trans}\right) - \frac{\p^2 \bT}{\p t^2}\bX - \frac{\p \bT}{\p t}\vV\right).
    \label{eqn:V_evol}
\end{equation}
In Eq.~\ref{eqn:V_evol}, we have rewritten the advective derivative equivalently in the reference frame, $\frac{\p}{\p t} + \vv\cdot \nab = \frac{\p}{\p t} + \vV\cdot\Nab$. The proof of this fact for an arbitrary transformation $\bT(t)$ is shown in Appendix~\ref{app:adv}. In Eq.~\ref{eqn:V_evol}, we have also defined the transformed stress tensor via the contravariant pullback,
\begin{equation}
    \bSig = \bT^{-1}\bsig\bT^{-\Trans}.
    \label{eqn:Sig}
\end{equation}
To derive an evolution equation for $\bSig$, we now use the linear elastic relation in Eq.~\ref{eqn:elas}. Taking an advective time derivative of the relation $\bsig = \bT\bSig\bT^\Trans$ and inverting, the transformed stress then obeys the transformed generalization of the linear elastic constitutive law. After expansion of the Truesdell rate,
\begin{align}
  \frac{\p \bSig}{\p t} &= -\left(\vV \cdot \Nab\right)\bSig - \tr(\bL)\bSig + \bSig\nabla_{\bX}\vV + \left(\nabla_{\bX}\vV\right)^{\Trans}\bSig + \bT^{-1}\left(\tC : \left(\tD - \Dpl\right)\right)\bT^{-\Trans}.
    \label{eqn:Sig_evol}
\end{align}
In Eq.~\ref{eqn:Sig_evol}, $\tD = \frac{1}{2}\left(\bL + \bL^\Trans\right)$ refers to the physical quantity. $\bL$ can be computed in terms of the transformed variables as
\begin{equation}
    \bL = \bT^{-\Trans}\frac{\p \bT^{\Trans}}{\p t} + \bT^{-\Trans}\Nab\vV\bT^{\Trans}.
    \label{eqn:L_trans}
\end{equation}
$\Dpl = D^{\text{pl}}\frac{\bsig_0}{\sbar}$ appears in Eq.~\ref{eqn:Sig_evol}, and its form depends on the plasticity model through the constant $D^{\text{pl}}$. As reviewed in Sec.~\ref{ssec:qshep}, the STZ theory provides an expression given by Eq.~\ref{eqn:stz_dpl}. $\Dpl$ is defined and must be computed in terms of the physical deviatoric stress $\bsig_0$. In line with the definition of $\bSig$, we can apply the contravariant pullback to $\bsig_0$ and write
\begin{equation}
    \bT^{-1} \bsig_0 \bT^{-\Trans} = \bSig - \frac{1}{3}\left(\bT^{-1} \tr\left(\bT\bSig \bT^\Trans\right)\bT^{-\Trans}\right)\mathbf{I}.
    \label{eqn:trans_dev}
\end{equation}
Using the natural definition $\bSig_0 = \bSig - \frac{1}{3} \tr\left(\bSig\right)\mathbf{I}$ and solving for $\bsig_0$, we can rewrite Eq.~\ref{eqn:trans_dev} as
\begin{equation}
    \bsig_0 = \bT \bSig_0 \bT^\Trans + \frac{1}{3}\left(\bT \tr\left(\bSig\right)\bT^\Trans - \tr\left(\bT \bSig \bT^\Trans\right)\right)\mathbf{I}.
    \label{eqn:dev_trans}
\end{equation}
Equation \ref{eqn:dev_trans} enables the computation of $\bsig_0$ entirely in terms of transformed quantities. We compute $\sbar$ by first computing the entire tensor $\bsig_0$ using Eq.~\ref{eqn:dev_trans} and then compute its Frobenius norm.

The equation for the effective temperature must also be transformed, though we do not define a transformed effective temperature. This can be accomplished by transforming the derivatives,
\begin{equation}
    c_0\frac{\p \chi}{\p t} = -c_0\left(\vV \cdot \Nab\right) \chi + \frac{\left(\Dpl : \bsig_0\right)}{s_Y}\left(\chi_\infty - \chi\right) + l^2 \bT^{-\Trans} \Nab \cdot \left(D^{\text{pl}}\bT^{-\Trans} \Nab \chi\right).
    \label{eqn:chi_trans}
\end{equation}
For brevity, $\Dpl$, $\bsig_0$ and $D^{\text{pl}}$ refer to the physical quantities in Eq.~\ref{eqn:chi_trans} and must be computed in terms of the transformed variables in an implementation. Transformation of the diffusive term ensures that diffusion occurs in the physical frame despite being implemented in the reference frame.

Equation~\ref{eqn:Sig_evol} demonstrates that our transformation methodology leaves the Truesdell rate invariant and only affects the deformation rate term $\tC : \left(\tD - \Dpl\right)$. This highlights a benefit of using the Truesdell rate, as opposed to using alternative rates (\textit{e.g.}~Green--Naghdi or Jaumann) that employ physical approximations to achieve a simpler form. For example, the Jaumann stress rate is based upon the approximation that the effect of material stretch is much smaller than the effect of rotation, so that the Jaumann formula only involves the material spin rather than the full velocity gradient tensor. If the Jaumann rate is used in the physical frame it will not perfectly transform into the Jaumann rate in the reference frame, as neglecting stretch in the physical frame is not equivalent to neglecting stretch in the reference frame.

The transformed system of equations has connections to the principle of material frame-indifference~\cite{truesdell65,speziale98} which states that ``the constitutive laws governing internal interactions between the parts of the system should not depend on whatever external frame of reference is used to describe them''~\cite{noll95}. Mathematically this is done by considering a transformation of the form $\bx= \mathbf{R}(t) (\bX - \bX_0(t))$ where $\bX_0(t)$ is a time-dependent vector and $\mathbf{R}(t)$ is a time-dependent rotation~\cite{gurtin10}. If we restrict our transformation in Eq.~\ref{eqn:X} to the case when $\bT(t)$ is a rotation, then Eq.~\ref{eqn:Sig_evol} is identical to Eq.~\ref{eqn:orig_sig}, but in terms of transformed variables. The first four terms of Eq.~\ref{eqn:Sig_evol} are always identical, and the final term involving $\tC:(\tD-\Dpl)$ is also identical in this case, since the rotation matrices cancel because $\tC$ is isotropic. Hence our coordinate transformation is consistent with material frame-indifference.

It is worth considering how the transformed system of equations differs from the original system. A particular case of interest is simple shear, given the immediate application to implementation of Lees--Edwards boundary conditions. This physical situation is described by the transformation
\begin{equation}
    \bT = \begin{pmatrix}1 & 0 & U_b t\\ 0 & 1 & 0 \\ 0 & 0 & 1 \end{pmatrix},
    \label{eqn:shear_trans}
\end{equation}
with $U_b$ a boundary shear velocity. Restriction to a two-dimensional plane-strain formulation reveals that the components of Eqs.~\ref{eqn:V_evol} \& \ref{eqn:Sig_evol} retain their original form with untransformed quantities replaced by transformed quantities, in addition to several new terms proportional to powers of $U_b t$.

\subsection{Transformed projection method}
We now formulate the projection method of Sec.~\ref{ssec:orig_proj} in the reference frame. This method enables solving for $\vV$ and $\bSig$ subject to the constraint in Eq.~\ref{eqn:qs_const}. In the first step (analogous to Eq.~\ref{eqn:sproj1}), the $\tC : \tD$ term in Eq.~\ref{eqn:Sig_evol} is neglected to compute the intermediate transformed stress $\bSig^*$,
\begin{align}
    \frac{\bSig^{*} - \bSig^n}{\dt} &= -\left(\vV^n \cdot \Nab\right)\bSig^n - \tr(\bL^n)\bSig^n + \bSig^n\left(\Nab\vV\right)^n \nonumber\\
    &\phantom{=} + \left(\Nab\vV\right)^n\bSig^n - \left(\bT^{-1}\right)^n\tC :\left(\Dpl\right)^n\left(\bT^{-\Trans}\right)^n
    \label{eqn:Sig_adv_step}
\end{align}
If the transformed velocity at the next timestep $\vV^{n+1}$ were known, we could compute $\bL^{n+1}$ via Eq.~\ref{eqn:L_trans}, compute $\tD^{n+1}$, and complete the transformed Euler step via
\begin{equation}
  \frac{\bSig^{n+1} - \bSig^*}{\dt} = \left(\bT^{-1}\right)^n \left(\tC : \tD^{n+1}\right) \left(\bT^{-\Trans}\right)^n,
    \label{eqn:Sig_corr}
\end{equation}
which is analogous to Eq.~\ref{eqn:sproj2}. To compute this correction, we need to use the physical constraint Eq.~\ref{eqn:qs_const}. Enforcing that $\nab \cdot \bsig^{n+1} = \vec{0}$ leads to a linear system for $\vv$ in the physical domain given by
\begin{equation}
    \Delta t \nab\cdot\left(\tC : \tD^{n+1}\right) = -\nab \cdot \bsig^*.
    \label{eqn:mg_sys_phys}
\end{equation}
Because $\bT^{-1}\bsig^*\bT^{-\Trans} = \bSig^*$ and $\nabla_\bx = \bT^{-\Trans}\nabla_\bX$, the right-hand side of Eq.~\ref{eqn:mg_sys_phys} transforms according to
\begin{equation}
    -\nab\cdot\bsig^* = -\bT^n \Nab \cdot \bSig^*.
    \label{eqn:trans_RHS}
\end{equation}
Similarly, the left-hand side of Eq.~\ref{eqn:mg_sys_phys} becomes
\begin{equation}
    \dt\left(\bT^{-\Trans}\right)^n\Nab\cdot \tC : \bigg(\left(\bT^{-\Trans}\right)^n\left(\Nab\vV\right)^{n+1}\left(\bT^{\Trans}\right)^n\bigg),
    \label{eqn:trans_LHS}
\end{equation}
where we have omitted $\bX$-independent terms as they will be eliminated by $\Nab$. Equations \ref{eqn:trans_RHS} \& \ref{eqn:trans_LHS} form a complicated linear system for the transformed velocity $\vV^{n+1}$. The appearance of the transformation $\bT$ in front of the gradient operator $\Nab$ ensures that all mixed spatial derivatives of all components of the velocity appear in each row of Eq.~\ref{eqn:trans_LHS}. Equation \ref{eqn:trans_LHS} is more complex than the linear system in the original quasi-static projection method, and it is dependent on the specific form of $\bT$. The components of Eq.~\ref{eqn:trans_LHS} in the specific cases of simple shear and pure shear are shown in App.~\ref{app:ss} and App.~\ref{app:ps} respectively.

The update for the effective temperature is handled through an explicit forward Euler step,
\begin{align}
  c_0\frac{\chi^{n+1} - \chi^n}{\Delta t} &= -c_0\left(\vV^n \cdot \Nab\right) \chi^n + \frac{\left((\Dpl)^n : \bsig_0^{n}\right)}{s_Y}\left(\chi_\infty - \chi^n\right)\nonumber\\
    &\phantom{=}+ l^2 (\bT^{-\Trans})^n \Nab \cdot \left((\dpl)^n(\bT^{-\Trans})^n \Nab \chi^n\right).\label{eqn:chi_exp_step}
\end{align}

\subsection{Numerical discretization, parallelization, and multigrid solver}
\label{ssec:num}
The explicit update for the transformed stress Eq.~\ref{eqn:Sig_adv_step} depends on transformed spatial derivatives of the transformed velocity through $\bL$. Similarly, the source term in the linear system for the transformed velocity Eq.~\ref{eqn:trans_RHS} depends on transformed spatial derivatives of the transformed stress. We exploit this structure through a staggered grid arrangement in the reference frame with uniform spacing $\Delta x = \Delta y = \Delta z = h$. The stress tensor $\bSig$ and effective temperature $\chi$ are stored at cell centers and indexed by half-integers, while the velocity $\vV$ is stored at cell corners and indexed by integers. Further discussion of the staggered grid arrangement can be found in \cite{boffi2019-1}.

Let $(\p f /\p X)_{i, j, k}$ denote the partial derivative of a field $f$ with respect to $X$ evaluated at grid point $(i, j, k)$. The staggered centered difference is
\begin{align}
  \left(\prX{f}\right)_{i+\half, j+\half, k+\half} &= \frac{1}{4h} \Big(f_{i+1, j, k} - f_{i, j, k} + f_{i+1, j+1, k} - f_{i, j+1, k} \nonumber \\
  &\phantom{=} \ \ \ \ \ \ \ \ \ \ \ \ + f_{i+1, j, k+1} - f_{i, j, k+1} + f_{i+1, j+1, k+1} - f_{i, j+1, k+1}\Big).
    \label{eqn:stag_stenc}
\end{align}
Equation \ref{eqn:stag_stenc} averages four edge-centered centered differences surrounding the cell center and has a discretization error of size $\bigO(h^2)$. The derivative at a cell corner is obtained by the replacement \smash{$(i, j, k) \rightarrow (i - \frac{1}{2}, j - \frac{1}{2}, k - \frac{1}{2})$}. The diffusive term appearing in the effective temperature update in Eq.~\ref{eqn:chi_exp_step} is computed by expanded the divergence term,
\begin{align}
    \bT^{-\Trans}\Nab \cdot\left(\dpl\bT^{-\Trans}\Nab\chi\right) &= \left(\Nab \dpl\right)\cdot\left[\left(\bT^{-1}\bT^{-\Trans}\right)\Nab\chi\right] \nonumber\\&\phantom{=}\ \ \ \ \ \ \ \ \ + \dpl\left[\left(\bT^{-1}\bT^{-\Trans}\right) : \left(\Nab \Nab \chi\right)\right].
    \label{eqn:trans_diff_num}
\end{align}
Equation \ref{eqn:trans_diff_num} is computed numerically by assembling the gradient vectors $\Nab \chi$ and $\Nab \dpl$ at cell centers using the standard centered difference formula,
\begin{align}
    \left(\prX{f}\right)_{i, j, k} &= \frac{1}{2h}\Big(f_{i + 1, j, k} - f_{i-1, j, k}\Big)
    \label{eqn:cent_stenc},
\end{align}
with analogous expressions for the other directions. We also must assemble the Hessian matrix $\Nab \Nab \chi$ using the second derivative stencils
\begin{align}
    \left(\frac{\p^2 f}{\p X^2}\right)_{i, j, k} &= \frac{1}{h^2}\Big(f_{i+1, j, k} - 2f_{i, j, k} + f_{i-1, j, k}\Big)
    \label{eqn:cent_2stenc},\\
    \left(\frac{\p^2 f}{\p X \p Y}\right)_{i, j, k} &= \frac{1}{h^2}\Big(f_{i+1, j+1, k} - f_{i+1, j-1, k} - f_{i-1, j+1, k} + f_{i-1, j-1, k}\Big).
    \label{eqn:cent_xy_stenc}
\end{align}
Analogous expressions for other second derivatives are obtained through Eqs.~\ref{eqn:cent_2stenc} \&~\ref{eqn:cent_xy_stenc} by suitable replacements. The matrix $\bT^{-1}\bT^{-\Trans}$ is computed from its definition.

The advective derivative in Eq.~\ref{eqn:Sig_adv_step} must be upwinded for stability; we use the second-order essentially non-oscillatory (ENO) scheme~\cite{shu88}. With $\left[f_{XX}\right]_{i, j, k}$ denoting the second derivative with respect to $X$ of the field $f$ at grid point $(i, j, k)$ computed using Eq.~\ref{eqn:cent_2stenc}, the ENO derivative is defined in the $X$ direction as
\begin{equation}
    \left(\prX{f}\right)_{i, j, k} = \frac{1}{2h}
    \begin{cases}
        -f_{i+2, j, k} + 4f_{i+1, j, k} - 3f_{i, j, k} &\text{if $U_{i, j, k} < 0$ and $\left|\left[f_{XX}\right]_{i, j, k}\right| > \left|\left[f_{XX}\right]_{i+1, j, k}\right|$,}\\
        3f_{i, j, k} - 4f_{i-1, j, k} + f_{i-2, j, k} &\text{if $U_{i, j, k} > 0$ and $\left|\left[f_{XX}\right]_{i, j, k}\right| > \left|\left[f_{XX}\right]_{i-1, j, k}\right|$,}\\
        f_{i+1, j, k} - f_{i-1, j, k} &\text{otherwise}.
    \end{cases}
    \label{eqn:ENO}
\end{equation}
Above, $U_{i, j, k}$ is the $X$ component of the transformed velocity at grid point $(i, j, k)$. Equation \ref{eqn:ENO} uses the curvature of $f$ to switch between an upwinded three-point derivative and a centered difference. Versions of Eq.~\ref{eqn:ENO} in the $Y$ and $Z$ coordinates are obtained analogously.

Despite its complexity, after spatial discretization of Eq.~\ref{eqn:trans_LHS}, the linear system is of the form $\mathbf{A}\mathbf{y} = \mathbf{b}$, and can be solved via standard techniques of numerical linear algebra. $\mathbf{b}$ is given in block form by the source term in Eq.~\ref{eqn:trans_RHS}, $\mathbf{b}_i = -\bT\Nab\cdot \bSig^*(\bX_i)$, where the index $i$ runs over all grid points. $\mathbf{y}$ is also given in block form, so that $\by$ contains the stacked values of $\vV$ across all grid points. The matrix $\mathbf{A}$ is sparse, and its degree of sparsity depends on the specific discretization scheme used. In the staggered centered difference scheme described here, grid point $(i, j, k)$ is only coupled to the 27 grid points in the surrounding $3\times 3\times 3$ cube.

$\mathbf{A}$ is thus most effectively reconstructed using submatrices \smash{$\mathbf{A}^{(i, j, k)}_{(k, l, m)}$}, which give the coefficients of velocity values \smash{$\vV_{(k, l, m)}$} appearing in the linear equation for \smash{$\vV_{(i, j, k)}$}. Each matrix \smash{$\mathbf{A}^{(i, j, k)}_{(k, l, m)}$} is symmetric. With this construction, we solve Eq.~\ref{eqn:trans_LHS} using a custom MPI-based parallel geometric multigrid solver; for further details of the solver, and how it interfaces with the explicit updates, the reader is referred to the non-transformed algorithm description \citep{boffi2019-1}. The explicit steps for $\chi$ and $\bSig$ in Eqs.~\ref{eqn:Sig_adv_step} \&~\ref{eqn:chi_exp_step} are also parallelized using MPI and domain decomposition, with further details in the non-transformed work \citep{boffi2019-1}.

A highlight of the transformation methodology is its flexibility and simplicity. Implementation of new boundary conditions, so-long as they can be specified in terms of a transformation $\bT(t)$, is only as difficult as writing the transformation down. The matrices \smash{$\mathbf{A}^{(i, j, k)}_{(k, l, m)}$} do, however, depend on the form of $\bT(t)$, and thus they need to be derived on a transformation-by-transformation basis. Furthermore, through their dependence on $\bT(t)$, these submatrices are time-dependent and thus need to be recomputed at each timestep.

For an arbitrary $3\times 3$ transformation with nine matrix elements, the analytical computation and hand-implementation of the \smash{$\mathbf{A}^{(i, j, k)}_{(k, l, m)}$} matrices is error-prone. To remedy this, we developed a metaprogramming scheme to auto-generate the relevant code. We used Mathematica to analytically calculate Eq.~\ref{eqn:trans_LHS} in terms of arbitrary matrix elements $T_{ij}(t)$, and subsequently to replace derivatives by their finite difference equivalents. Collecting coefficients accordingly in the resulting equation gives $191$ non-zero coefficients comprising the $27$ submatrices \smash{$\mathbf{A}^{(i, j, k)}_{(k, l, m)}$}. We used Python to write a skeleton file that contained function primitives for $191$ \texttt{C++} functions to compute each of these coefficients individually. We then used the Mathematica function \texttt{splice} to fill in valid \texttt{C++} code that calculates the resulting coefficients in each of these functions. Finally, we again used Python to write \texttt{C++} code that calls the auto-generated \texttt{C++} functions to populate the submatrices. The metaprogramming scheme only needs to be run once to generate the needed code, and it does not take any meaningful amount of time to run. In this way, the multigrid system is automatically generated at each timestep, and new simulation conditions can be immediately constructed by providing the matrix $\bT(t)$ as a $3\times 3$ matrix class implemented in \texttt{C++}.

\section{Numerical convergence tests}
\label{sec:numerical_tests}
\begin{table}
    \caption{Material parameters used in this study, for both linear elasticity and the STZ model of amorphous plasticity. The Boltzmann constant $k_B$ is used to convert energetic values to temperatures.}
    \vspace{5mm}
    \label{table:params}
    \centering
    \begin{tabular}{ll}
    \hline
        Parameter & Value \\ \hline
        Young's modulus $E$ & 101~GPa\\
        Poisson ratio $\nu$ &  0.35\\
        Bulk modulus $K$ & 122~GPa\\
        Shear modulus $\mu$ & 37.4~GPa\\
        Density $\rho_0$ & $6125\text{~kg~m}^{-3}$\\
        Shear wave speed $c_s$ & $2.47\text{~km~s}^{-1}$\\
        Yield stress $s_Y$ & 0.85~GPa\\
        Molecular vibration timescale $\tau_0$ & $10^{-13}$~s\\
        Typical local strain $\epsilon_0$ & 0.3\\
        Effective heat capacity $c_0$ & 0.4\\
        Typical activation barrier $\Delta/k_B$ & 8000~K\\
        Typical activation volume $\Omega$ & 300~\AA${}^3$\\
        Thermodynamic bath temperature $T$ & 400~K\\
        Steady state effective temperature  $\chi_\infty$ & 900~K\\
        STZ formation energy $e_z/k_B$ & 21000~K
    \end{tabular}
\end{table}
In this section, we demonstrate convergence of the transformed projection method to the non-transformed method in physically equivalent situations. In all simulations, a periodic domain in $X$ and $Y$ is considered, $-L \leq X < L$, $-L \leq Y < L$ with $L = 1\text{~cm}$. We consider both periodic and non-periodic boundary conditions in $Z$, corresponding to domains $Z\in [-\gamma L,\gamma L)$ and $Z\in [-\gamma L, \gamma L]$, respectively. \smash{$\gamma = \frac{1}{2}$} in all simulations. We measure time in terms of the natural unit $t_s = L/c_s$ with \smash{$c_s = \sqrt{\mu/\rho}$} the material shear wave speed. Boundary conditions in the non-periodic case are given by
\begin{equation}
    \vV(X, Y, \pm \gamma L, t) = (0, 0, 0).
    \label{bdry:dirich}
\end{equation}
Elasticity and plasticity parameters are provided in Table~\ref{table:params}, and for these parameters, $t_s = 4.05 \text{~\textmu{}s}$. All simulations in this section are run with $32$ processes on an Ubuntu Linux computer with dual 14-core 1.70~GHz Intel Xeon E5-2650L v4 processors.

The global three-dimensional grid has spacing $h$ in each direction. The cell-cornered grid points are indexed according to $i = 0, \hdots, Q-1$, $j = 0, \hdots, M-1$ in the $X$ and $Y$ directions. In the $Z$ direction, the grid points are indexed according to $k = 0, \hdots, N$ and $k = 0, \hdots, N-1$ for non-periodic and periodic boundary conditions, respectively. The cell-centered grid points run according to \smash{$i = \frac{1}{2}, \frac{3}{2}, \hdots Q - \frac{1}{2}$}, \smash{$j = \frac{1}{2}, \frac{3}{2}, \hdots M - \frac{1}{2}$}, and \smash{$k = \frac{1}{2}, \frac{3}{2}, \hdots M - \frac{1}{2}$}. As described in Sec.~\ref{ssec:num}, $\bSig$ and $\chi$ are stored at cell centers while $\vV$ is stored at cell corners. The additional grid points $(i, j, k=N)$ in the $Z$ direction in the non-periodic case are ghost points used for enforcing the Dirichlet boundary conditions $\vV = \mathbf{0}$.

The cell-centered grid points on the top boundary $(i, j, N+\frac{1}{2})$ contain linearly extrapolated $\bSig$ and $\chi$ values to ensure that $\bSig$ and $\chi$ remain free on the top boundary. In the periodic case, the grid points $(i, j, k=N)$ hold the velocity values $\vV_{(i, j, 0)}$, and the corresponding cell-centered grid points are used to hold the wrapped values of $\bSig_{(i, j, \frac{1}{2})}$ and $\chi_{(i, j, \frac{1}{2})}$. At the simulation boundaries in the $X$ and $Y$ directions, ghost points leaving the simulation domain are filled with values that wrap around, so that the ghost point corresponding to grid point $(Q, j, k)$ is filled with the real values from grid point $(0, j, k)$. Similarly, values at points $(i, M, k)$ are filled using values from $(i,0,k)$.

\subsection{Qualitative comparison between the transformed and non-transformed methods}
\label{ssec:qual_compare}
We now demonstrate the qualitative similarity of solutions computed with the transformed and the standard quasi-static methods. In the following subsection, this comparison is made quantitatively rigorous. To visualize the results three-dimensionally, we use a custom opacity function,
\begin{equation}
    O(\bx) =
    \begin{cases}
        \left(\frac{\chi(\bx) - \chi_{bg}}{\chi_\infty - \chi_{bg}}\right) & \qquad \text{if $\frac{\chi(\bx) - \chi_{bg}}{\chi_\infty - \chi_{bg}} > \frac{3}{4},$} \\
        \exp\left(-a \left(\frac{\chi_\infty - \chi_{bg}}{\chi(\bx) - \chi_{bg}}\right)^\eta\right) & \qquad \text{otherwise,}
    \end{cases}
    \label{eqn:opac}
\end{equation}
where $\chi_{bg}$ is a background $\chi$ value. By choice of $a$ and $\eta$, the most physically relevant features in three-dimensional visualizations of the $\chi$ field can be revealed.

\begin{figure*}
    \centering
    \fcolorbox{black}{black}{\includegraphics[width=\subpanelwid]{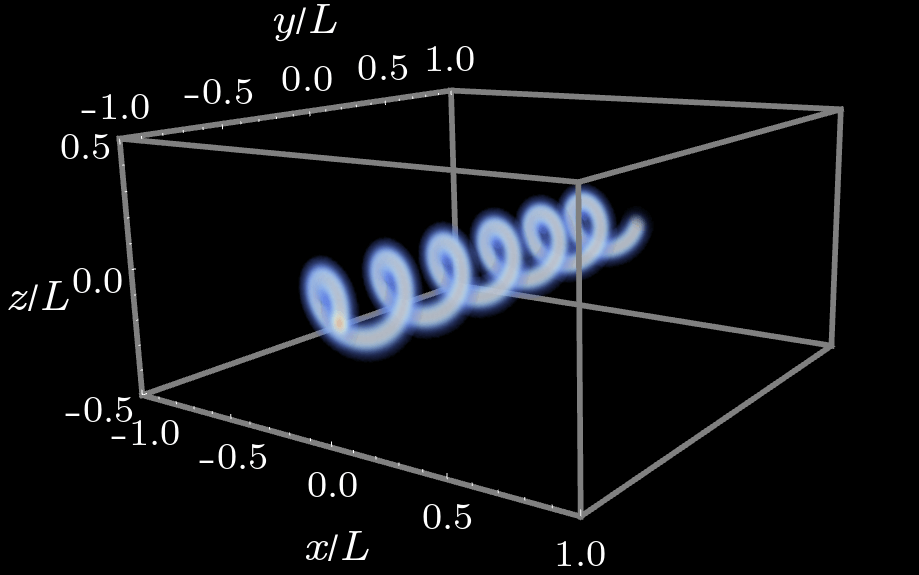}}
    ~ \vspace{5mm} \\
    \includegraphics[width=.75\textwidth]{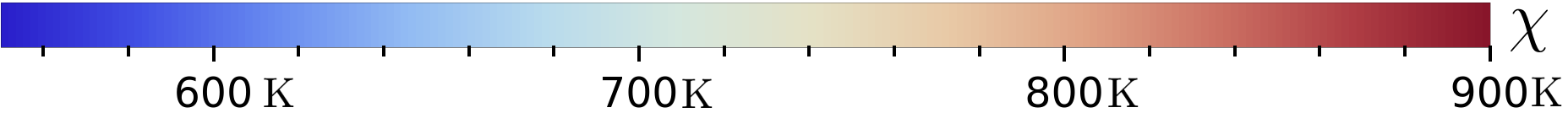}
    \caption{The initial configuration for the transformed to non-transformed comparison. Here, $a = 0.3$ and $\eta = 1.2$ in the opacity function, and $\chi_{bg} = 550\text{~K}$.}
    \label{fig:tr_ph_init}
\end{figure*}

To compare the transformed and non-transformed methods, a physically equivalent situation is now constructed. We employ non-periodic Dirichlet boundary conditions in the $Z$-direction and enforce $\vV(X, Y, \pm \gamma L) = (0, 0, 0)$. To impose deformation, we use a shear transformation $\bT(t)$ corresponding to
\begin{equation}
    \bT = \begin{pmatrix}1 & 0 & \frac{U_b}{\gamma L}t\\ 0 & 1 & 0 \\ 0 & 0 & 1 \end{pmatrix}.
    \label{eqn:shear_trans_comp}
\end{equation}
Boundary conditions in the non-transformed simulation correspond to shearing between two parallel plates, $\vv(x, y, \pm \gamma L) = (U_b, 0, 0)$. An initial linear velocity gradient is imposed in the non-transformed frame, so that
\begin{equation}
    \vv\left(\bx, t=0\right) = \left(\frac{U_B z}{\gamma L}, 0, 0\right).
    \label{eqn:phys_vel_grad}
\end{equation}
Equation \ref{eqn:phys_vel_grad} ensures equivalent initial conditions in both methodologies, and also prevents the introduction of large gradients in the deformation rate near the boundary. To create interesting dynamics, an initial condition in $\chi$ corresponding to a helix oriented perpendicular to the direction of shear is considered. This is represented as
\begin{align}
    \delta_x &= \frac{x}{L} - \left(\frac{\cos\left(6\pi\left(\frac{y}{L} + 1\right)\right)}{8} - \frac{1}{16}\right),\nonumber\\
    \delta_z &= \frac{z}{L} - \left(\frac{\cos\left(6\pi\left(\frac{y}{L} + 1\right)\right)}{8} - \frac{1}{16}\right),\nonumber\\
    \chi\left(\bx, t=0\right) &= 600\text{~K} + \left(200\text{~K}\right)e^{-750\left(\delta_x^2 + \delta_z^2\right)}.
    \label{eqn:chi_init_helix}
\end{align}
Equation \ref{eqn:chi_init_helix} is written in the non-transformed simulation, but the same initial conditions are used in the transformed simulation with the substitution $\bx \rightarrow \bX$. The configuration is visualized in the reference frame in Fig.~\ref{fig:tr_ph_init}.

\begin{figure*}
\fcolorbox{black}{black}{
    \begin{tabular}{cc}
        \includegraphics[width=\subpanelwid]{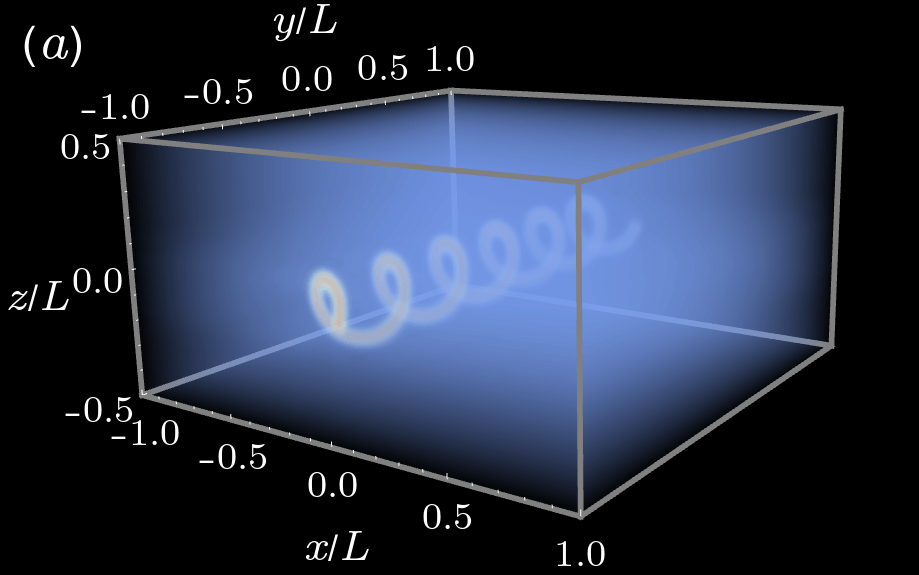} &
        \includegraphics[width=\subpanelwid]{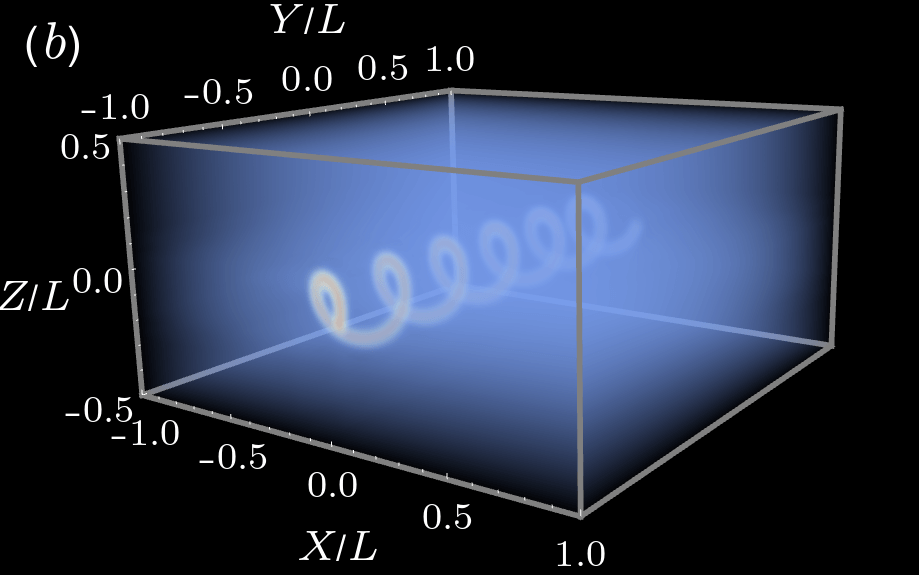} \\
        \includegraphics[width=\subpanelwid]{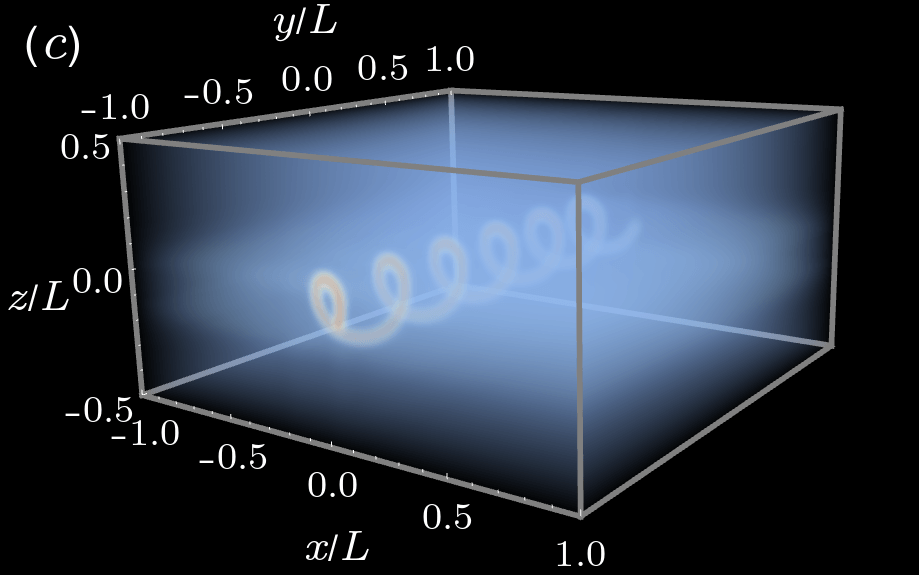} &
        \includegraphics[width=\subpanelwid]{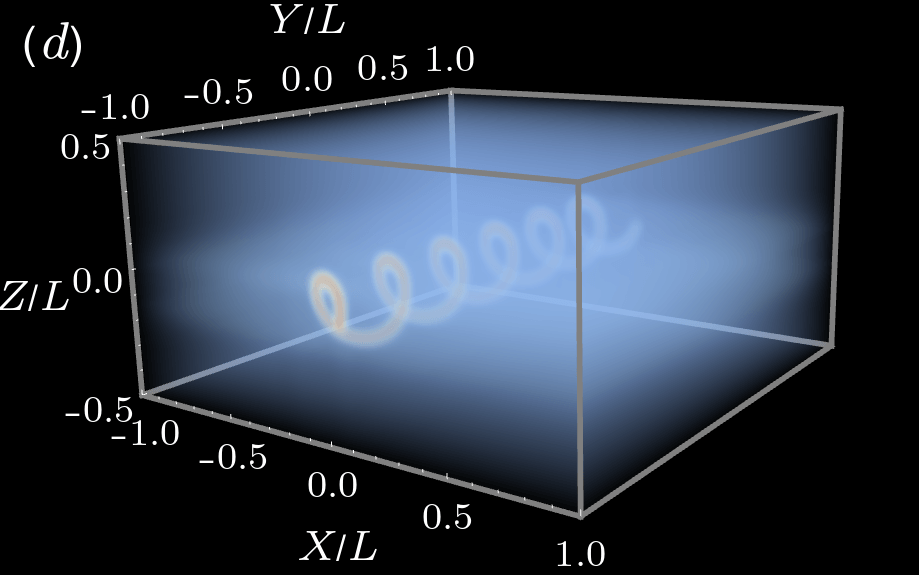} \\
        \includegraphics[width=\subpanelwid]{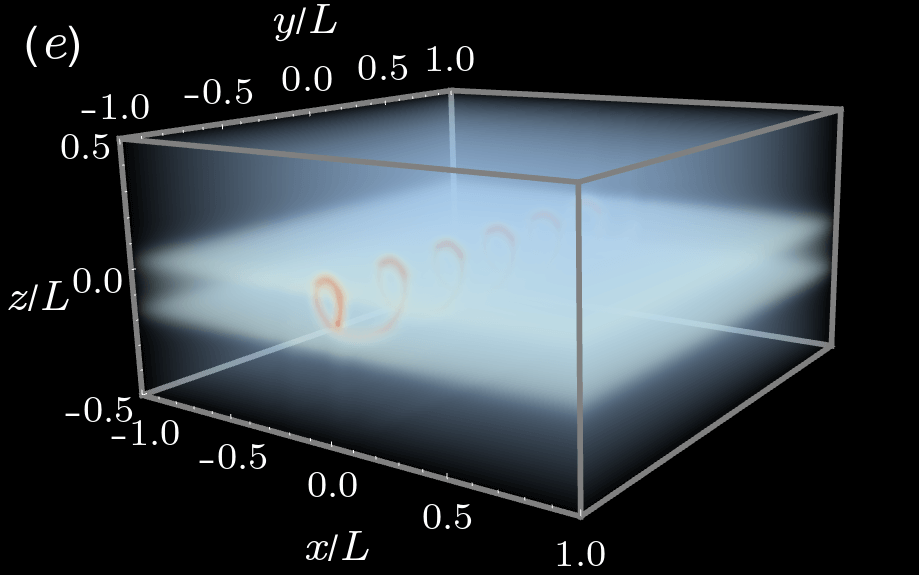} &
        \includegraphics[width=\subpanelwid]{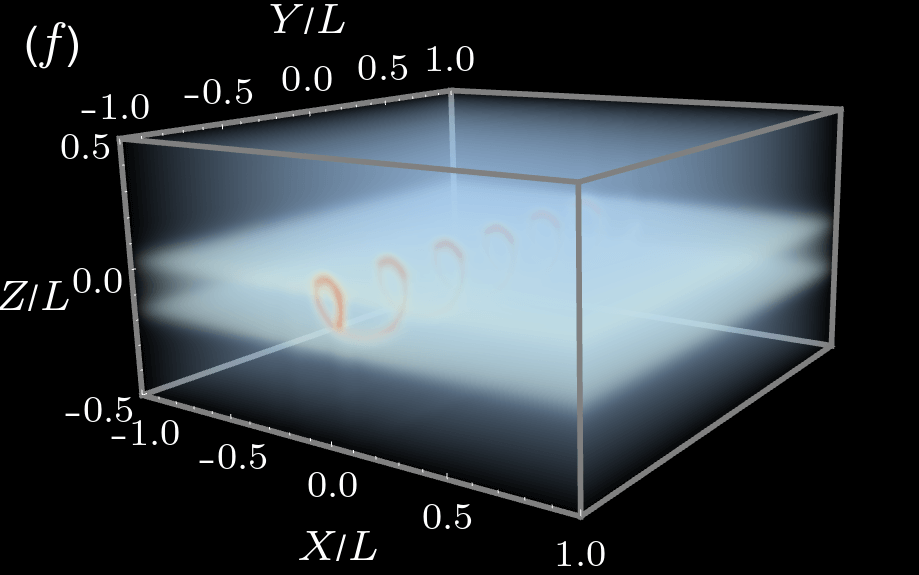}
    \end{tabular}}
    \vspace{5mm} \\
    \includegraphics[width=.75\textwidth]{imgs/rslt_figs/colorbar_2019}
    \caption{Snapshots of the effective temperature field $\chi(\bx, t)$ for the (a,c,e) non-transformed and (b,d,f) transformed simulation . $\chi_{bg} = 550\text{~K}$ in the opacity function in all panels. (a,b) $t = 2.88\times 10^5 t_s$, $a = 0.7$, $\eta = 1.25$. (c,d) $t = 4.02\times 10^5 t_s$, $a=0.8$, $\eta = 1.35$. (e,f) $t = 6\times 10^5 t_s$, $a = 0.9$, $\eta = 1.5$.}
    \label{fig:tr_ph_qual}
\end{figure*}

The simulations are conducted on two grids of size $256\times 256\times 128$ with a quasi-static timestep $\Delta t = 31.25 t_s$ and with a value of \smash{$U_b = 10^{-7} \frac{L}{t_s}$}. Snapshots at three representative time points are shown in Fig.~\ref{fig:tr_ph_qual}. In Fig.~\ref{fig:tr_ph_qual}~(a,b) at $t = 2.88\times 10^5 t_s$, shear band nucleation has not begun, and there is an increase in the $\chi$ field across the entire domain. At $t = 4.02\times 10^5$ in Fig.~\ref{fig:tr_ph_qual}~(c,d), shear bands have begun to nucleate along the top and bottom planes of the helices. At $t = 6 \times 10^5 t_s$ in Fig.~\ref{fig:tr_ph_qual}(e,f), the bands have grown sharper, stronger, and span the system. In all cases, the qualitative agreement is very good.

\subsection{Quantitative comparison between the transformed and non-transformed methods}
\label{sec:quan_compare}
Having demonstrated the qualitative similarity between the solutions computed by the transformed and non-transformed methods, we now present a rigorous quantitative comparison. We utilize the same simulation geometry, boundary conditions, shear transformation, and initial conditions as in Sec.~\ref{ssec:qual_compare}. We introduce a norm over simulation fields,
\begin{equation}
    \Vert\mathbf{f}\Vert(t) = \sqrt{\frac{1}{8\gamma L^3}\int_{-\gamma L}^{\gamma L}dZ\int_{-L}^LdY\int_{-L}^LdX |\mathbf{f}(\bX, t)|^2},
    \label{eqn:norm}
\end{equation}
where the integral in Eq.~\ref{eqn:norm} runs over the entire simulation domain and is numerically computed using the trapezoid rule. The appearance of $|\cdot|$ in Eq.~\ref{eqn:norm} is interpreted as the two-norm for vectors, absolute value for scalars, and the Frobenius norm for matrices.  With subscript NT denoting ``non-transformed'' and subscript T denoting ``transformed'', Eq.~\ref{eqn:norm} is applied to the quantities $\vv(\bX, t)_{\text{NT}} - \vv(\bX, t)_{\text{T}}$, $\bsig(\bX, t)_{\text{NT}} - \bsig(\bX, t)_{\text{T}}$ and $\chi(\bX, t)_{\text{NT}} - \chi(\bX, t)_{\text{T}}$. The \textit{physical} field values are compared across the \textit{reference} grid, a procedure that involves two subtleties.

In the transformed simulation, this comparison requires computing $\bsig$ from $\bSig$ and $\vv$ from $\vV$ using Eqs.~\ref{eqn:Sig} and \ref{eqn:V} respectively at all reference grid points. In the non-transformed simulation, it is necessary to compute the non-transformed field values at reference grid points. Because the reference grid maps to a sheared physical grid, these values may not be defined in the non-transformed simulation. We handle this via the following procedure. The non-transformed simulation grid point $\bx(\bX)$ corresponding to the reference grid point $\bX$ is first computed. If $\bx(\bX)$ does not lie on the non-transformed grid, adjacent grid points are linearly interpolated to compute an approximate field value at $\bx$. This procedure incurs an $\mathcal{O}(h^2)$ error, which is the same order of accuracy as the centered differences used for spatial discretization in the two methods. As the sizes of the simulation grids are increased, the discrepancy in solutions will decrease.

To ensure that issues with temporal discretization do not affect the comparison, it is also necessary to scale the quasi-static timestep as the grid size is decreased. Because the spatial order of accuracy is $\mathcal{O}(h^2)$, we keep the ratio $\Delta t/h^2$ fixed across all simulations. We perform comparisons across grids of size $N\times N \times \frac{N}{2}$ with $N = 64$, $96$, $128$, $160$, $192$, and $256$. Respectively, these correspond to grid spacings $L/32$, $L/48$, $L/64$, $L/80$, $L/96$, and $L/128$. The quasi-static timestep is taken to be $\Delta t = 500 t_s$ for the coarsest simulation, leading to quasi-static timesteps $\Delta t = 222.14, 125, 80, 55.55$, and $31.25$ respectively for the finer simulations. The diffusion length scale in the effective temperature equation is taken to be zero in all simulations for the purpose of the comparison.

\begin{figure*}
    \begin{tabular}{cc}
        \includegraphics[width=\subpanelwid]{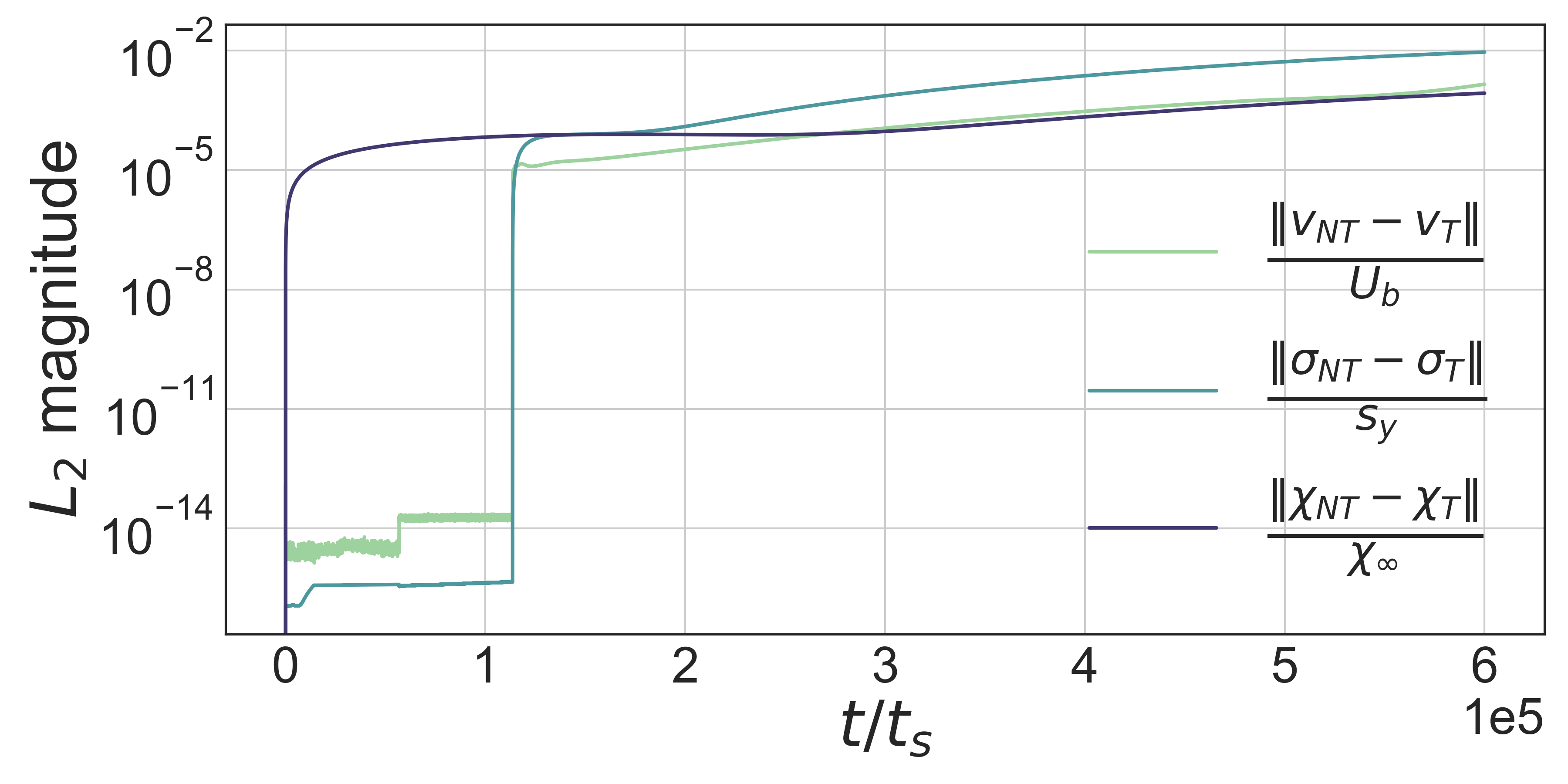} &
        \includegraphics[width=\subpanelwid]{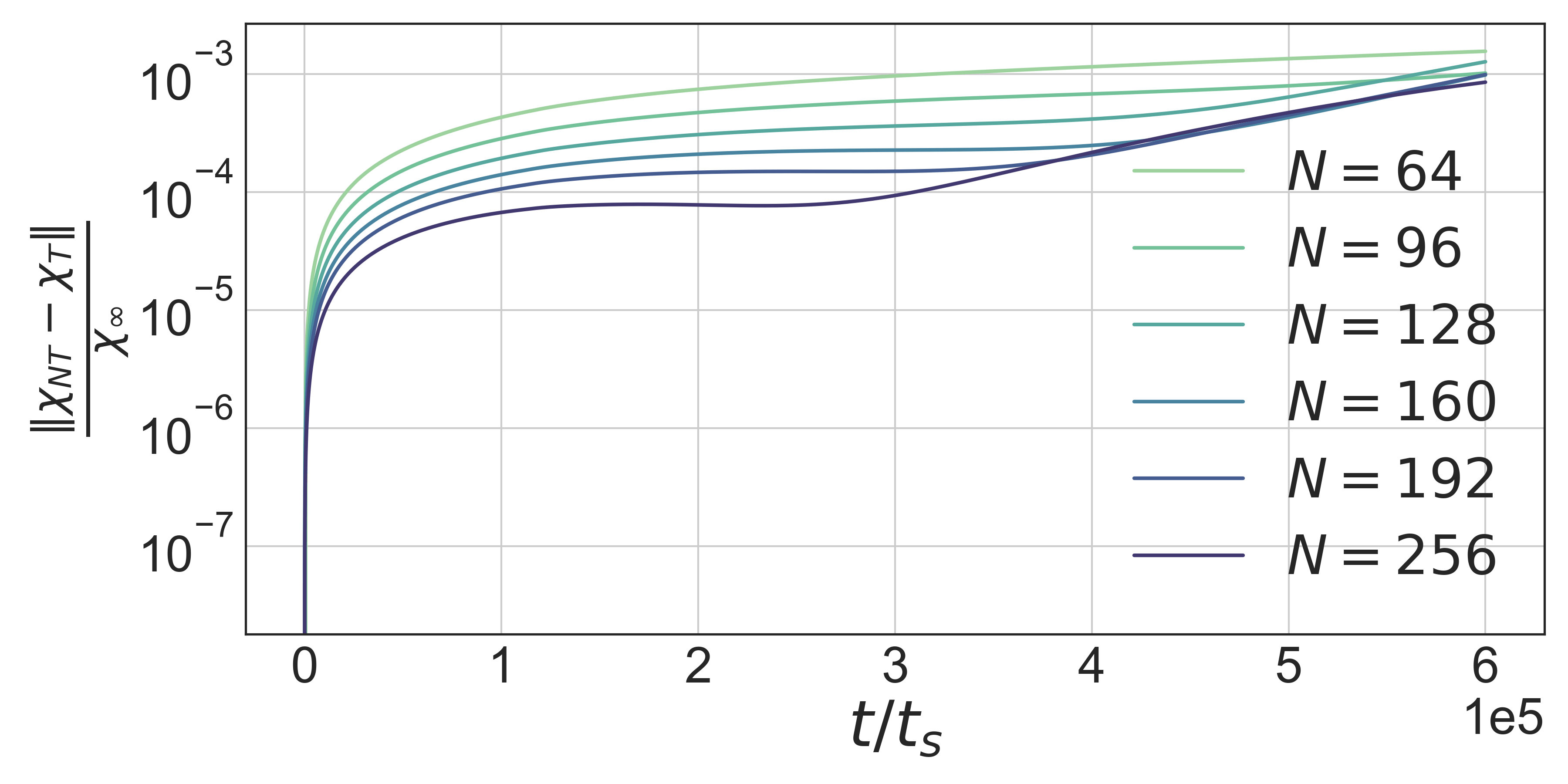} \\
        \includegraphics[width=\subpanelwid]{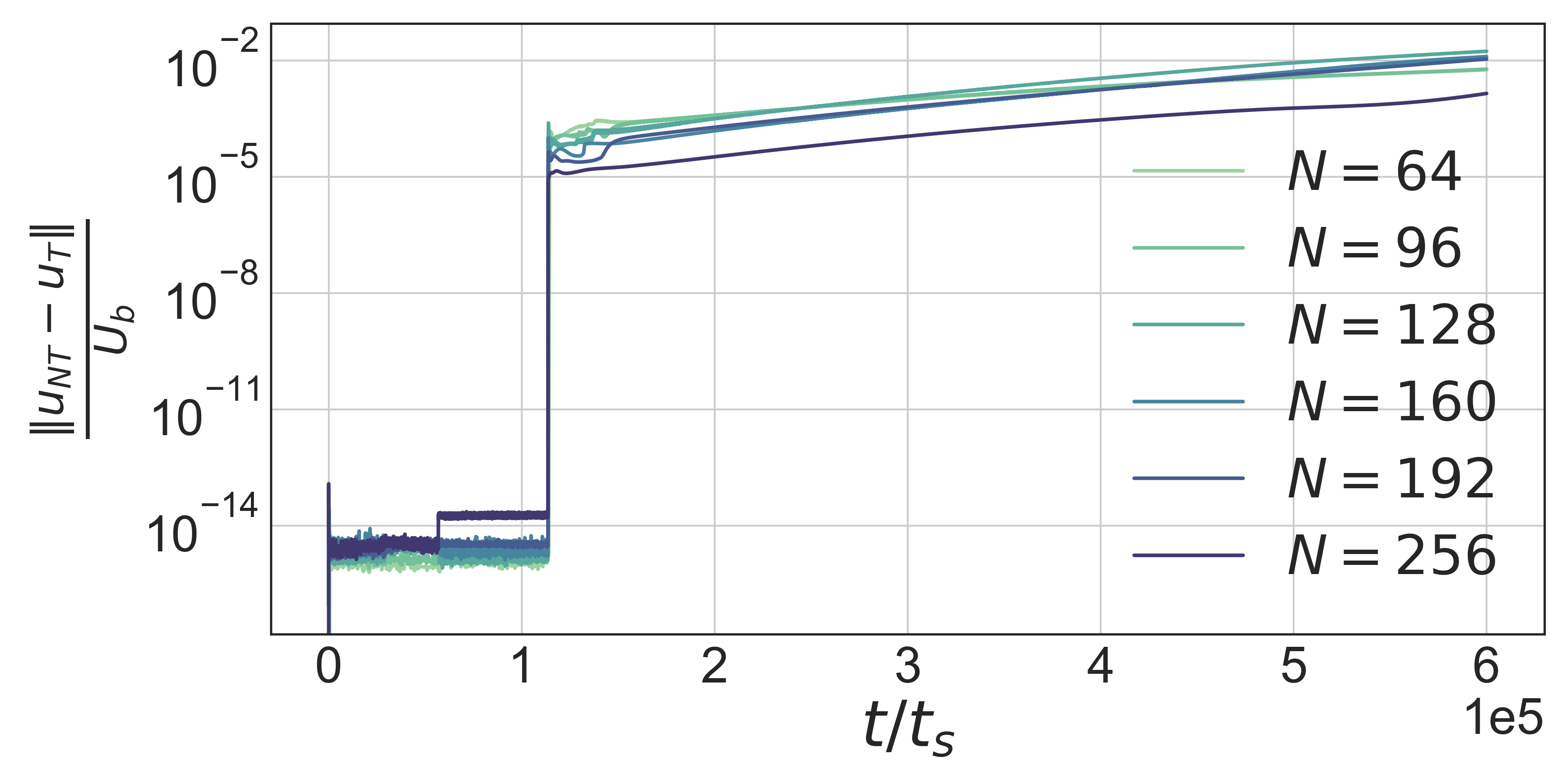} &
        \includegraphics[width=\subpanelwid]{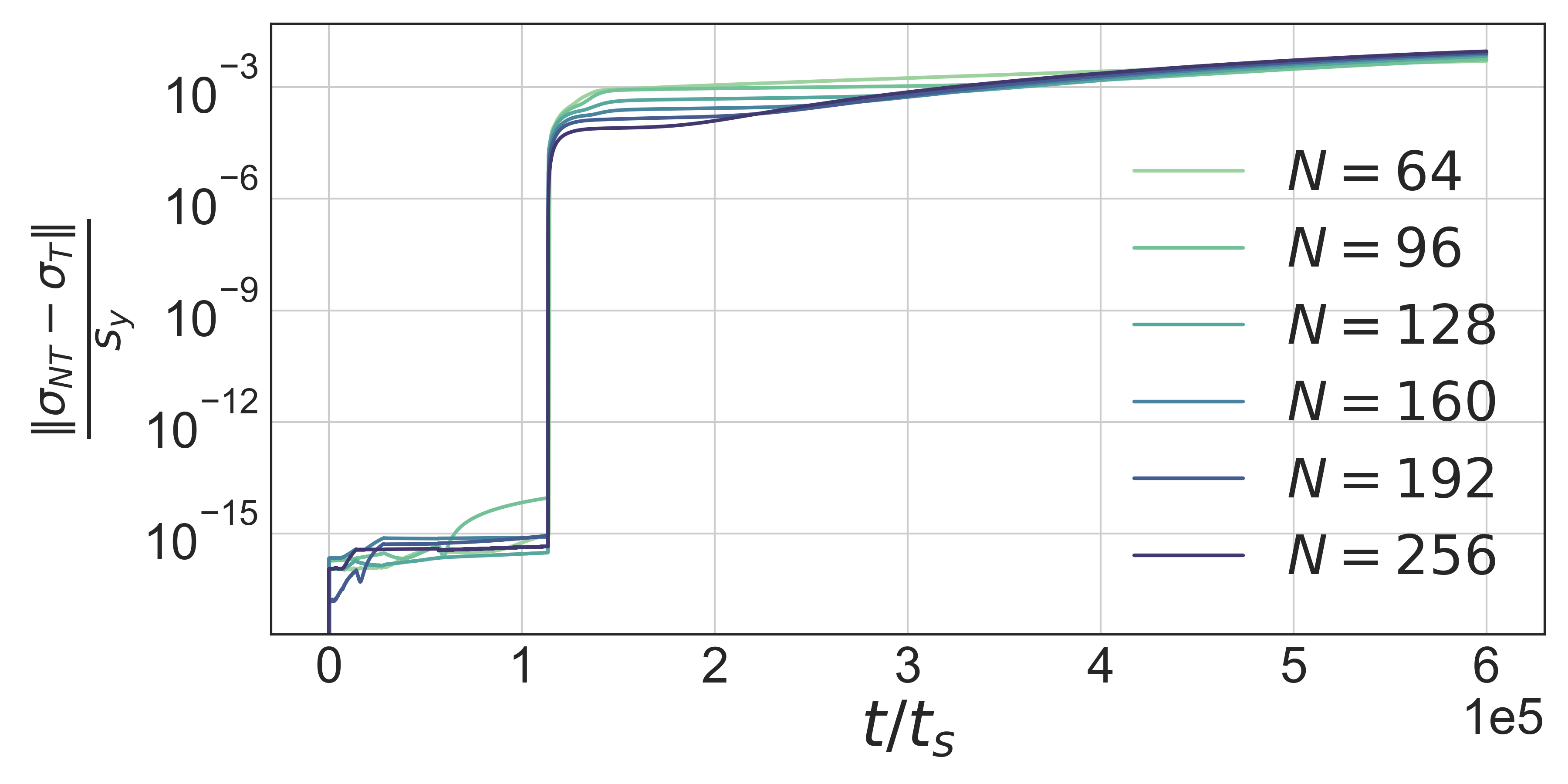}
    \end{tabular}
    \caption{$L_2$ norm of the $\chi$, $\vv$, and $\bsig$ simulation field differences between the transformed and non-transformed methods computed using Eq.~\ref{eqn:norm} in a simple shear simulation. (top left) A comparison of the three different field norms on a grid of size $256\times 256\times 128$. (top right), (bottom left), (bottom right) The velocity, effective temperature, and stress norm differences respectively for varying levels of discretization $N = N_x = N_y = 2N_z$.}
    \label{fig:tr_ph}
\end{figure*}

The results for the quantitative comparisons are shown in Fig.~\ref{fig:tr_ph}. In Fig.~\ref{fig:tr_ph} (top left), the three $L_2$ norm curves are plotted together for a value of $N = 256$, where each curve is normalized by a representative value in order to plot on a comparable dimensionless scale. The effective temperature norm increases rapidly early on in the simulation, but then saturates around $10^{-4}$. The $\bsig$ norm stays around machine precision until the onset of plasticity, when it rapidly increases and then saturates around $10^{-3}$. Similarly, the $\vv$ norm stays below $10^{-13}$ until the onset of plasticity, when it rapidly increases and then saturates around $10^{-4}$. The agreement up to machine precision prior to the onset of plasticity is expected, and validates the accuracy of the derivation of the equations in the reference frame.

In Fig.~\ref{fig:tr_ph} (top right), the effective temperature norm curves are shown for all values of $N$. Here, there is a steady increase in the discrepancy before the onset of plasticity due to advection across the grid. After plasticity is activated around $t = 1.2\times 10^5 t_s$, there is a period of saturation in all curves, followed by a period of increase beginning around $t \approx 3\times 10^5 t_s$, where some simulation curves cross and end at roughly equal values. As expected, the discrepancies generally decrease as the grid spacing is decreased.

In Fig.~\ref{fig:tr_ph} (bottom left), the velocity norm curves are shown as a function of time for all discretization levels. In all cases, the difference between the simulation methods is on the order of machine precision until the onset of plasticity, when there is a sharp and immediate jump. The size of the jump decreases with the discretization level as expected.

In Fig.~\ref{fig:tr_ph} (bottom right), the stress norm curves are shown. These curves display a combination of the trends in the velocity and effective temperature plots. Before the onset of plasticity, the error in all simulations is very low - on the order of machine precision. After the onset of plasticity, there is a sharp jump in all simulations, and the size of the jump decreases with higher resolution. Past around $t \approx 2 \times 10^5 t_s$, the curves begin to cross, all ending at roughly equivalent values.

\begin{table}
\centering
\resizebox{\textwidth}{!}{%
\begin{tabular}{|c|c|c|c|c|c|c|c|c|c|c|c|c|}
\hline
 & \multicolumn{2}{c|}{$N=64$} & \multicolumn{2}{c|}{$N=96$} & \multicolumn{2}{c|}{$N=128$} & \multicolumn{2}{c|}{$N=160$} & \multicolumn{2}{c|}{$N=192$} & \multicolumn{2}{c|}{$N=256$} \\ \hline
 & T & NT & T & NT & T & NT & T & NT & T & NT & T & NT \\ \hline
    Total time (hours) & 0.0633 & 0.0343 & 0.5623 & 0.2863 & 2.4283 & 1.1981 & 3.3058 & 1.7890 & 8.5285 & 4.5787 & 33.3239 & 20.0178 \\ \hline
V-cycle time (hours) & 0.0452 & 0.0280 & 0.4130 & 0.2434 & 1.7845 & 1.0365 & 2.1976 & 1.3251 & 5.7242 & 3.4398 & 21.2268 & 15.2432 \\ \hline
\# of V-cycles & 5544 & 3603 & 12481 & 8106 & 22181 & 14402 & 34658 & 22502 & 49913 & 32404 & 73164 & 57600 \\ \hline
Time/V-cycle (seconds) & 0.0294 & 0.0280 & 0.1191 & 0.1081 & 0.2896 & 0.2591 & 0.2282 & 0.2120 & 0.4129 & 0.3810 & 1.0444 & 0.9527 \\ \hline
\end{tabular}%
}
\caption{Data describing the total time taken, the total amount of time spent in multigrid V-cycles, the total number of multigrid V-cycles, and the average time per V-cycle for the two simulation approaches. ``T'' specifies the transformed simulation and ``NT'' the non-transforemd simulation. The transformed method takes longer than the non-transformed method in general due to an increased number of multigrid V-cycles required to achieve convergence. The average time spent per V-cycle is roughly the same in the two approaches. Each simulation uses $32$ processes.}
\label{tab:time_data}
\end{table}

To compare the computational efficiency of the two methods, we have reported timing statistics for all simulations in Tab.~\ref{tab:time_data}. Displayed are the total time, the total number of multigrid V-cycles, the total time spent in multigrid V-cycles, and the average time per V-cycle for the transformed (T) and non-transformed (NT) methods. As is clear from the table, the transformed method incurs a mild increase in computational expensive. The average time spent per V-cycle is roughly the same, but the total number of multigrid V-cycles is higher for the transformed method. This is likely due to the increased complexity of the linear system required for the stress projection in the transformed method when compared to the non-transformed method.

\section{Numerical examples}
\label{sec:examples}

\subsection{Simple shear and the effect of Lees--Edwards boundary conditions}
\label{ssec:LE}
As a first example application of the transformation method, we consider connecting a continuum-scale model to typical discrete molecular dynamics simulations. A significant difference between continuum simulation and molecular dynamics is in the boundary conditions. Molecular dynamics simulations commonly employ Lees--Edwards boundary conditions, where periodic copies of the system are placed above and below with a prescribed horizontal velocity. Continuum-scale boundary conditions usually set a shear velocity on the top and bottom boundaries to achieve the same effect.

Lees--Edwards boundary conditions can be implemented in the continuum through the use of the coordinate transformation methodology presented here, by combining a shear transformation $\bT(t)$ as in Eq.~\ref{eqn:shear_trans} with periodicity in the $Z$ direction. In the following sections, we present several numerical examples using Lees--Edwards and non-periodic boundary conditions. Particular attention is paid to differences in shear banding dynamics produced by these two choices of boundary conditions.

\subsubsection{Cylindrical inclusion}
\label{sssec:cyl}
\begin{figure*}
    \centering
    \fcolorbox{black}{black}{\includegraphics[width=\subpanelwid]{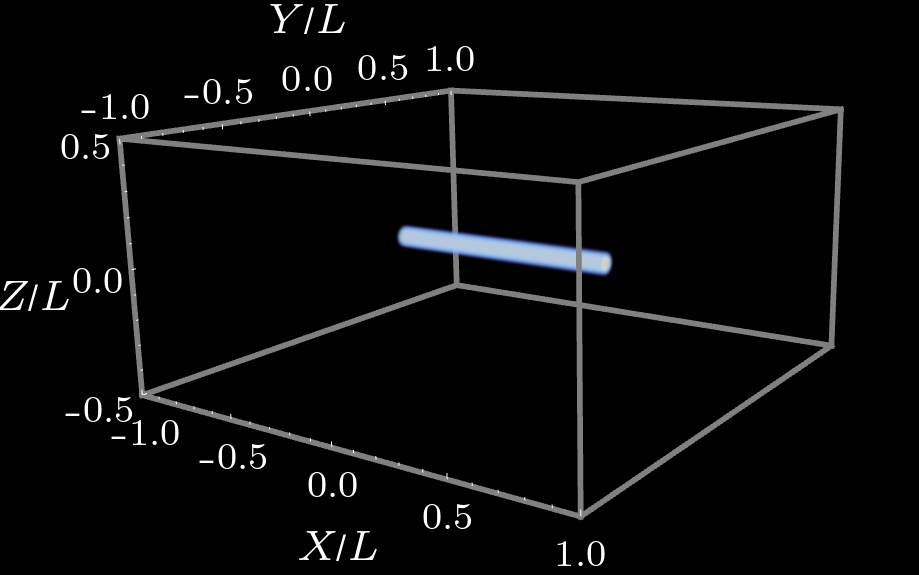}}
    ~\vspace{5mm}\\
    \includegraphics[width=.75\textwidth]{imgs/rslt_figs/colorbar_2019}
    \caption{The initial conditions for the cylindrical inclusion numerical experiments. $\chi_{bg} = 600\text{~K}$, $a = 0.3$, and $\eta = 1.2$ in the opacity function in Eq.~\ref{eqn:opac}.}
    \label{fig:cyl_shear_init}
\end{figure*}

We first consider an initial condition corresponding to a cylindrical defect in the material. Accordingly, the effective temperature field is initially elevated throughout a cylinder of finite length oriented along the direction of shear,
\begin{equation}
    \chi(\bX, t=0) =
    \begin{cases}
      550\text{~K} + \left(200\text{~K}\right)e^{-500\left(\frac{Z^2}{L^2} + \frac{Y^2}{L^2}\right)} & \text{if $\frac{X}{L} \in \left[\frac{a_X}{2}, \frac{b_X}{2}\right]$,} \\
        0 & \text{otherwise.}
    \end{cases}
\end{equation}
The initial condition is shown in Fig.~\ref{fig:cyl_shear_init}. The diffusion lengthscale is set to $l = \frac{3}{2}h$ and the quasi-static timestep is set to $\Delta t = 200 t_s$. The grid is of size $256\times 256\times 128$. The simulation is performed to a final value of $t = 2\times 10^6 t_s$. To induce shear banding, a shear transformation of the form Eq.~\ref{eqn:shear_trans} is used with a value of $U_b = 10^{-7} L/t_s$, and both clamped and Lees--Edwards boundary conditions are considered. The clamped simulation takes 13.851 total hours when run with 32 processes on an Ubuntu Linux computer with dual 10-core 2.20~GHz Intel Xeon E5-2630 v4 processors. 10.452 hours are spent in multigrid V-cycles and 28293 total V-cycles are required. The Lees--Edwards simulation takes 10.082 total hours when run with 32 processes on an Ubuntu Linux computer with dual 10-core 2.20~GHz Intel Xeon Silver 4114 processors. The total time spent in multigrid V-cycles is 7.393 hours and 28293 total V-cycles are required.

\begin{figure*}
    \centering
\fcolorbox{black}{black}{
    \begin{tabular}{cc}
        \includegraphics[width=\subpanelwid]{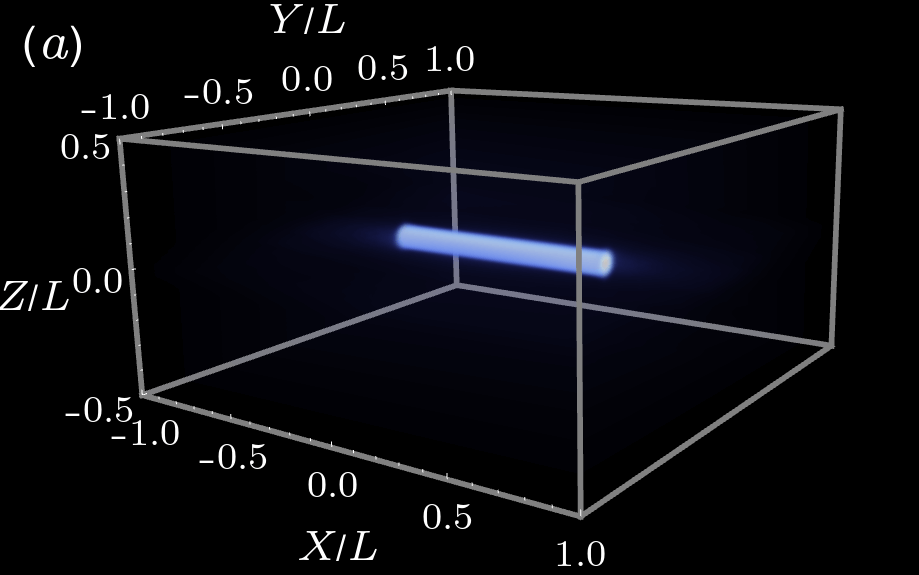} &
        \includegraphics[width=\subpanelwid]{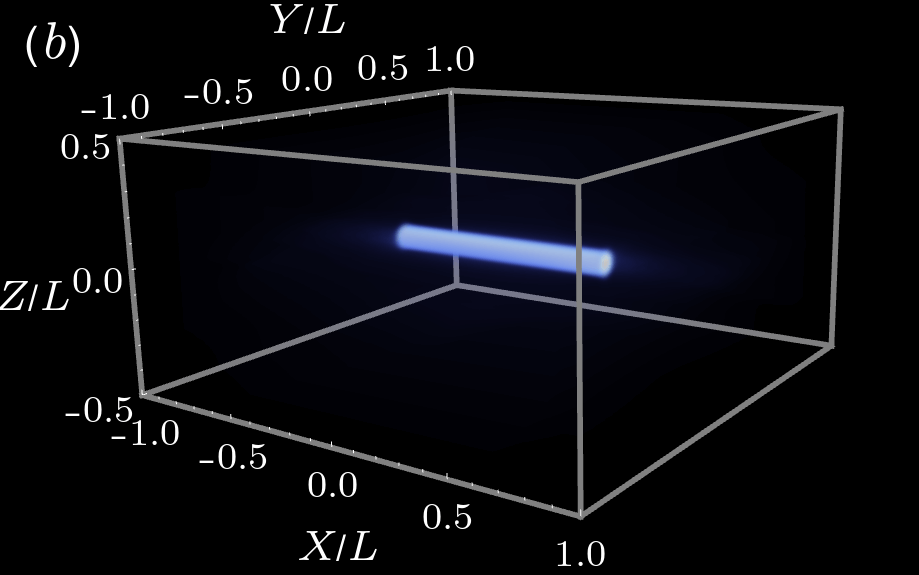} \\
        \includegraphics[width=\subpanelwid]{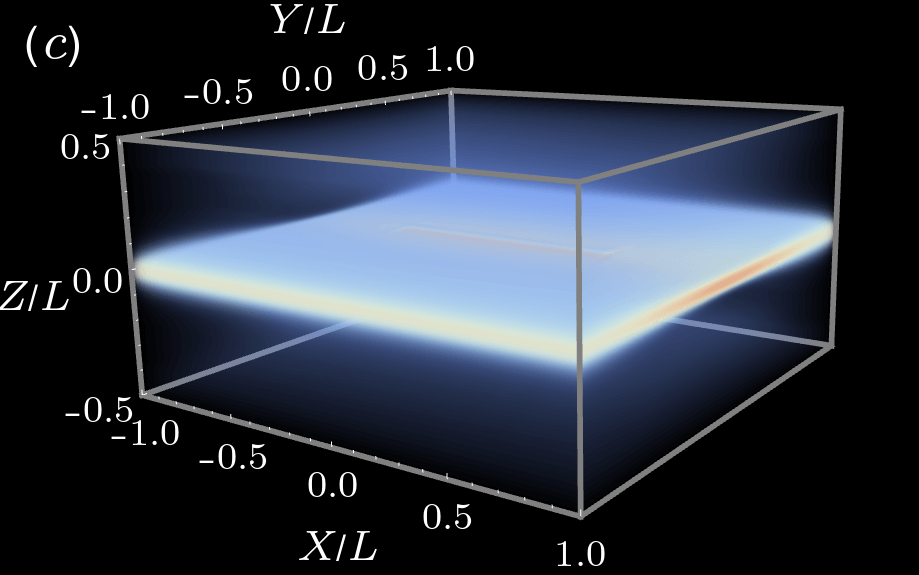} &
        \includegraphics[width=\subpanelwid]{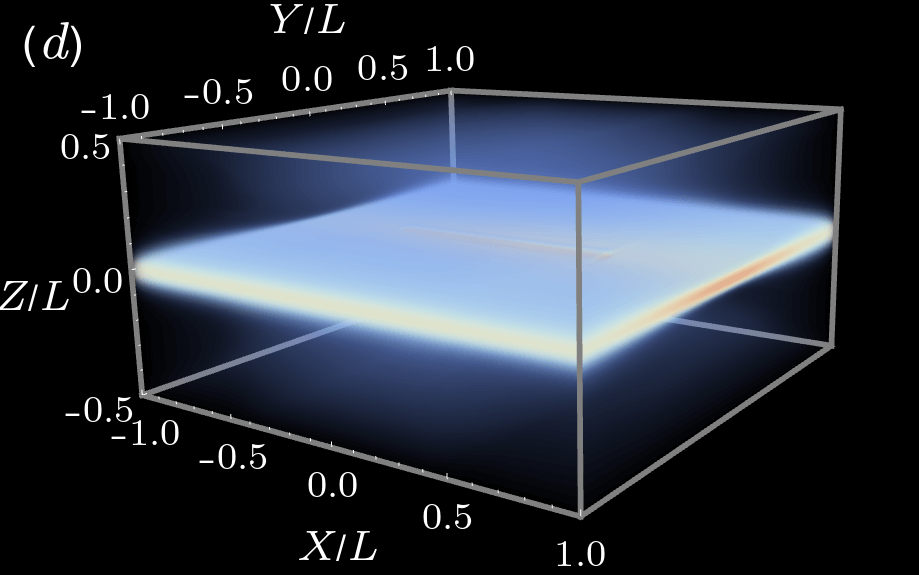} \\
        \includegraphics[width=\subpanelwid]{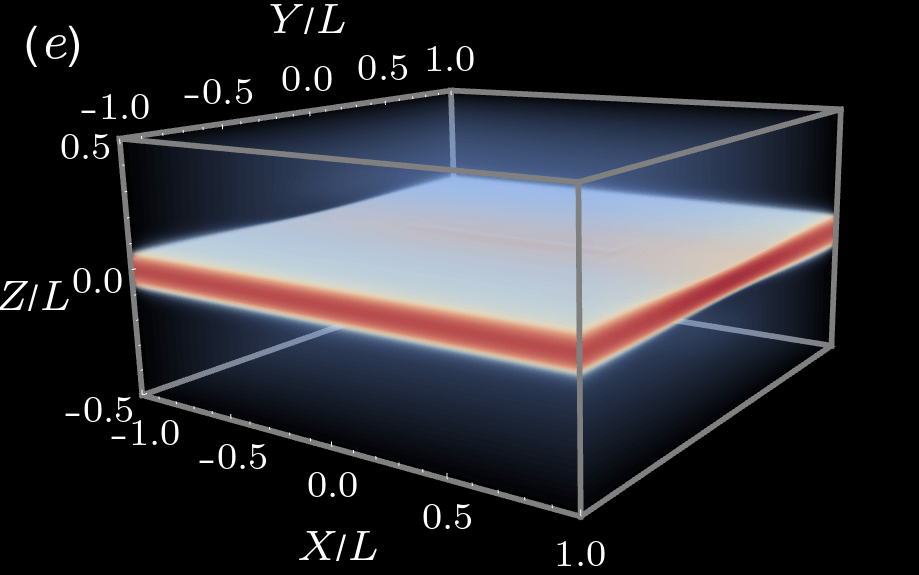} &
        \includegraphics[width=\subpanelwid]{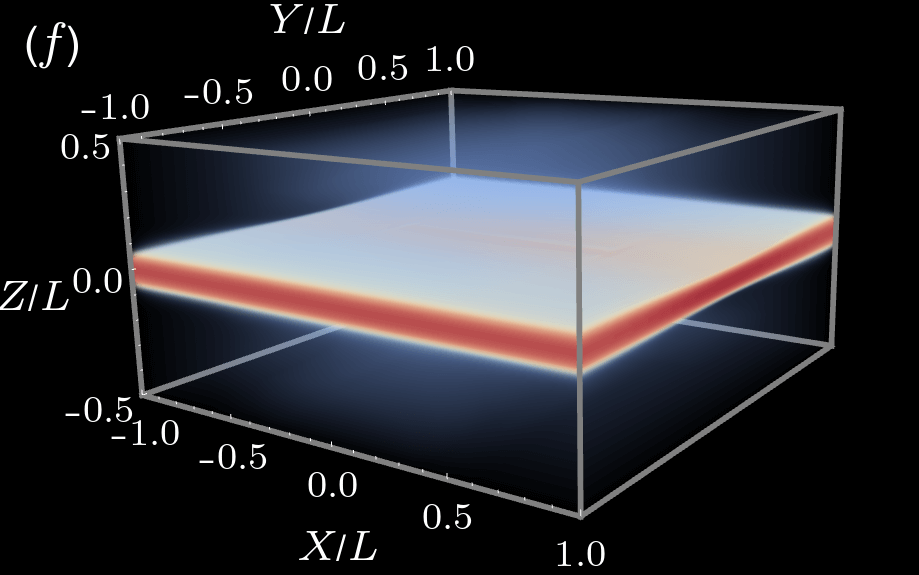}
    \end{tabular}}
    ~\vspace{5mm}\\
    \includegraphics[width=.75\textwidth]{imgs/rslt_figs/colorbar_2019}
    \caption{Snapshots of the effective temperature distribution $\chi(\bX, t)$. Simple shear deformation is imposed via a domain transformation. The initial condition in $\chi$ corresponds to a cylindrical inclusion as described in Sec.~\ref{sssec:cyl} and shown in Fig.~\ref{fig:cyl_shear_init}. On the left, clamped boundary conditions in $Z$ are used, while on the right, Lees--Edwards boundary conditions are used. $\chi_{bg} = 600\text{~K}$ in the opacity function in all subfigures. (a,b) $t = 5\times 10^5 t_s$. $a = 0.3$ and $\eta = 1.2$ in the opacity function. (c,d) $t = 1.25\times 10^6 t_s$. $a = 0.45$ and $\eta = 1.55$ in the opacity function. (e,f) $t = 2\times 10^6 t_s$. $a = 0.55$ and $\eta = 1.6$ in the opacity function.}
    \label{fig:cyl_shear_both}
\end{figure*}

Results for Lees--Edwards and non-periodic boundary conditions are shown in Fig.~\ref{fig:cyl_shear_both}, on the right and left respectively. The shear banding dynamics in this case are simple, and correspond to outward nucleation of a single band from the localized cylinder. At $t= 5 \times 10^5 t_s$ in Fig.~\ref{fig:cyl_shear_both} (a,b), nucleation of the shear band has begun, and there is some spreading in the $\chi$ field visible at the caps of the cylinder. By $t = 1.25\times 10^6 t_s$, a prominent system-spanning shear band has formed, as displayed in Fig.~\ref{fig:cyl_shear_both} (c,d). In Fig.~\ref{fig:cyl_shear_both} (e,f) at $t = 2\times 10^6 t_s$, the shear band continues to grow stronger and thicker. In this case, the dynamics are virtually identical for the Lees--Edwards and nonperiodic boundary conditions.

\subsubsection{A randomly fluctuating effective temperature field}
\label{sssec:rndm_ss}
We now consider a randomly distributed initial condition in the effective temperature field $\chi(\bX, t=0)$. We first populate the grid and a shell of ghost points with random variables $\chi_\zeta(\bX)$ using the Box--Muller algorithm. With $\mu_\chi$ and $\sigma_\chi$ respectively denoting the desired mean and standard deviation, we perform the convolution
\begin{equation}
    \chi(\bX) = \frac{\sigma_\chi}{N} \sum_{\mathbf{R} \in V'} e^{-\frac{\left\|\bX - \mathbf{R}\right\|^2}{l_c^2}}\chi_\zeta(\mathbf{R}) + \mu_\chi, \qquad
    N = \sqrt{\sum_{\mathbf{R} \in V} e^{-2\frac{\|\mathbf{R}\|^2}{l_c^2}}},
    \label{eqn:convolve}
\end{equation}
where $V$ denotes the set of grid points and $V'$ denotes the extended set of grid points and ghost points. Equation \ref{eqn:convolve} ensures that the effective temperature value at each point is normally distributed with mean $\mu_\chi$ and standard deviation $\sigma_\chi$. In practice, the sums in Eq.~\ref{eqn:convolve} are performed with a cutoff length scale specified as a multiplicative factor of the convolution length scale $l_c$, and the number of ghost points in $V'$ is set by the choice of cutoff length scale. For computational feasibility, we choose a cutoff length of $5l_c$, so that the Gaussian kernel is considered to be zero past this point. In the following studies, a value of $l_c = 5h$ is used, leading to an additional $25$ ghost points padding the grid for the purpose of the convolution.

Simulations are performed for mean values $\mu_\chi = 450\text{~K}, 500\text{~K}, 525\text{~K}, 550\text{~K}, 575 \text{~K}, 600\text{~K}$ with a fixed value of $\sigma_\chi = 15 \text{~K}$ for both non-periodic and Lees--Edwards boundary conditions. The diffusion length scale is set to $l = \frac{3}{2}h$, and the quasi-static timestep is set to $\Delta t = 200 t_s$. The simulations are all conducted on a $512\times 512\times 256$ cell grid to a final value of $t = 1\times 10^6 t_s$. To induce shear banding, a shear transformation of the form Eq.~\ref{eqn:shear_trans} with a value of $U_b = 10^{-7} L/t_s$ is imposed on the domain. Timing details for the non-periodic simulations are shown in Table~\ref{tab:clamp_shear_time}, while timing details for the Lees--Edwards simulations are shown in Table~\ref{tab:LE_shear_time}.

\begin{table}
\centering
\resizebox{\textwidth}{!}{%
\begin{tabular}{|c|c|c|c|c|c|c|}
\hline
 & $\mu_\chi=450\text{~K}$ & $\mu_\chi=500\text{~K}$ & $\mu_\chi=525\text{~K}$ & $\mu_\chi=550\text{~K}$ & $\mu_\chi=575\text{~K}$ & $\mu_\chi=600\text{~K}$ \\ \hline
Total time (hours) & 95.2948 & 89.7239 & 76.9704 & 82.7853 & 71.7865 & 69.1593 \\ \hline
V-cycle time (hours) & 65.7853 & 60.5694 & 48.6470 & 53.4612 & 41.7683 & 40.4283 \\ \hline
\# of V-cycles & 34846 & 30663 & 26628 & 24991 & 22649 & 20735 \\ \hline
Time/V-cycle (seconds) & 6.7964 & 7.1111 & 6.5749 & 7.7012 & 6.6390 & 7.0195 \\ \hline
\multicolumn{1}{|l|}{Processor details} & \multicolumn{1}{l|}{\begin{tabular}[c]{@{}l@{}}Dual 10-core \\ 2.20~GHz Intel Xeon\\ E5-2630 v4\end{tabular}} & \multicolumn{1}{l|}{\begin{tabular}[c]{@{}l@{}}Dual 10-core \\ 2.20~GHz Intel Xeon\\ E5-2630 v4\end{tabular}} & \multicolumn{1}{l|}{\begin{tabular}[c]{@{}l@{}}Dual 10-core \\ 2.20~GHz Intel Xeon\\ Silver 4114\end{tabular}} & \multicolumn{1}{l|}{\begin{tabular}[c]{@{}l@{}}Dual 14-core \\ 1.70~GHz Intel Xeon\\ E5-2650L v4\end{tabular}} & \multicolumn{1}{l|}{\begin{tabular}[c]{@{}l@{}}Dual 8-core\\ 2.40~GHz Intel Xeon \\ E5-2630 v3\end{tabular}} & \begin{tabular}[c]{@{}l@{}}Dual 10-core \\ 2.20~GHz Intel Xeon\\ E5-2630 v4\end{tabular} \\ \hline
\end{tabular}%
}
\caption{Data describing the total time, total time spent in multigrid V-cycles, total number of multigrid V-cycles, average time spent per multigrid V-cycle, and processor details for each randomly initialized simulation. This data applies to the randomly initialized simulations with non-periodic boundary conditions and simple shear deformation. The number of required multigrid V-cycles decreases as the background $\chi$ field increases, likely due to more homogeneous dynamics. Each simulation uses $32$ processes.}
\label{tab:clamp_shear_time}
\end{table}

\begin{figure*}
\fcolorbox{black}{black}{
    \begin{tabular}{cc}
        \includegraphics[width=\subpanelwid]{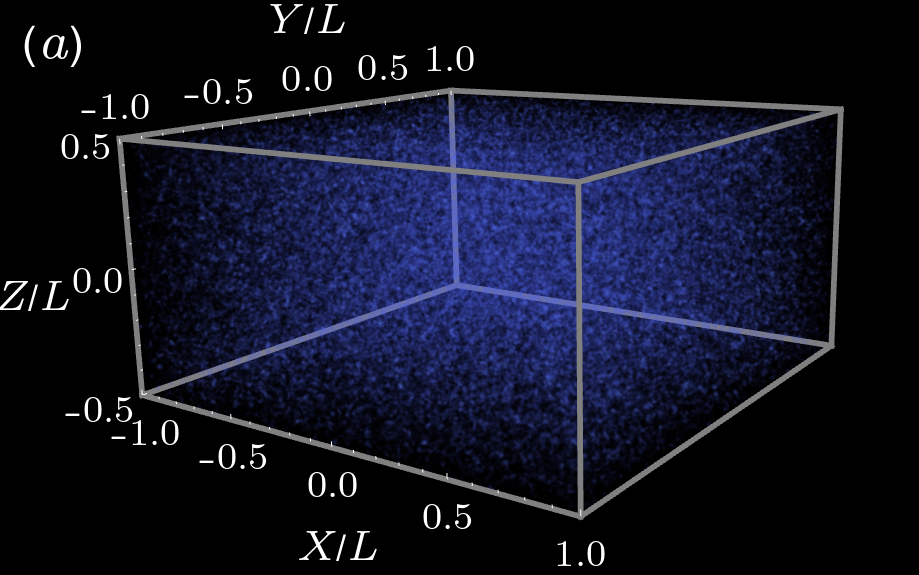} &
        \includegraphics[width=\subpanelwid]{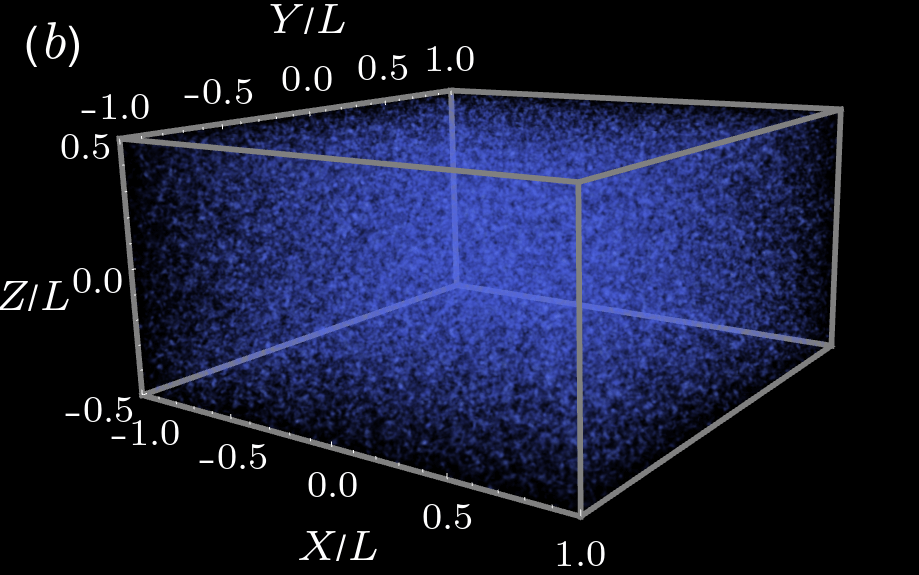} \\
        \includegraphics[width=\subpanelwid]{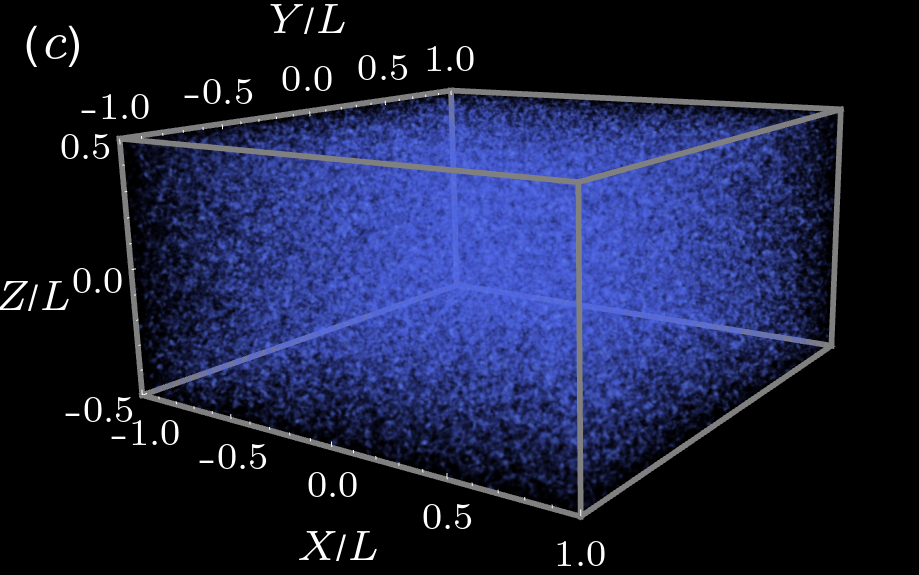} &
        \includegraphics[width=\subpanelwid]{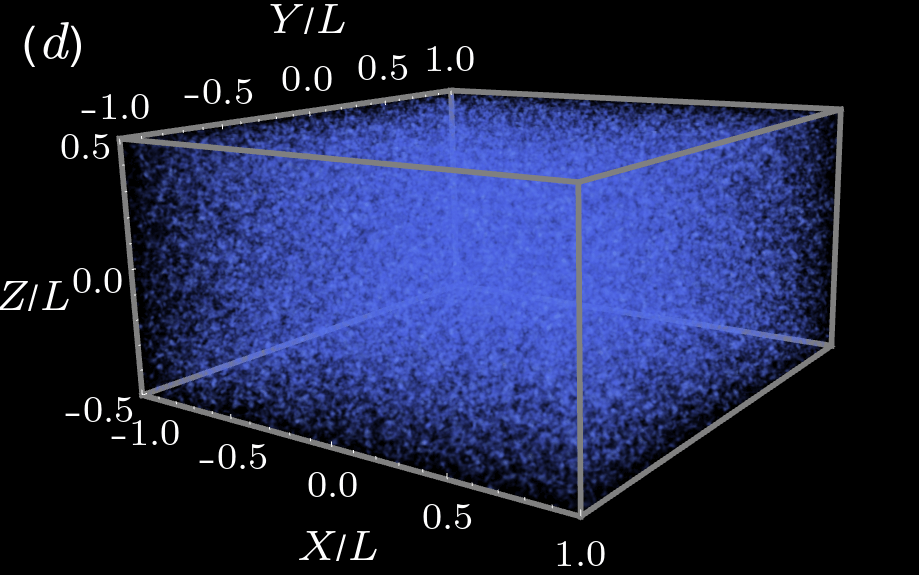} \\
        \includegraphics[width=\subpanelwid]{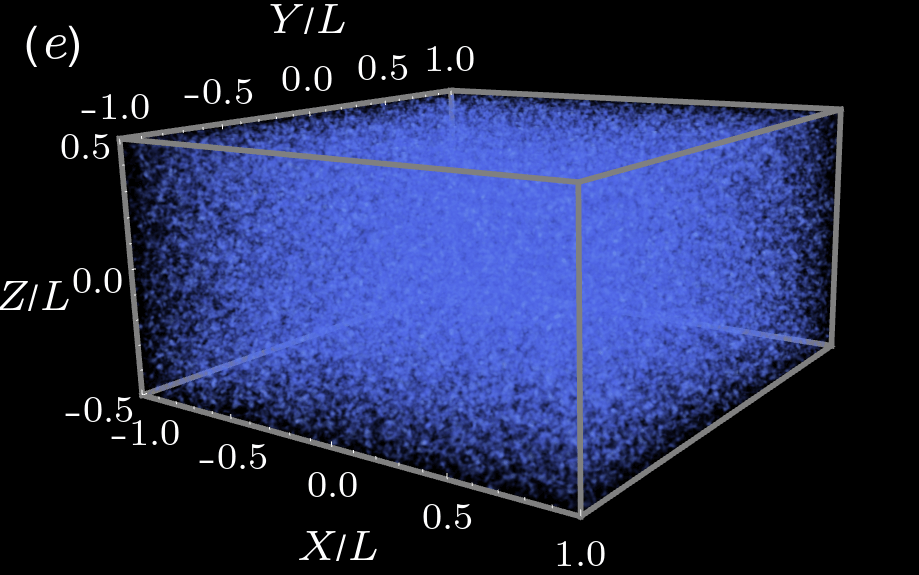} &
        \includegraphics[width=\subpanelwid]{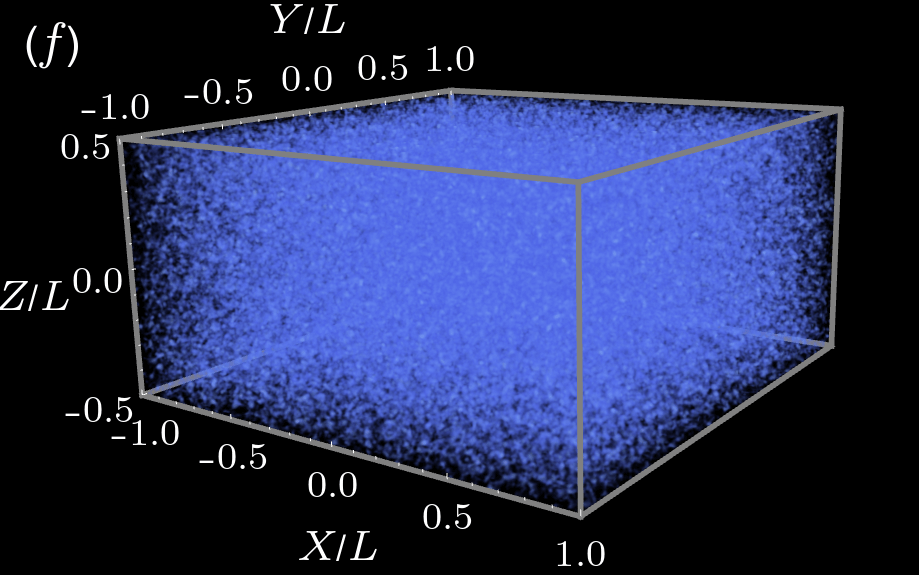}
    \end{tabular}}
    ~\vspace{5mm}\\
    \includegraphics[width=.75\textwidth]{imgs/rslt_figs/colorbar_2019}
    \caption{Snapshots of the effective temperature field at $t = 0t_s$. All simulations use non-periodic boundary conditions in $Z$ and apply simple shear deformation. For all plots, a value of $a = 0.25$ and $\eta = 1.3$ is used in the opacity function. $\chi_{\text{bg}}$ is set to $\mu_\chi - 25\text{~K}$ in each pane in the opacity function. Figures (a)--(f) have $\mu_\chi = 450\tK, 500\tK, 525\tK, 550\tK, 575\tK$, and $600\tK$ respectively.}
    \label{fig:clamp_0}
\end{figure*}

\begin{figure*}
\fcolorbox{black}{black}{
    \begin{tabular}{cc}
        \includegraphics[width=\subpanelwid]{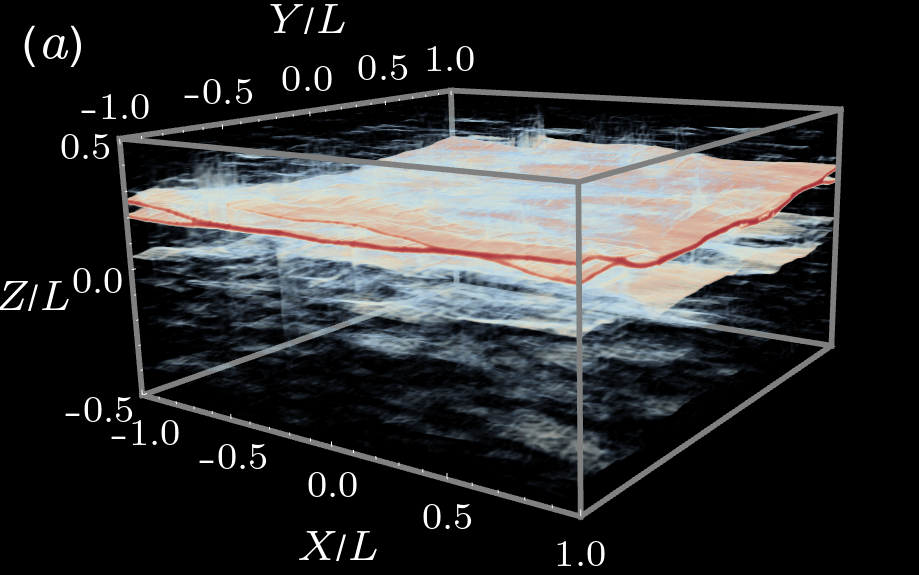} &
        \includegraphics[width=\subpanelwid]{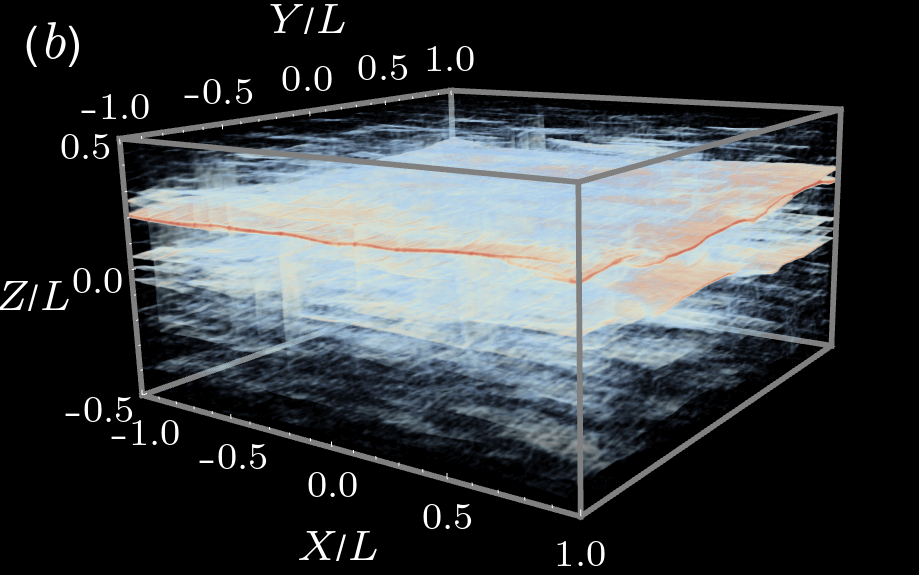} \\
        \includegraphics[width=\subpanelwid]{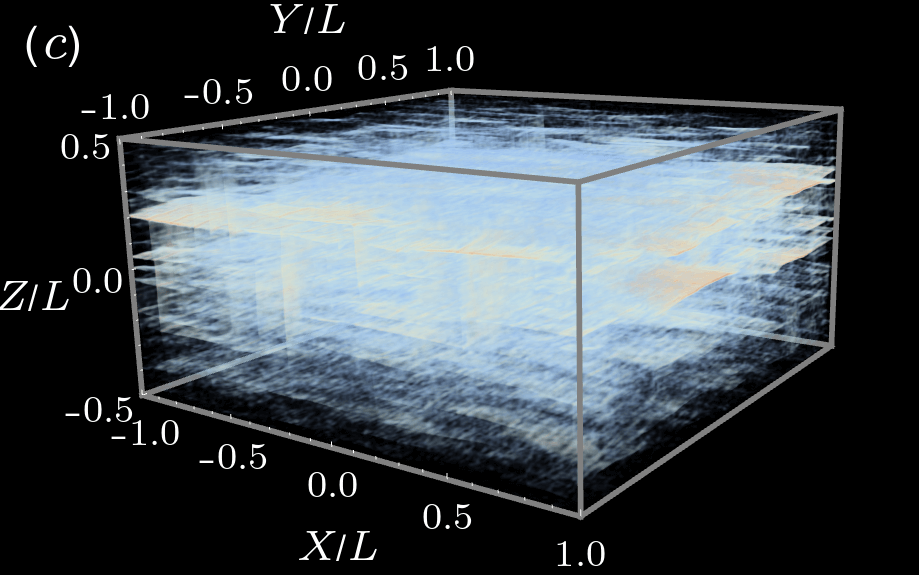} &
        \includegraphics[width=\subpanelwid]{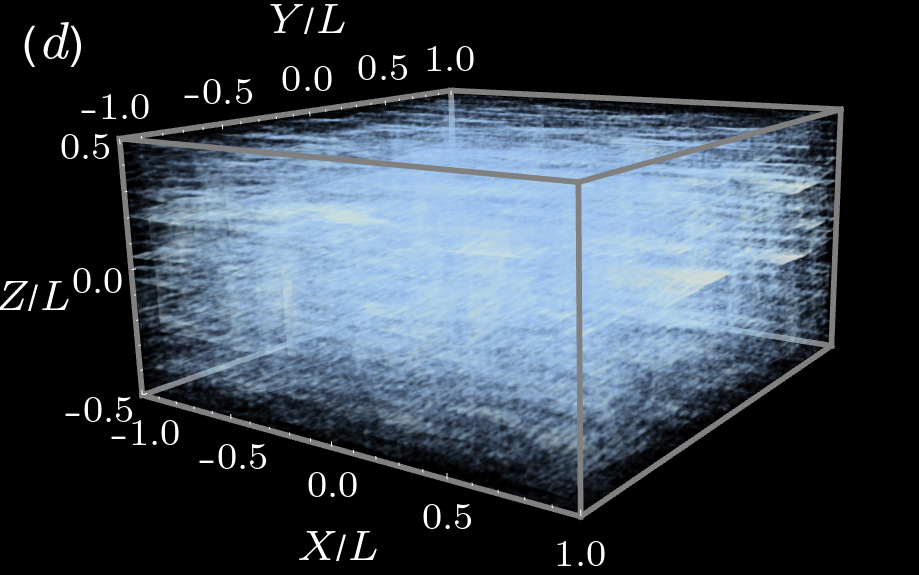} \\
        \includegraphics[width=\subpanelwid]{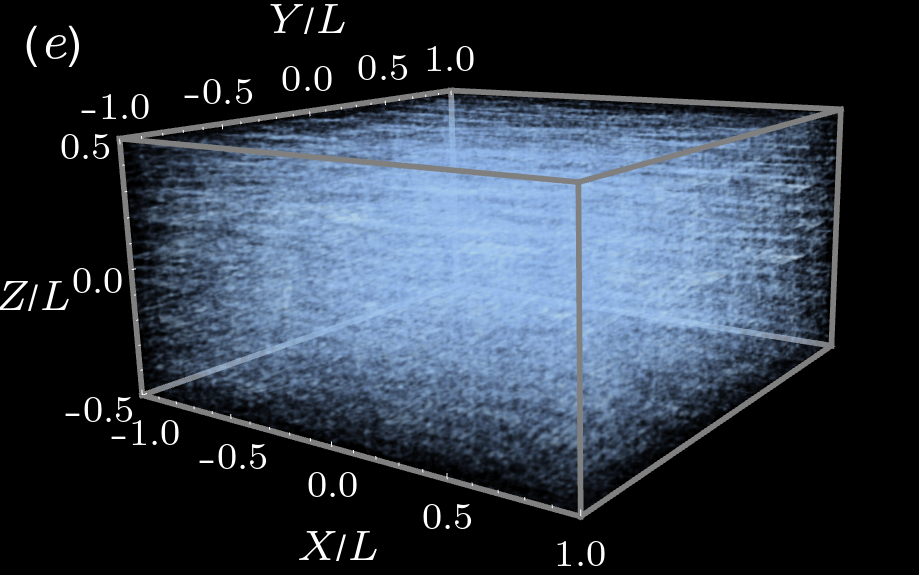} &
        \includegraphics[width=\subpanelwid]{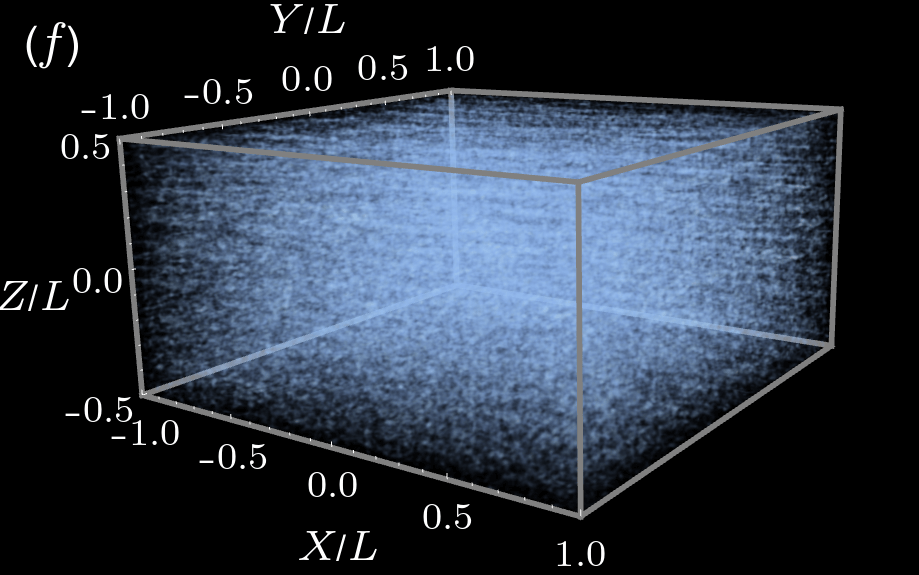}
    \end{tabular}}
    ~\vspace{5mm}\\
    \includegraphics[width=.75\textwidth]{imgs/rslt_figs/colorbar_2019}
    \caption{Snapshots of the effective temperature field at $t = 4\times 10^5t_s$. All simulations use non-periodic boundary conditions in $Z$ and apply simple shear deformation. For all plots, a value of $a = 0.45$ and $\eta = 1.75$ is used in the opacity function. $\chi_{\text{bg}}$ is set to $\mu_\chi - 25\text{~K}$ in each pane in the opacity function. Figures (a)--(f) have $\mu_\chi = 450\tK, 500\tK, 525\tK, 550\tK, 575\tK$, and $600\tK$ respectively.}
    \label{fig:clamp_40}
\end{figure*}

\begin{figure*}
\fcolorbox{black}{black}{
    \begin{tabular}{cc}
        \includegraphics[width=\subpanelwid]{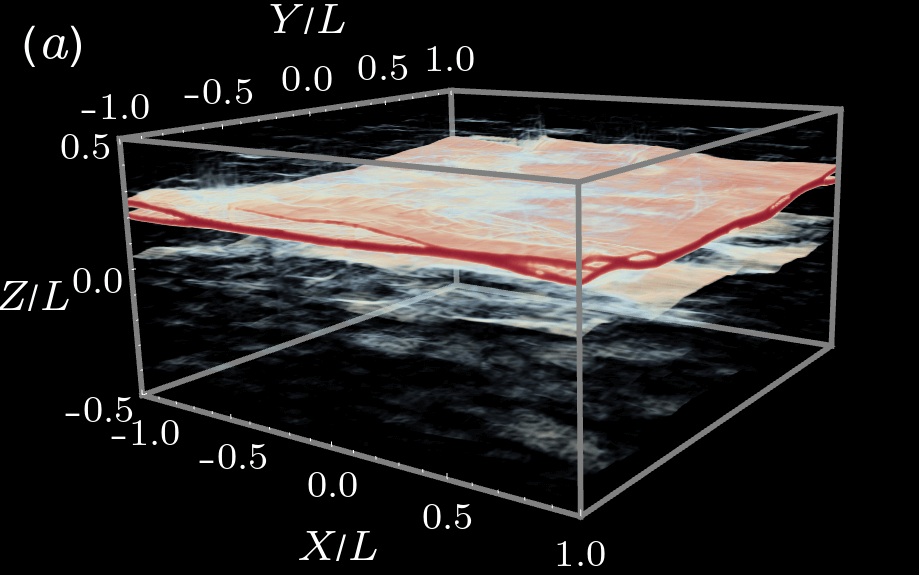} &
        \includegraphics[width=\subpanelwid]{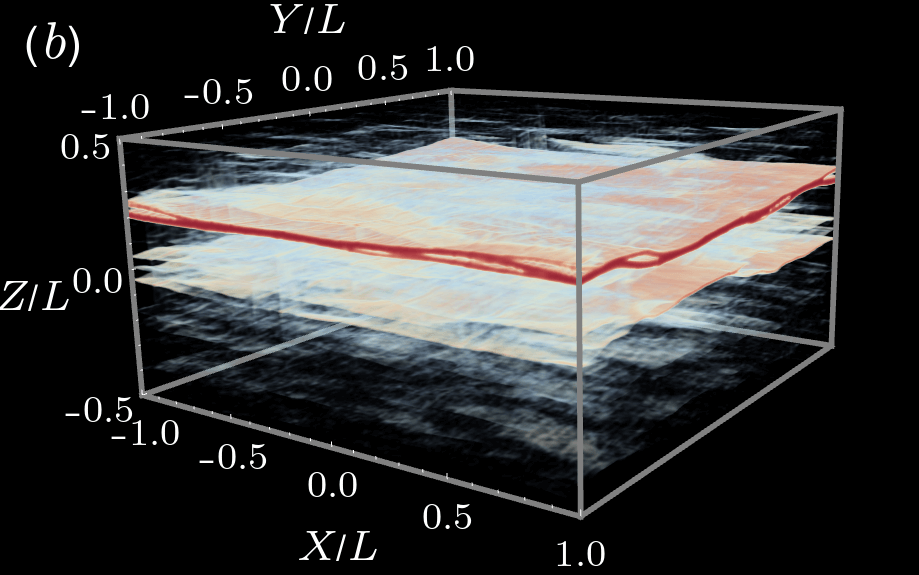} \\
        \includegraphics[width=\subpanelwid]{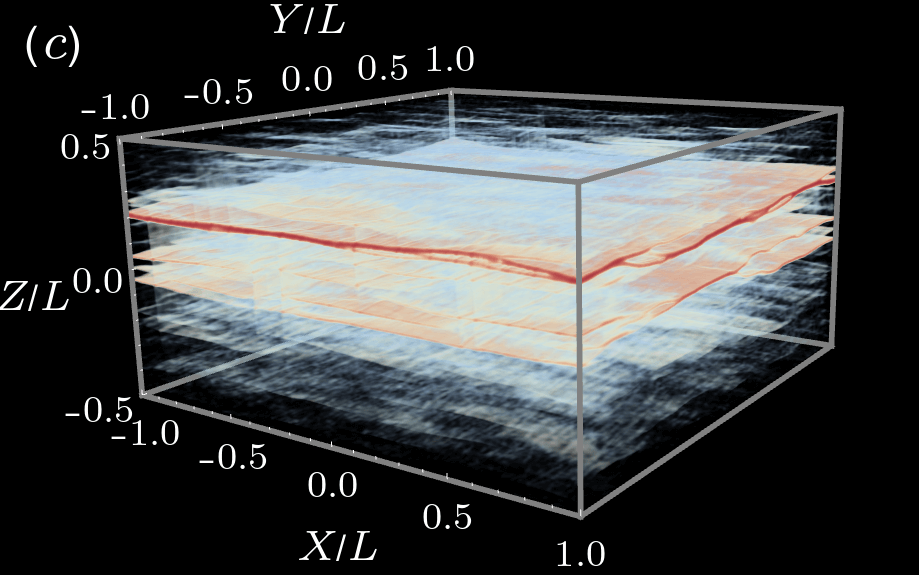} &
        \includegraphics[width=\subpanelwid]{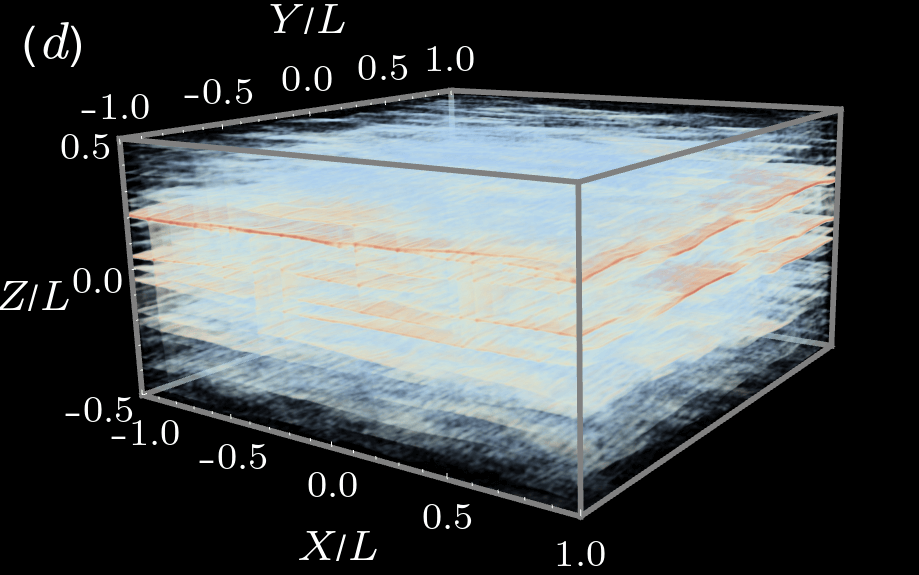} \\
        \includegraphics[width=\subpanelwid]{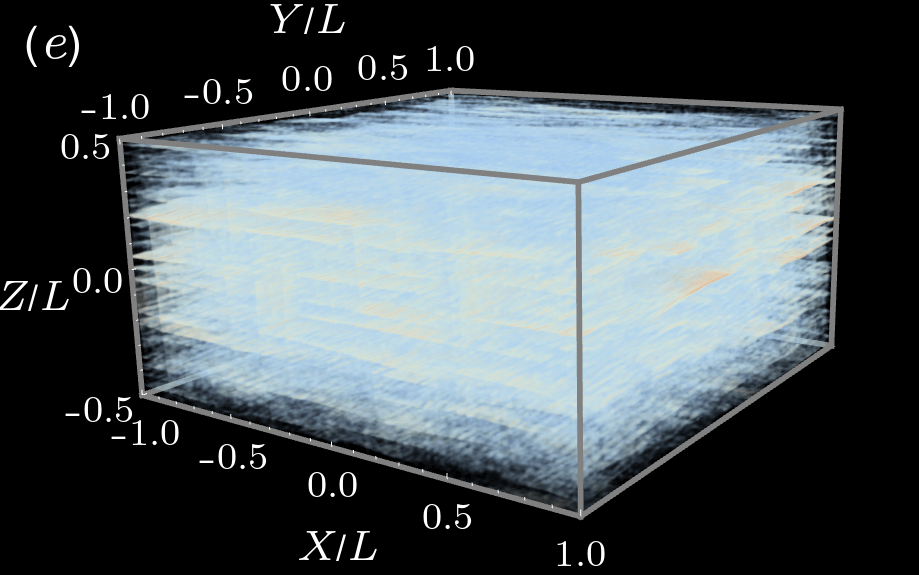} &
        \includegraphics[width=\subpanelwid]{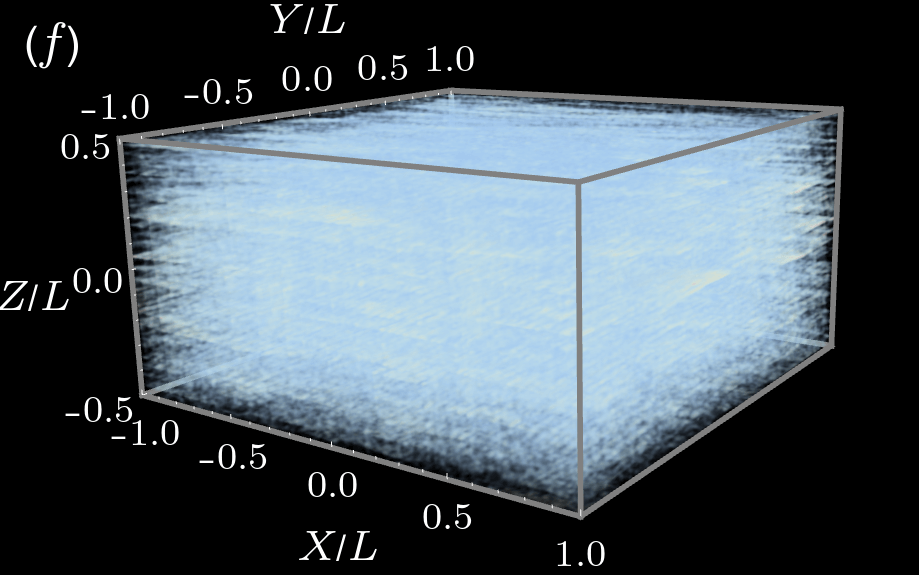}
    \end{tabular}}
    ~\vspace{5mm}\\
    \includegraphics[width=.75\textwidth]{imgs/rslt_figs/colorbar_2019}
    \caption{Snapshots of the effective temperature field at $t = 6\times 10^5t_s$. All simulations use non-periodic boundary conditions in $Z$ and apply simple shear deformation. For all plots, a value of $a = 0.45$ and $\eta = 1.75$ is used in the opacity function. $\chi_{\text{bg}}$ is set to $\mu_\chi - 25\text{~K}$ in each pane in the opacity function. Figures (a)--(f) have $\mu_\chi = 450\tK, 500\tK, 525\tK, 550\tK, 575\tK$, and $600\tK$ respectively.}
    \label{fig:clamp_60}
\end{figure*}

\begin{figure*}
\fcolorbox{black}{black}{
    \begin{tabular}{cc}
        \includegraphics[width=\subpanelwid]{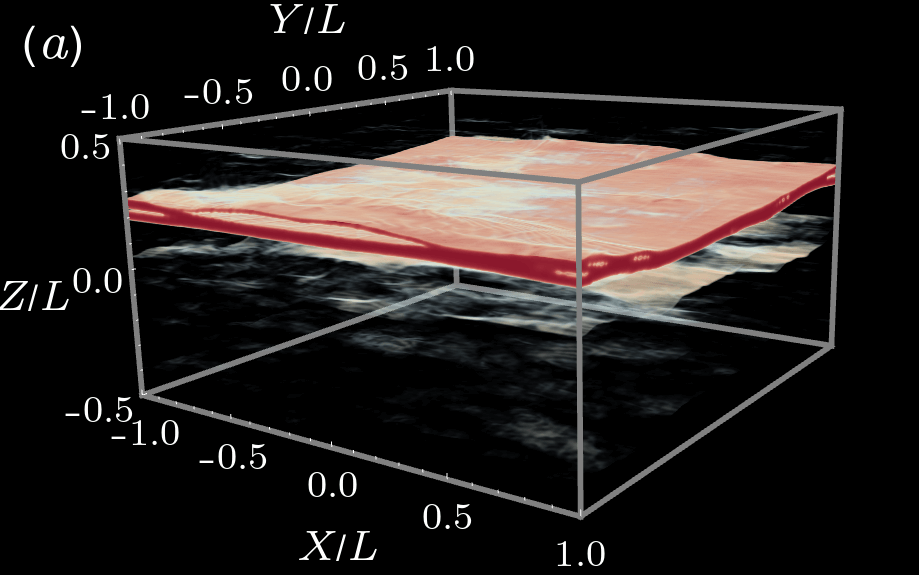} &
        \includegraphics[width=\subpanelwid]{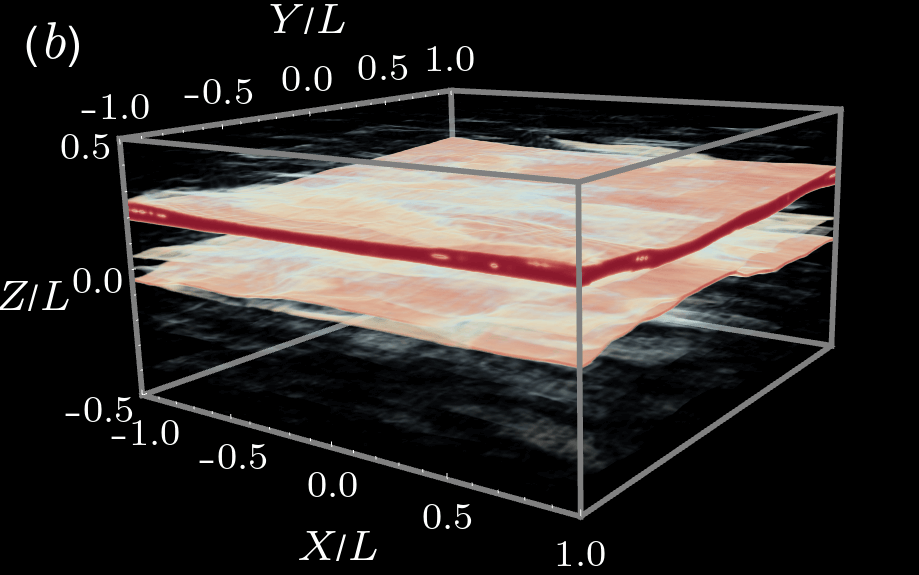} \\
        \includegraphics[width=\subpanelwid]{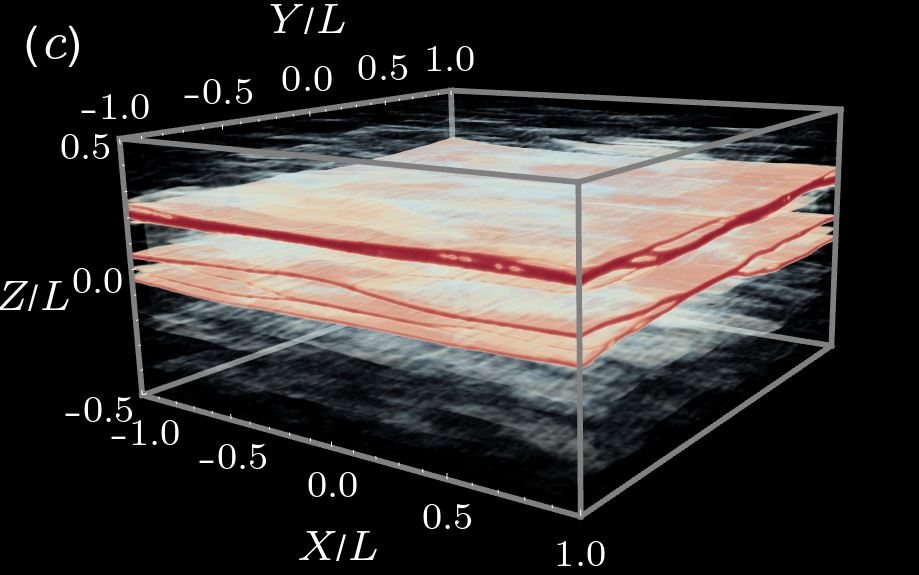} &
        \includegraphics[width=\subpanelwid]{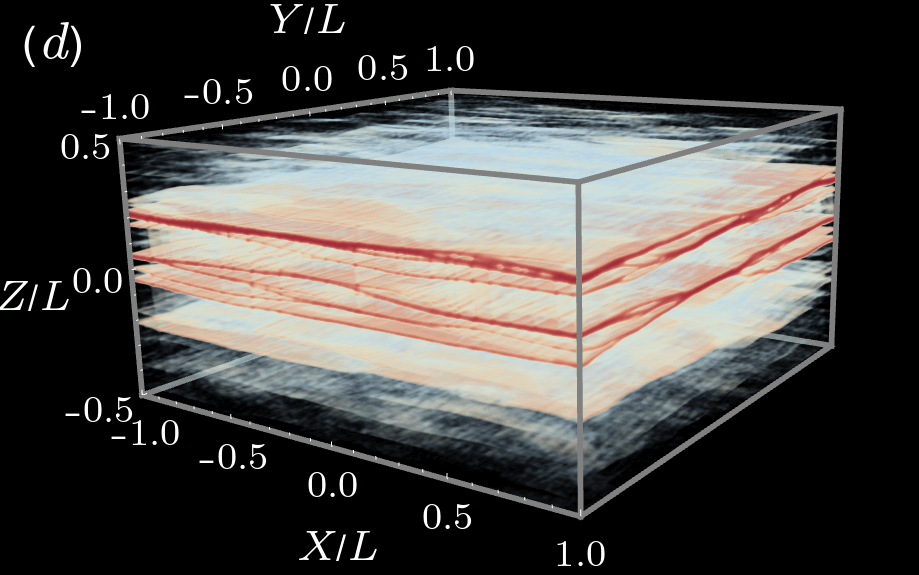} \\
        \includegraphics[width=\subpanelwid]{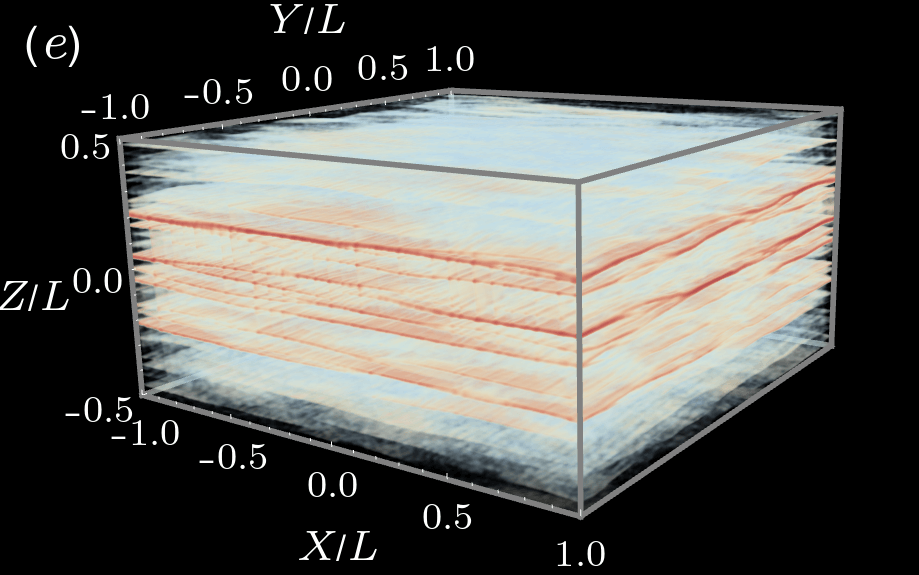} &
        \includegraphics[width=\subpanelwid]{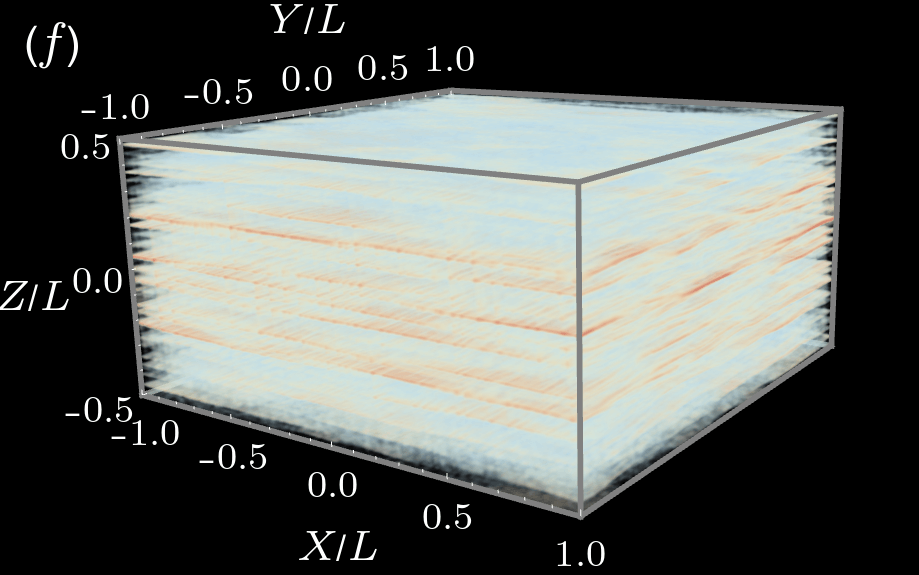}
    \end{tabular}}
    ~\vspace{5mm}\\
    \includegraphics[width=.75\textwidth]{imgs/rslt_figs/colorbar_2019}
    \caption{Snapshots of the effective temperature field at $t = 10^6t_s$. All simulations use non-periodic boundary conditions in $Z$ and apply simple shear deformation. For all plots, a value of $a = 0.75$ and $\eta = 2$ is used in the opacity function. $\chi_{\text{bg}}$ is set to $\mu_\chi - 25\text{~K}$ in each pane in the opacity function. Figures (a)--(f) have $\mu_\chi = 450\tK, 500\tK, 525\tK, 550\tK, 575\tK$, and $600\tK$ respectively.}
    \label{fig:clamp_100}
\end{figure*}

The results for this sequence of simulations in the case of non-periodic boundary conditions are shown in Figs.~\ref{fig:clamp_0}--\ref{fig:clamp_100}. Each figure corresponds to a single snapshot in time, and the mean increases with the alphabetical labeling. The value of $\chi_{bg}$ used in the opacity function in each case is given by $\mu_\chi - 25\text{~K}$. The initial conditions for the effective temperature field are shown in Fig.~\ref{fig:clamp_0}. At $t = 0$, all simulations look essentially the same. The realization of the noise in each configuration is identical, and each pane is obtained from the previous by a constant shift in $\chi$.

By $t = 4\times 10^5 t_s$ in Fig.~\ref{fig:clamp_40}, the simulations with the two lowest values of $\mu_\chi$ exhibit clear shear bands with curvature in both the $X$ and $Y$ directions. The simulation with $\mu_\chi = 450\text{~K}$. in Fig.~\ref{fig:clamp_40}(a) displays two shear bands that cross each other diagonally near $\frac{X}{L} = 0.5$. The simulation with $\mu_\chi = 500\text{~K}$ in Fig.~\ref{fig:clamp_40}(b) displays only one of these bands, though the second has begun to nucleate. This single band is also apparent in Fig.~\ref{fig:clamp_40}(c), but it is significantly weaker. A third nascent band near $\frac{Z}{L} = 0$ may also be observed.

More details are clear at $t = 6\times 10^5 t_s$ in Fig.~\ref{fig:clamp_60}. Figure \ref{fig:clamp_60}(a) is similar to Fig.~\ref{fig:clamp_40}(a), whereas Fig.~\ref{fig:clamp_60}(b) shows further development of the shear bands in Fig.~\ref{fig:clamp_40}(b). Figures \ref{fig:clamp_60}(c) and (d) show the development of several flat and thin shear bands centered around $\frac{Z}{L} = 0$.

Figure \ref{fig:clamp_100} ($t = 10^6 t_s$) displays clear shear banding across all values of $\mu_\chi$, and makes clear the dependence of shear banding structure on  $\mu_\chi$. There is one primary band in Fig.~\ref{fig:clamp_100}(a), with a split near around $\frac{X}{L}\approx -0.5$ not present in Fig.~\ref{fig:clamp_100}(b). Figure \ref{fig:clamp_100}(b) also displays an additional thin band near $\frac{Z}{L} \approx 0.0$ that has not formed in Fig.~\ref{fig:clamp_100}(a). Figure \ref{fig:clamp_100}(c) displays several additional bands near $\frac{Z}{L} = 0$ that form a complex branching pattern. Figure \ref{fig:clamp_100}(d) resolves more fine-scale structure in the band near $\frac{Z}{L} \approx 0.25$ when compared to Figs.~\ref{fig:clamp_100}(a)-(c) as if a single band has begun to split, and has more bands near lower values of $\frac{Z}{L}$. Figs.~\ref{fig:clamp_100}(e) and (f)  show several additional thin bands when compared to the previous panels, but they are earlier in their formation and less prominently displayed.

Taken together, Figs.~\ref{fig:clamp_0}--\ref{fig:clamp_100} provide qualitative insight into how macroscopic shear banding dynamics and structure reflect the underlying effective temperature distribution. In a simulation with small mean, there are few regions susceptible shear band nucleation, most clearly displayed in the formation of only a single band in the lowest mean simulation. These nucleation points must connect to form a band, as indicated by the mild curvature seen in the bands in Figs.~\ref{fig:clamp_100}(a) and (b). As $\mu_\chi$ is increased, additional regions of sufficiently high $\chi$ exist for band nucleation, curvature decreases, and the number of bands increases. This first presents itself, as seen in Figs.~\ref{fig:clamp_0}(d)--\ref{fig:clamp_100}(d), as an existing band splitting into multiple. The gap in the split grows with $\mu_\chi$, as seen in Figs.~\ref{fig:clamp_0}(d)-\ref{fig:clamp_100}(d) and Figs.~\ref{fig:clamp_0}(e)-\ref{fig:clamp_100}(e), until it eventually breaks off into its own band. With high $\mu_\chi$ as in Figs.~\ref{fig:clamp_0}(e)-\ref{fig:clamp_100}(e) and Figs.~\ref{fig:clamp_0}(f)-\ref{fig:clamp_100}(f), shear bands can nucleate in many different locations without curvature. The timescale for shear band development is also more rapid in simulations with low background $\chi$ field.

\begin{table}
\centering
\resizebox{\textwidth}{!}{%
\begin{tabular}{|c|c|c|c|c|c|l|}
\hline
 & $\mu_\chi=450\text{~K}$ & $\mu_\chi=500\text{~K}$ & $\mu_\chi=525\text{~K}$ & $\mu_\chi=550\text{~K}$ & $\mu_\chi=575\text{~K}$ & $\mu_\chi=600\text{~K}$ \\ \hline
Total time (hours) & 91.4188 & 93.9238 & 81.7296 & 94.3549 & 72.8395 & 68.4994 \\ \hline
V-cycle time (hours) & 63.5750 & 64.8320 & 51.2489 & 62.8451 & 44.4758 & 39.9291 \\ \hline
\# of V-cycles & 34915 & 30467 & 26658 & 24697 & 22256 & 20495 \\ \hline
Time/V-cycle (seconds) & 6.5551 & 7.6606 & 6.9209 & 9.1607 & 7.1941 & 7.0137 \\ \hline
\multicolumn{1}{|l|}{Processor details} & \multicolumn{1}{l|}{\begin{tabular}[c]{@{}l@{}}Dual 10-core \\ 2.20~GHz Intel Xeon\\ Silver 4114 v4\end{tabular}} & \multicolumn{1}{l|}{\begin{tabular}[c]{@{}l@{}}Dual 14-core \\ 1.70~GHz Intel Xeon\\ E5-2650L v4\end{tabular}} & \multicolumn{1}{l|}{\begin{tabular}[c]{@{}l@{}}Dual 8-core\\ 2.40~GHz Intel Xeon \\ E5-2630 v3\end{tabular}} & \multicolumn{1}{l|}{\begin{tabular}[c]{@{}l@{}}Dual 10-core \\ 2.20~GHz Intel Xeon\\ E5-2630 v4\end{tabular}} & \multicolumn{1}{l|}{\begin{tabular}[c]{@{}l@{}}Dual 14-core \\ 1.70~GHz Intel Xeon\\ E5-2650L v4\end{tabular}} & \begin{tabular}[c]{@{}l@{}}Dual 10-core \\ 2.20~GHz Intel Xeon\\ E5-2630 v4\end{tabular} \\ \hline
\end{tabular}%
}
\caption{Data describing the total time, total time spent in multigrid V-cycles, total number of multigrid V-cycles, average time spent per multigrid V-cycle, and the processor details for each simulation. This data applies to the randomly initialized simulations with Lees--Edwards boundary conditions. The number of required multigrid V-cycles decreases as the background $\chi$ field increases, likely due to more homogeneous dynamics. Each simulation uses $32$ processes.}
\label{tab:LE_shear_time}
\end{table}

\begin{figure*}
\fcolorbox{black}{black}{
    \begin{tabular}{cc}
        \includegraphics[width=\subpanelwid]{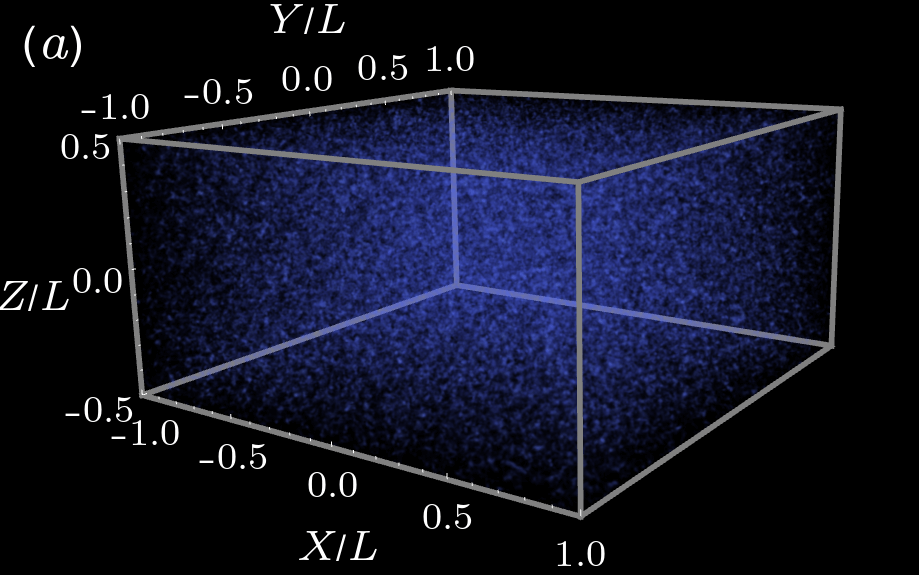} &
        \includegraphics[width=\subpanelwid]{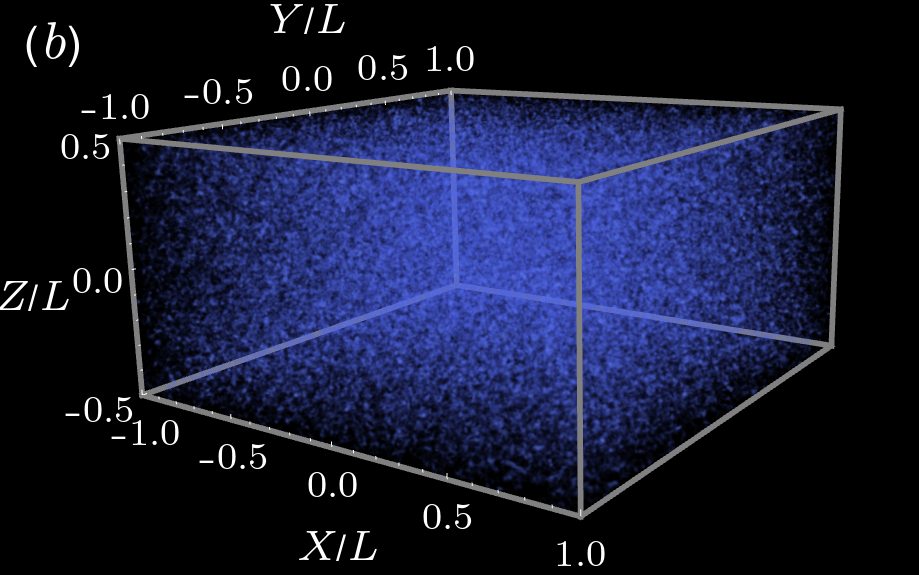} \\
        \includegraphics[width=\subpanelwid]{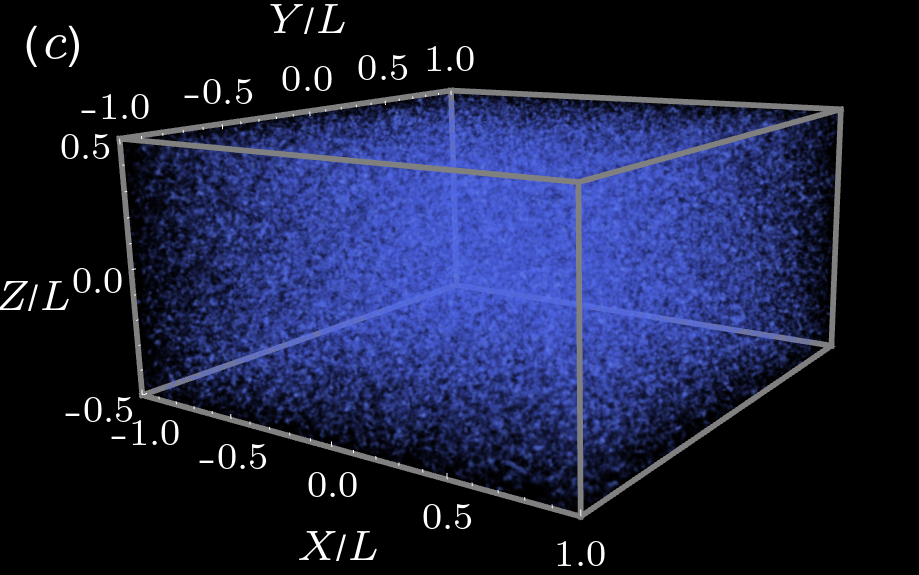} &
        \includegraphics[width=\subpanelwid]{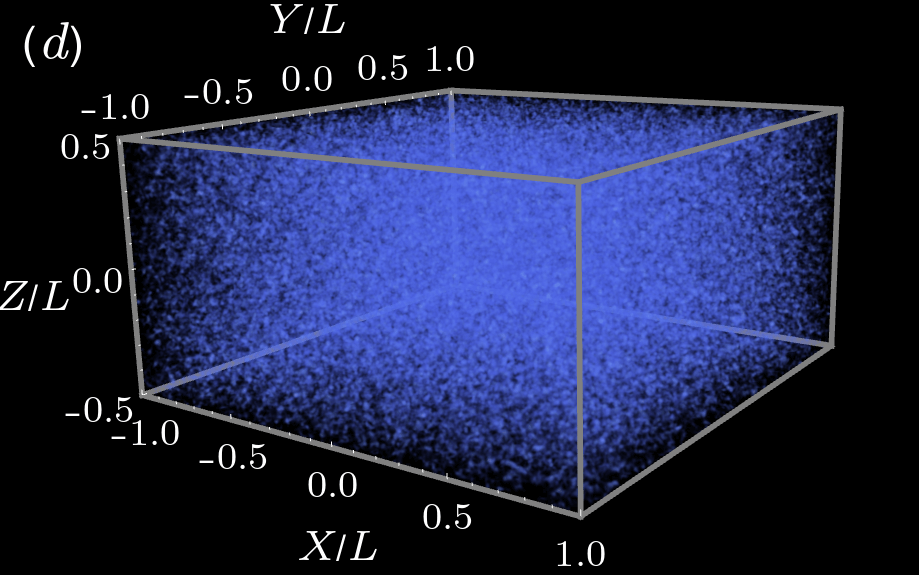} \\
        \includegraphics[width=\subpanelwid]{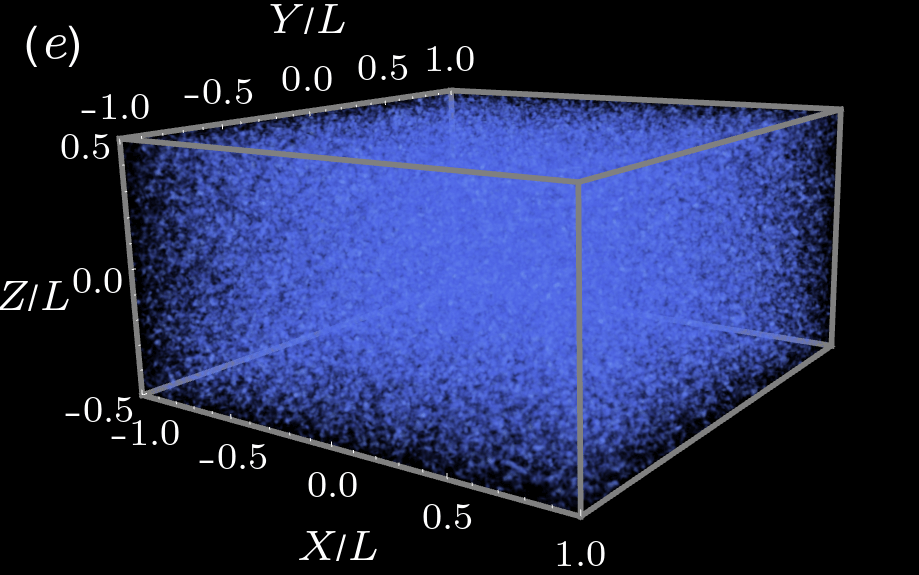} &
        \includegraphics[width=\subpanelwid]{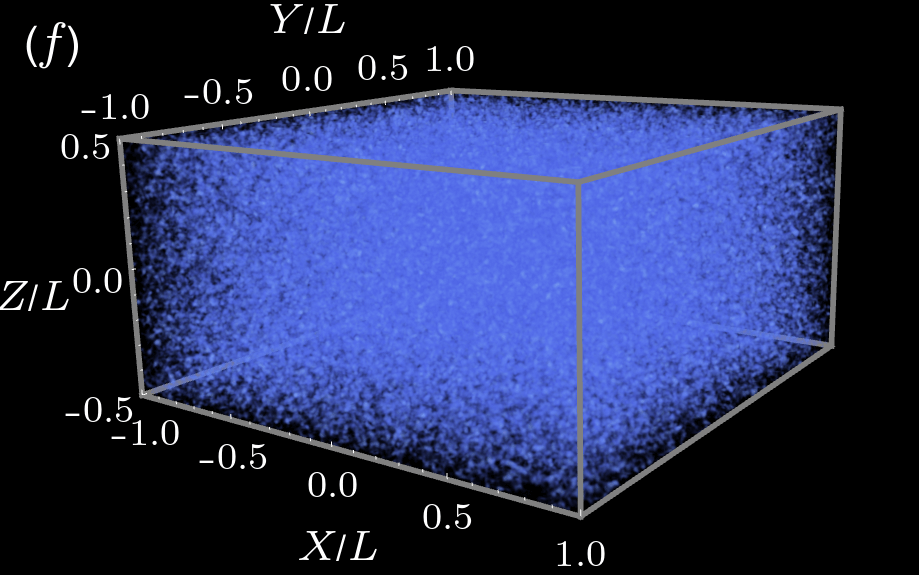}
    \end{tabular}}
    ~\vspace{5mm}\\
    \includegraphics[width=.75\textwidth]{imgs/rslt_figs/colorbar_2019}
    \caption{Snapshots of the effective temperature field at $t = 0t_s$. All simulations use Lees--Edwards boundary conditions. For all plots, a value of $a = 0.25$ and $\eta = 1.3$ is used in the opacity function. $\chi_{\text{bg}}$ is set to $\mu_\chi - 25\text{~K}$ in each pane in the opacity function. Figures (a)--(f) have $\mu_\chi = 450\tK, 500\tK, 525\tK, 550\tK, 575\tK$, and $600\tK$ respectively.}
    \label{fig:le_0}
\end{figure*}

\begin{figure*}
    \centering
\fcolorbox{black}{black}{
    \begin{tabular}{cc}
        \includegraphics[width=\subpanelwid]{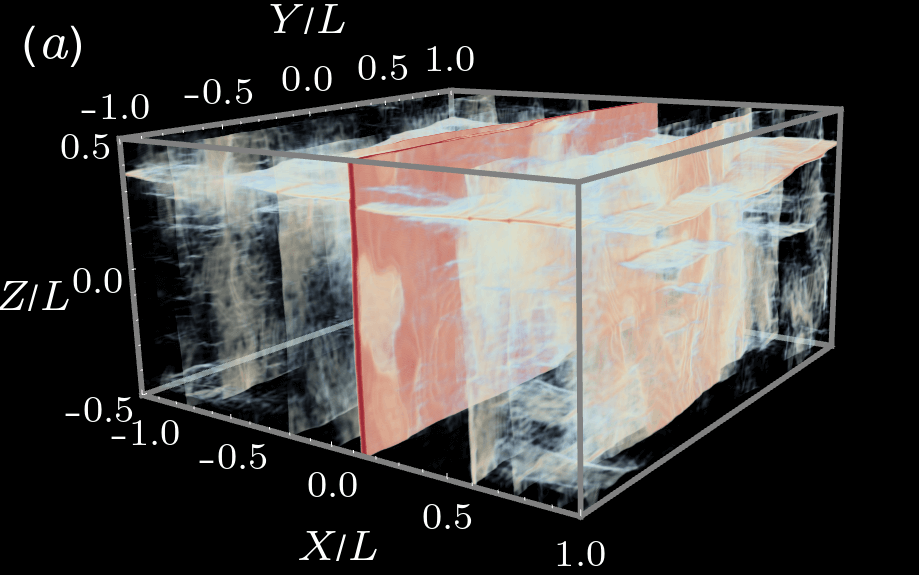} &
        \includegraphics[width=\subpanelwid]{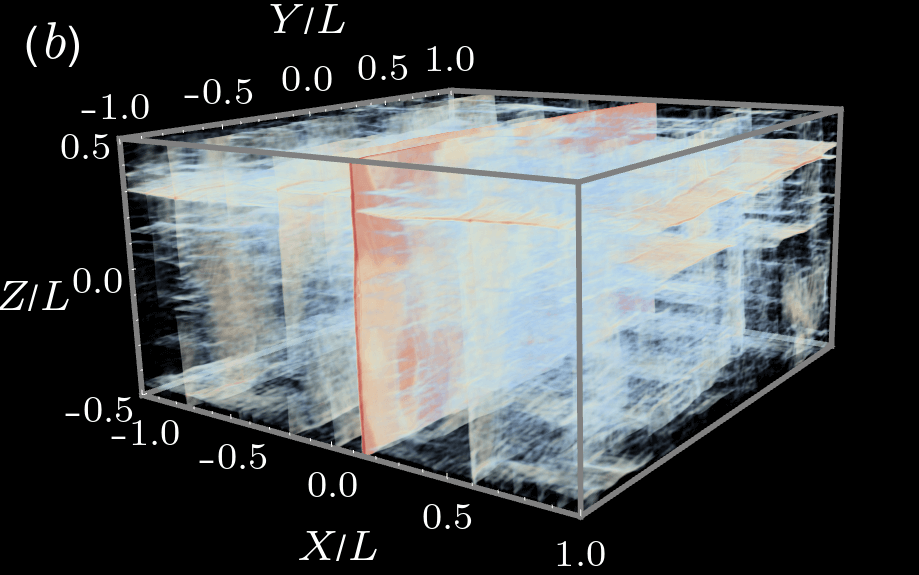} \\
        \includegraphics[width=\subpanelwid]{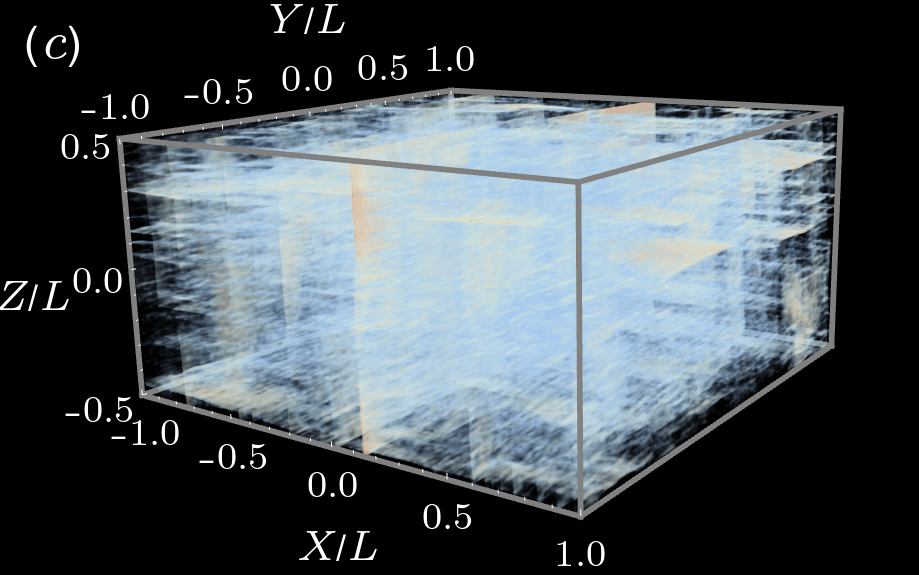} &
        \includegraphics[width=\subpanelwid]{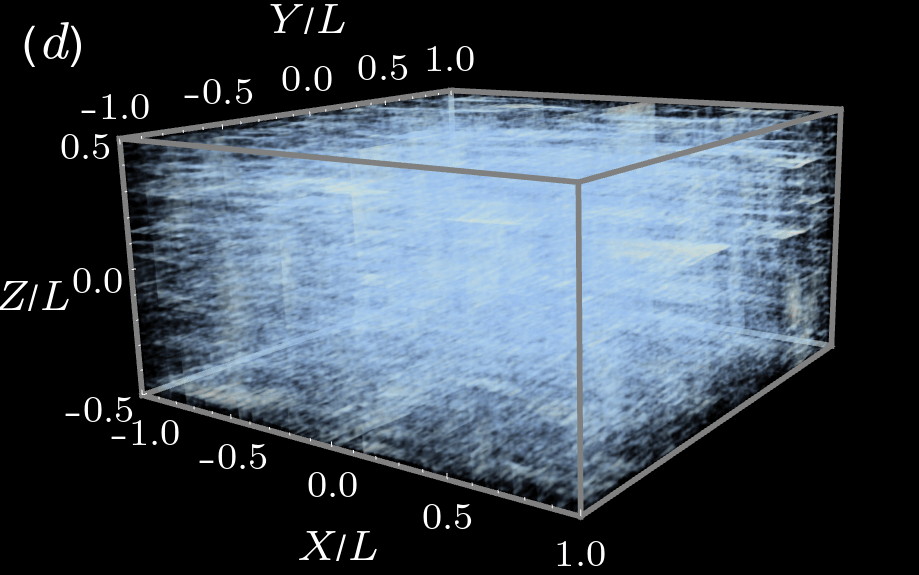} \\
        \includegraphics[width=\subpanelwid]{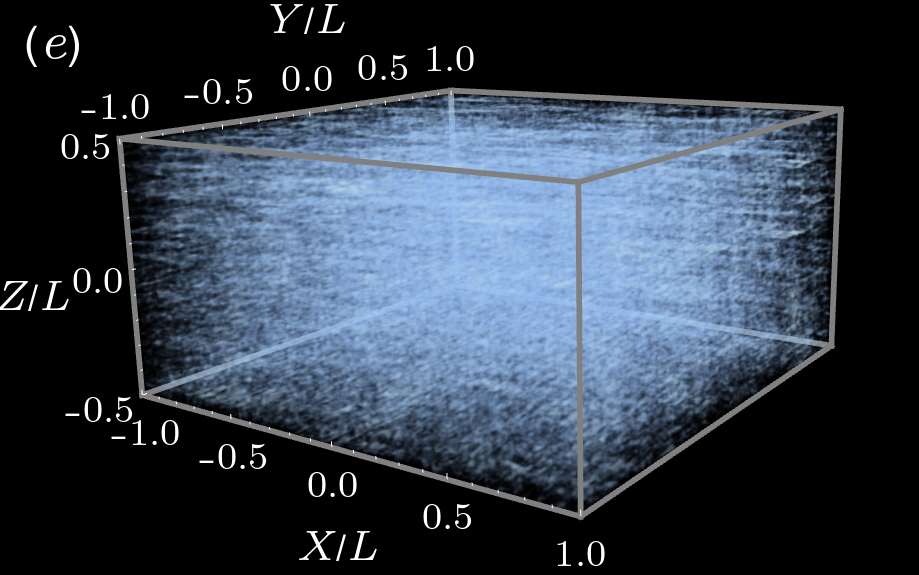} &
        \includegraphics[width=\subpanelwid]{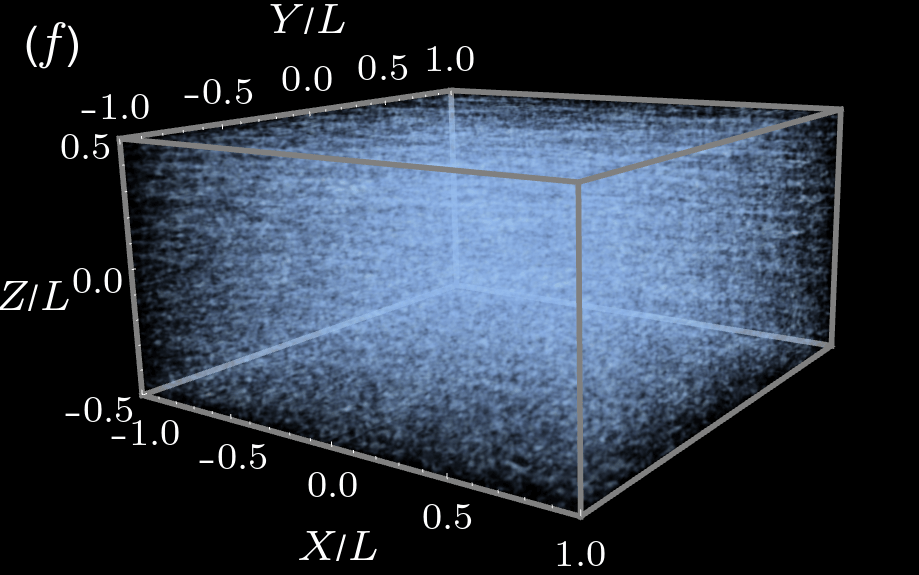}
    \end{tabular}}
    ~\vspace{5mm}\\
    \includegraphics[width=.75\textwidth]{imgs/rslt_figs/colorbar_2019}
    \caption{Snapshots of the effective temperature field at $t = 4\times 10^5t_s$. All simulations use Lees--Edwards boundary conditions. For all plots, a value of $a = 0.45$ and $\eta = 1.75$ is used in the opacity function. $\chi_{\text{bg}}$ is set to $\mu_\chi - 25\text{~K}$ in each pane in the opacity function. Figures (a)--(f) have $\mu_\chi = 450\tK, 500\tK, 525\tK, 550\tK, 575\tK$, and $600\tK$ respectively.}
    \label{fig:le_40}
\end{figure*}

\begin{figure*}
    \centering
\fcolorbox{black}{black}{
    \begin{tabular}{cc}
        \includegraphics[width=\subpanelwid]{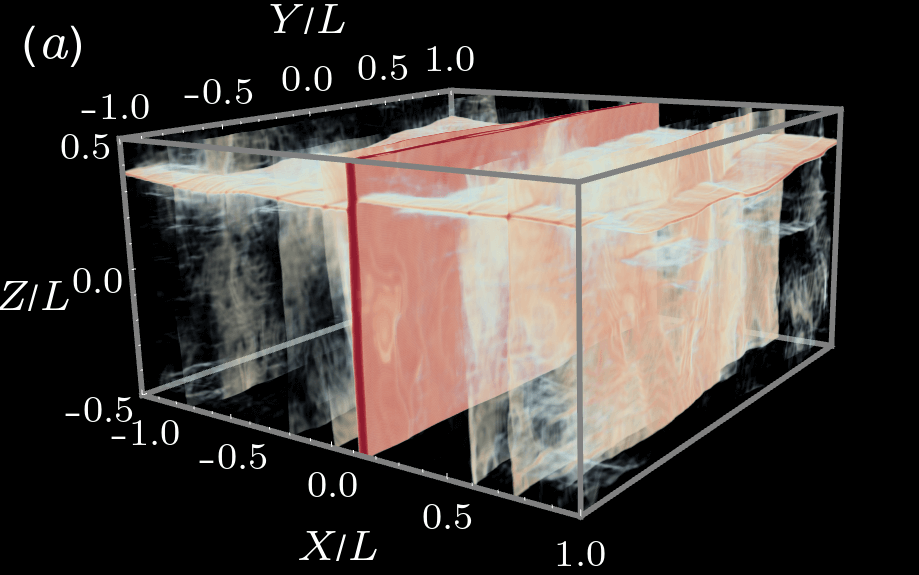} &
        \includegraphics[width=\subpanelwid]{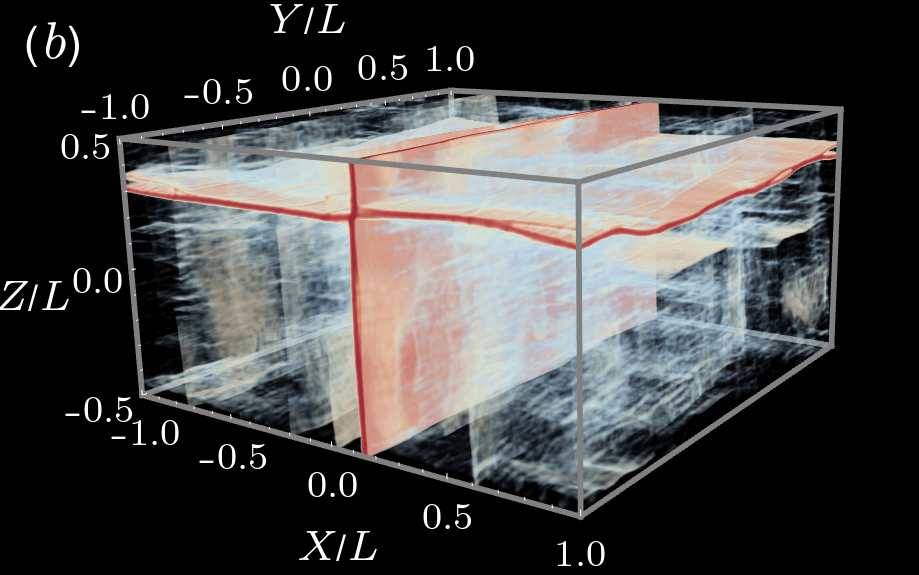} \\
        \includegraphics[width=\subpanelwid]{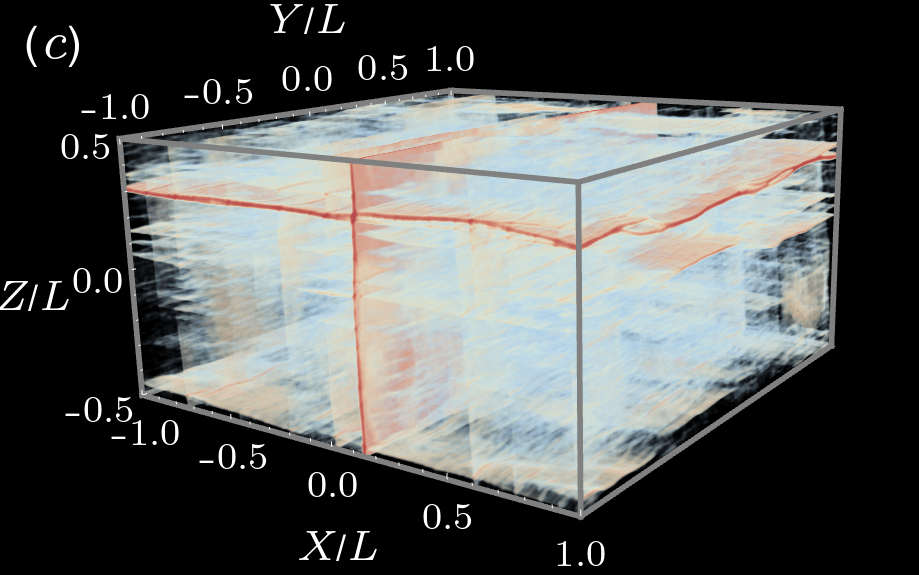} &
        \includegraphics[width=\subpanelwid]{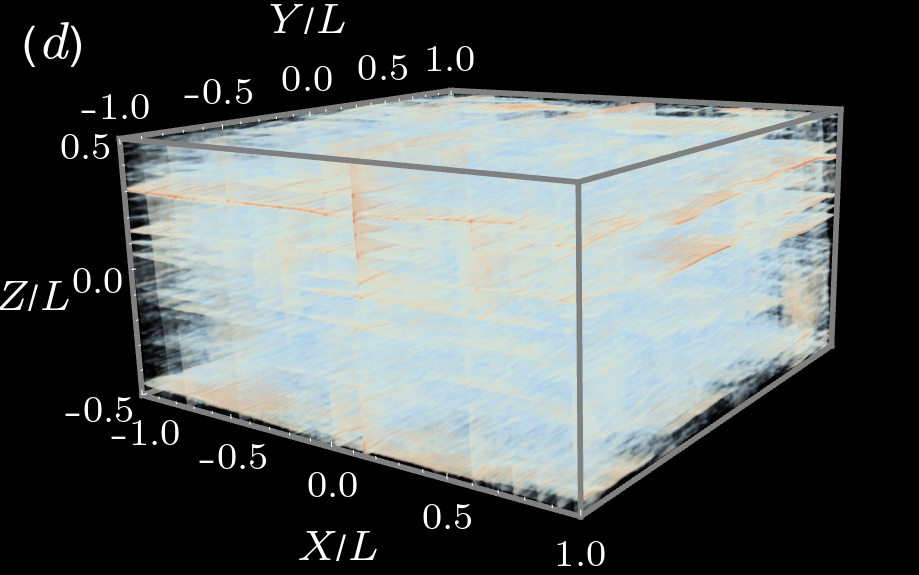} \\
        \includegraphics[width=\subpanelwid]{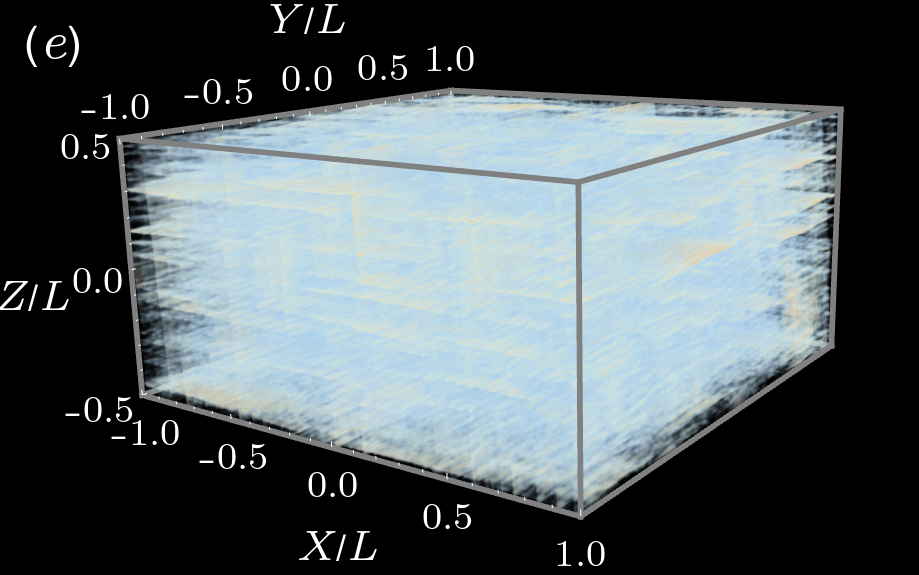} &
        \includegraphics[width=\subpanelwid]{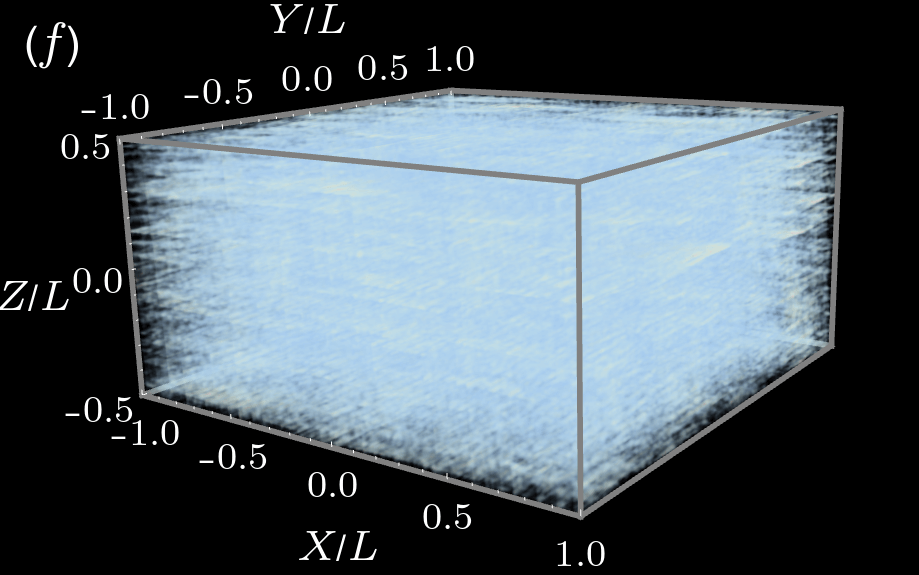}
    \end{tabular}}
    ~\vspace{5mm}\\
    \includegraphics[width=.75\textwidth]{imgs/rslt_figs/colorbar_2019}
    \caption{Snapshots of the effective temperature field at $t = 6\times 10^5t_s$. All simulations use Lees--Edwards boundary conditions. For all plots, a value of $a = 0.45$ and $\eta = 1.75$ is used in the opacity function. $\chi_{\text{bg}}$ is set to $\mu_\chi - 25\text{~K}$ in each pane in the opacity function. Figures (a)--(f) have $\mu_\chi = 450\tK, 500\tK, 525\tK, 550\tK, 575\tK$, and $600\tK$ respectively.}
    \label{fig:le_60}
\end{figure*}

\begin{figure*}
\fcolorbox{black}{black}{
    \begin{tabular}{cc}
        \includegraphics[width=\subpanelwid]{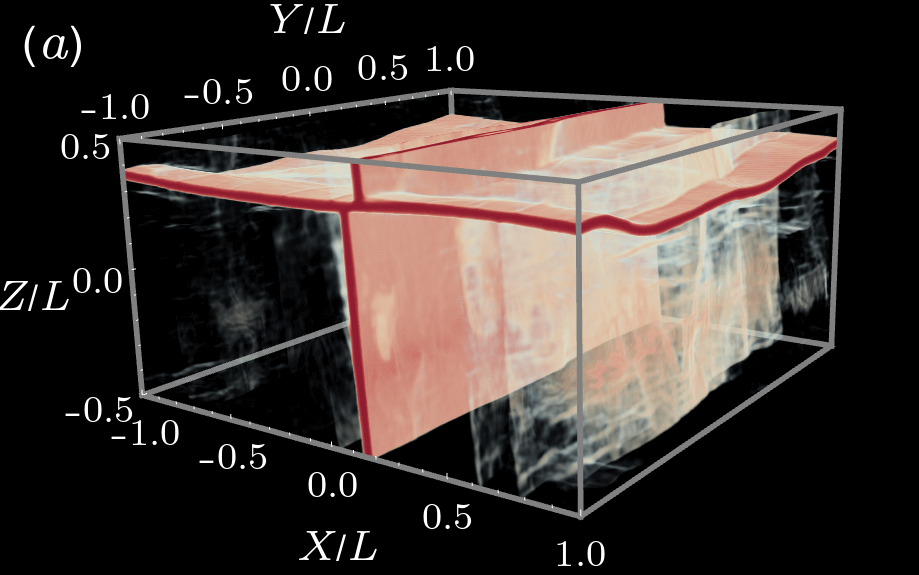} &
        \includegraphics[width=\subpanelwid]{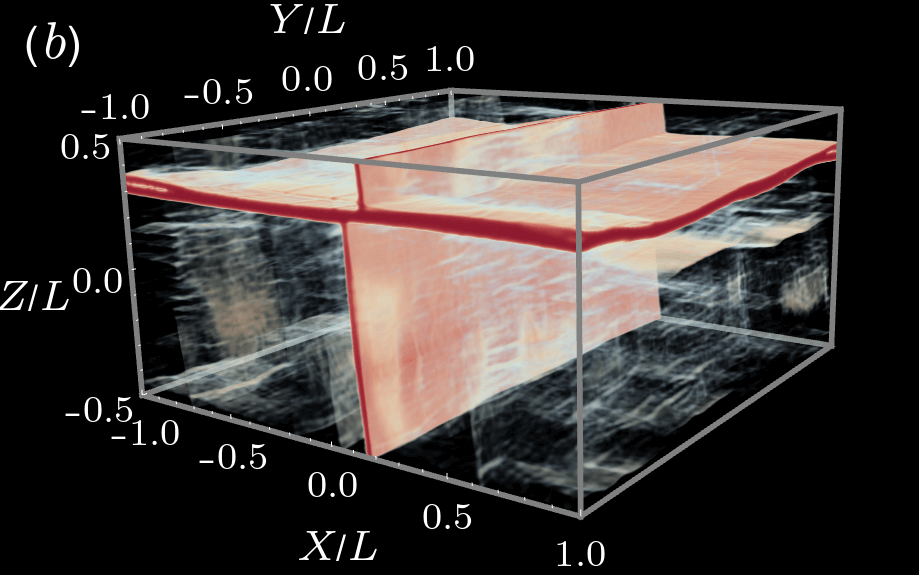} \\
        \includegraphics[width=\subpanelwid]{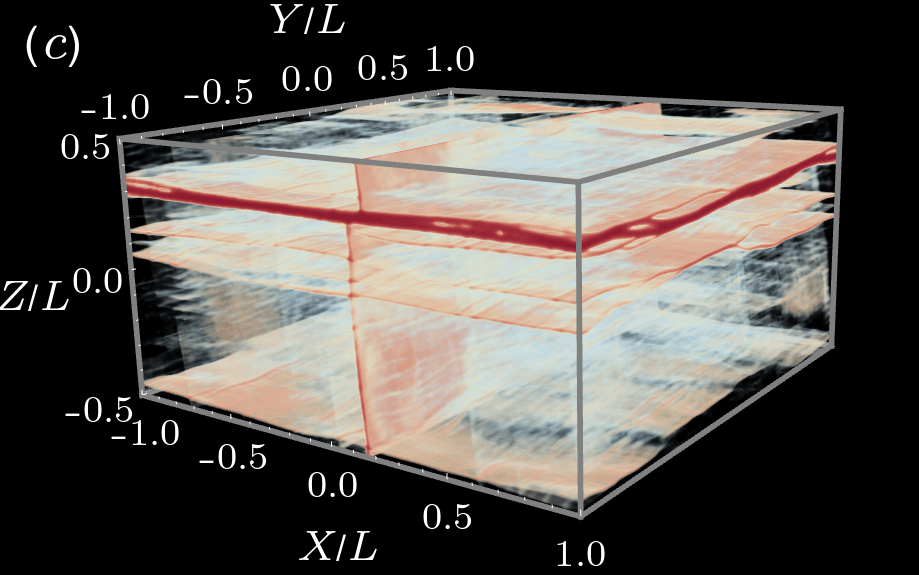} &
        \includegraphics[width=\subpanelwid]{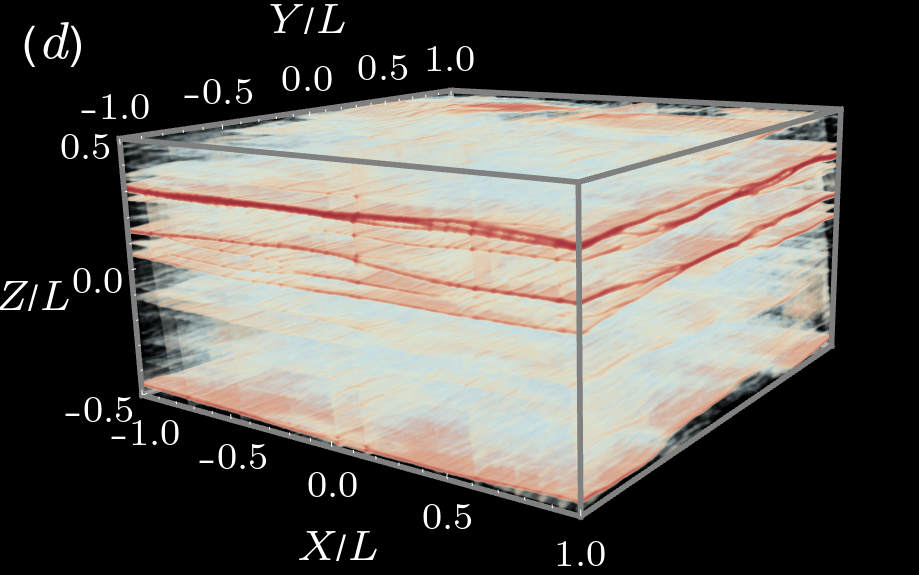} \\
        \includegraphics[width=\subpanelwid]{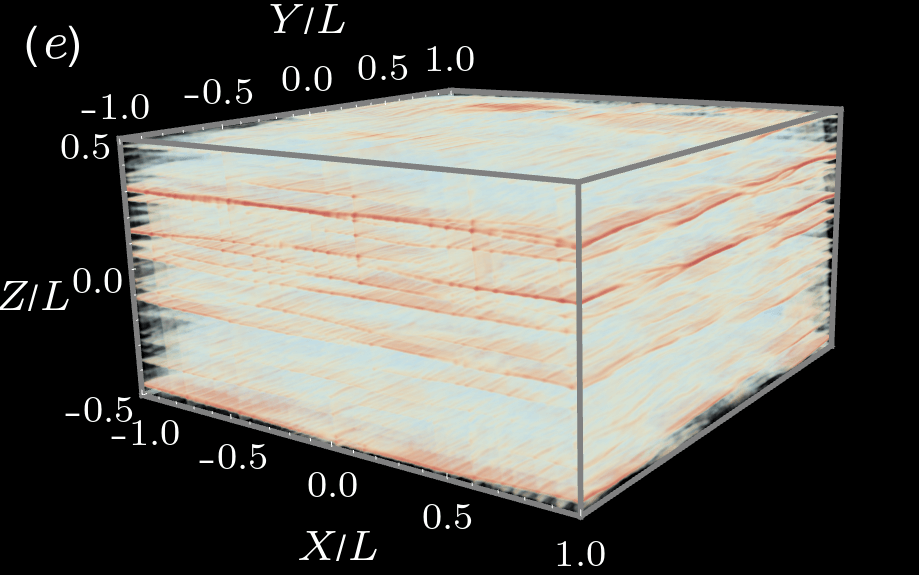} &
        \includegraphics[width=\subpanelwid]{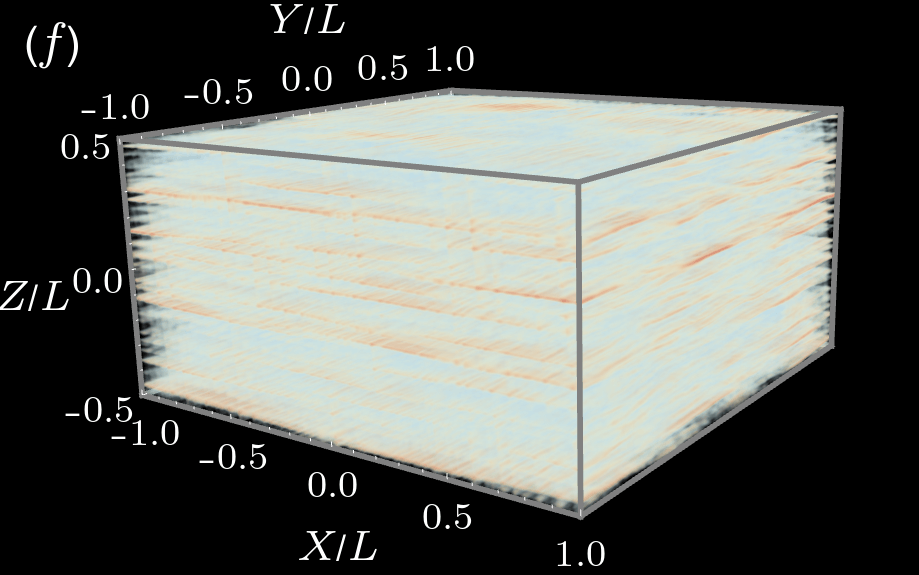}
    \end{tabular}}
    ~\vspace{5mm}\\
    \includegraphics[width=.75\textwidth]{imgs/rslt_figs/colorbar_2019}
    \caption{Snapshots of the effective temperature field at $t = 10^6t_s$. All simulations use Lees--Edwards boundary conditions. For all plots, a value of $a = 0.75$ and $\eta = 2$ is used in the opacity function. $\chi_{\text{bg}}$ is set to $\mu_\chi - 25\text{~K}$ in each pane in the opacity function. Figures (a)--(f) have $\mu_\chi = 450\tK, 500\tK, 525\tK, 550\tK, 575\tK$, and $600\tK$ respectively.}
    \label{fig:le_100}
\end{figure*}

The results for an identical sequence of simulations in the case of Lees--Edwards boundary conditions are displayed in Figs.~\ref{fig:le_0}--\ref{fig:le_100}. The initial conditions are displayed in Fig.~\ref{fig:le_0}, which differ from those in Fig.~\ref{fig:clamp_0}, as the convolution used to generate the initial distribution wraps around over the boundary in $Z$ to enforce periodicity.

By $t = 4\times 10^5 t_s$ in Fig.~\ref{fig:le_40}(a), a single vertical shear band has formed, along with an additional, weaker vertical band and a similar horizontal band. These bands are also visible in Fig.~\ref{fig:le_40}(b) earlier in their development. Vertical shear bands do not typically form in continuum simulations with non-periodic boundary conditions in $Z$, but are frequently seen in MD simulations~\citep{falk98, Shi2005}, indicating that the orientation of shear bands could be strongly related to the specific boundary conditions used.

Further progression is clear at $t = 6\times 10^5t_s$ in Fig.~\ref{fig:le_60}. Figure \ref{fig:le_60}(a) is similar to Fig.~\ref{fig:le_40}(a). Figure \ref{fig:le_60}(b) displays significant strengthening of the early-stage bands in Fig.~\ref{fig:le_40}(b). Figs.~\ref{fig:le_60}(c) and (d) show the initiation of several shear bands.

Figure \ref{fig:le_100} shows the results for $t = 10^6 t_s$. Figure \ref{fig:le_100}(a) displays a horizontal band perpendicular to the vertical band that exhibits significant curvature. Figure \ref{fig:le_100}(b) shows a similar result, with a thinner vertical band and a thicker, flatter horizontal band. Figure \ref{fig:le_100}(c) shows similar features, but also displays the development of several additional horizontal bands extending to the bottom of the simulation domain. Furthermore, the thick horizontal band in Fig.~\ref{fig:le_100}(b) can be seen to split and fracture. In Fig.~\ref{fig:le_100}(d), the vertical band has been almost entirely washed out, and a complex branching pattern of horizontal bands is seen. Figs.~\ref{fig:le_100}(e) and (f) are similar to Fig.~\ref{fig:le_100}(d), but they are earlier in their development and some of the fine-scale features are washed out due to the high background $\chi$ field. The agreement with the non-periodic simulations increases strongly as $\mu_\chi$ is increased.

Figure \ref{fig:le_100} clearly demonstrates the effect of increasing $\mu_\chi$ with periodic boundary conditions. In the simulations with lower $\mu_\chi$, nucleation of vertical shear bands is more likely, and curved horizontal bands develop later in the simulation than vertical bands. As $\mu_\chi$ is increased, the vertical bands begin to disappear. As in the non-periodic case, the curvature in the horizontal bands decreases with $\mu_\chi$. As $\mu_\chi$ is increased further, the vertical bands disappear altogether. In this regime, increasing $\mu_\chi$ increases the number of horizontal bands, and the qualitative agreement with the nonperiodic results is good. These results suggest that, for higher $\mu_\chi$, the effect of periodicity in the $Z$ direction is less significant.

\subsection{Pure shear}
\label{ssec:ps}
As a second example transformation, we now consider pure shear deformation. In metallic glasses, experimental evidence indicates that pure shear is the primary failure mode under compressive stress, and several recent experiments have been conducted probing BMGs under pure shear conditions~\citep{Chen2016, Zhang2003, Wright2001, Sun2014}. Pure shear is particularly interesting due to the simplicity of its implementation in the transformation methodology. To simulate pure shear on a physical grid, it would be necessary to impose traction boundary conditions on the top, bottom, and sides, which poses computational difficulties. Within the transformation framework, pure shear can be implemented using the transformation
\begin{equation}
    \bT(t) = \begin{pmatrix} A(t) & 0 & 0 \\ 0 & 1 & 0 \\ 0 & 0 & \frac{1}{A(t)} \end{pmatrix}.
    \label{eqn:ps_trans}
\end{equation}
$A(t)$ can be chosen as any monotonically increasing function of time. In the following studies, we choose $A(t) = e^{\xi t}$, where $\xi$ is a simulation parameter that sets the rate of extension and compression of the $x$ and $z$ axes respectively. For numerical stability, it is important to choose $\xi$ small, so that large stresses do not cause divergences in the simulation fields. In our simulations, we choose $\xi$ as a fraction of $t_f$, which effectively sets the strain at the end of the simulation.

\subsubsection{Gaussian defects}
\label{sssec:defects_ps}
To gain some physical intuition about shear banding dynamics with pure shear boundary conditions, we first consider an example initial condition in $\chi$ corresponding to localized defects in the material. It is expected that diagonal shear bands will nucleate outwards from the imperfections. We first define the quantities
\begin{align*}
    \mathbf{X}_1 &= L\times(-0.3, -0.3, 0.2), \\
    \mathbf{X}_2 &= L\times(0.3, 0.3, -0.2), \\
    \mathbf{X}_3 &= L\times(-0.1, -0.1, 0.1), \\
    \mathbf{X}_4 &= L\times(0.1, 0.1, -0.1), \\
    \mathbf{X}_5 &= L\times(0, 0, 0), \\
    \delta_1 &= \delta_2 = \delta_5 = 200, \\
    \delta_3 &= \delta_4 = 150,
\end{align*}
and then take the initial condition in $\chi$ to be
\begin{equation}
    \chi\left(\mathbf{X}, t=0\right) = 550 \text{~K} + \left(200 \text{~K}\right) \sum_{i = 1}^5 e^{-\delta_i\left\Vert\frac{\mathbf{X}}{L} - \frac{\mathbf{X}_i}{L}\right\Vert^2}.
    \label{eqn:super_diag}
\end{equation}
Simulations are performed with periodic and non-periodic boundary conditions in $Z$ on grids of size $256\times 256 \times 128$. The $X$ and $Y$ dimensions use periodic boundary conditions in both cases. The diffusion length scale is set to $l = 3h$ and the quasi-static timestep is $\Delta t = 200 t_s$. $\xi$ in Eq.~\ref{eqn:ps_trans} is set to be $\frac{1}{4t_f}$ with $t_f = 4\times 10^5 t_s$ the total simulation duration, so that $A(t_f) = e^{1/4} \approx 1.284$. Both periodic and non-periodic boundary conditions are considered. The periodic simulation is run with 32 processes on an Ubuntu Linux computer with dual 14-core 1.70~GHz Intel Xeon E5-2650L processors. The total time is 4.721 hours, the total time spent in multigrid V-cycles is 3.283, and the total number of V-cycles is 11596. The non-periodic simulation is run with 32 processes on an Ubuntu Linux computer with dual 10-core 2.20~GHz Intel Xeon E5-2630 processors. The total time is 8.832 hours, the total time spent in multigrid V-cycles is 6.506 hours, and the total number of V-cycles is 11075.

\begin{figure*}
\fcolorbox{black}{black}{
    \begin{tabular}{cc}
        \includegraphics[width=\subpanelwid]{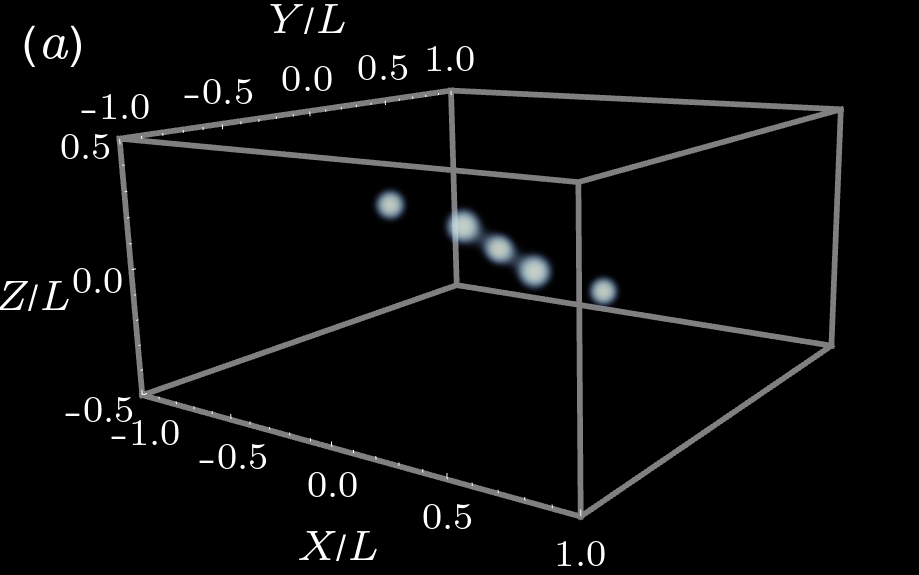} &
        \includegraphics[width=\subpanelwid]{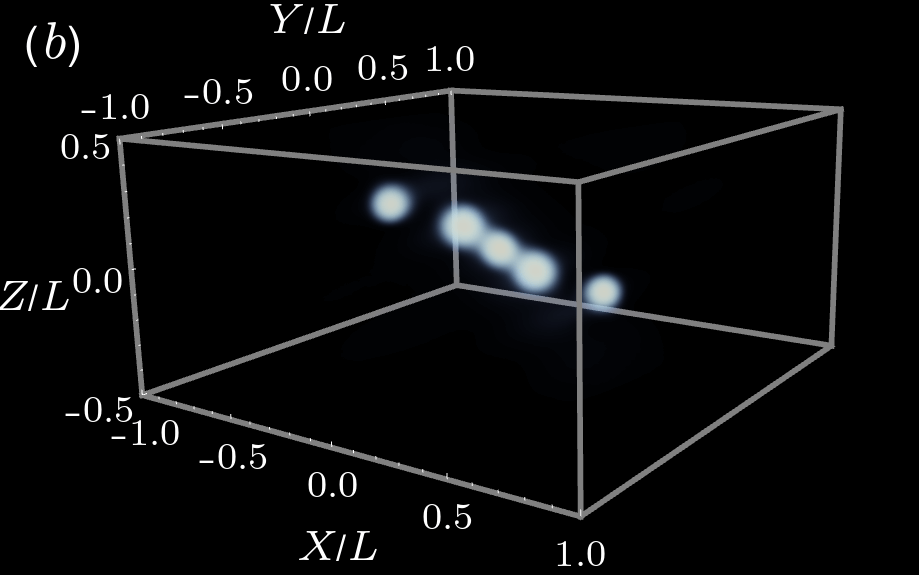} \\
        \includegraphics[width=\subpanelwid]{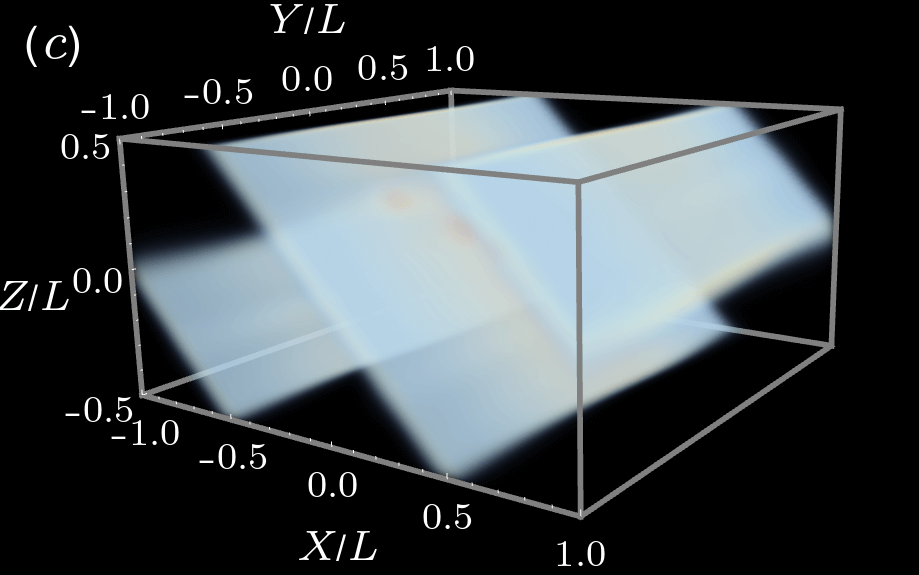} &
        \includegraphics[width=\subpanelwid]{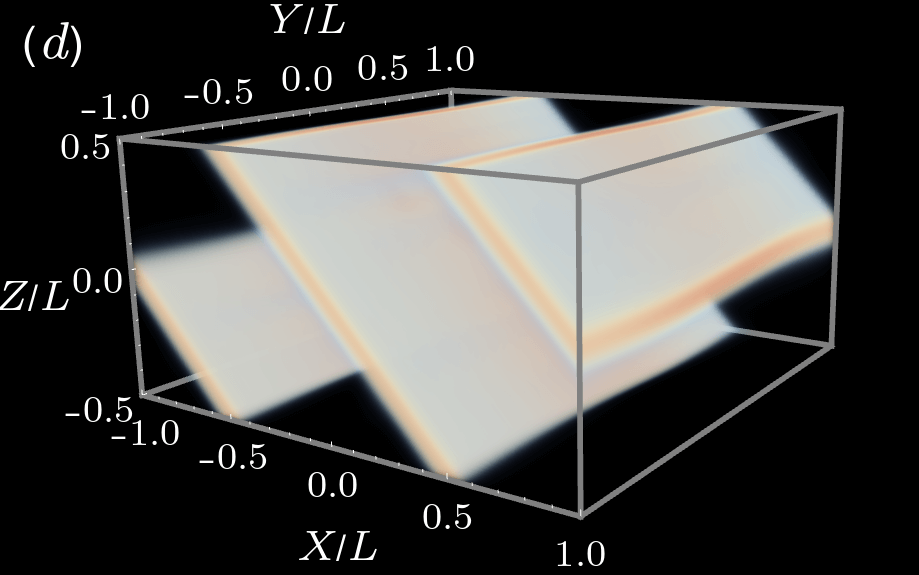} \\
        \includegraphics[width=\subpanelwid]{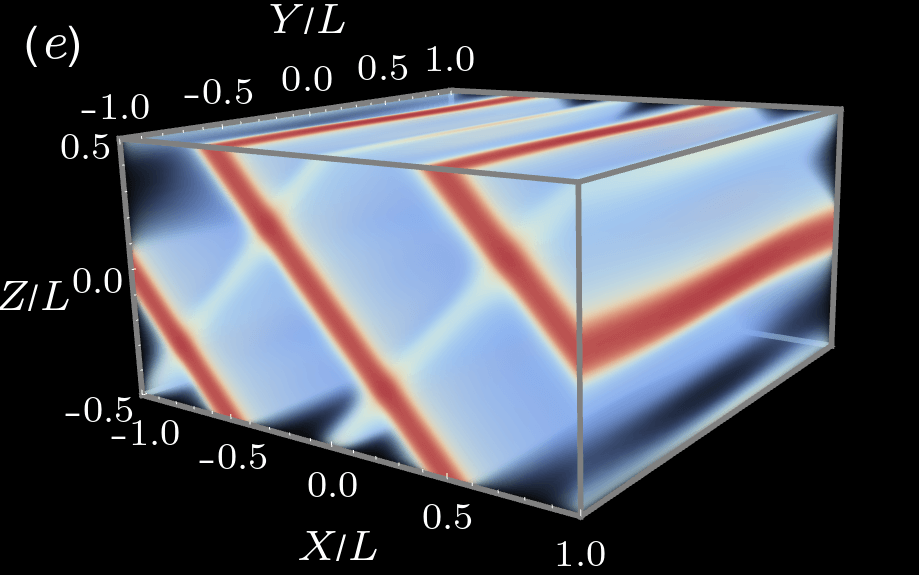} &
        \includegraphics[width=\subpanelwid]{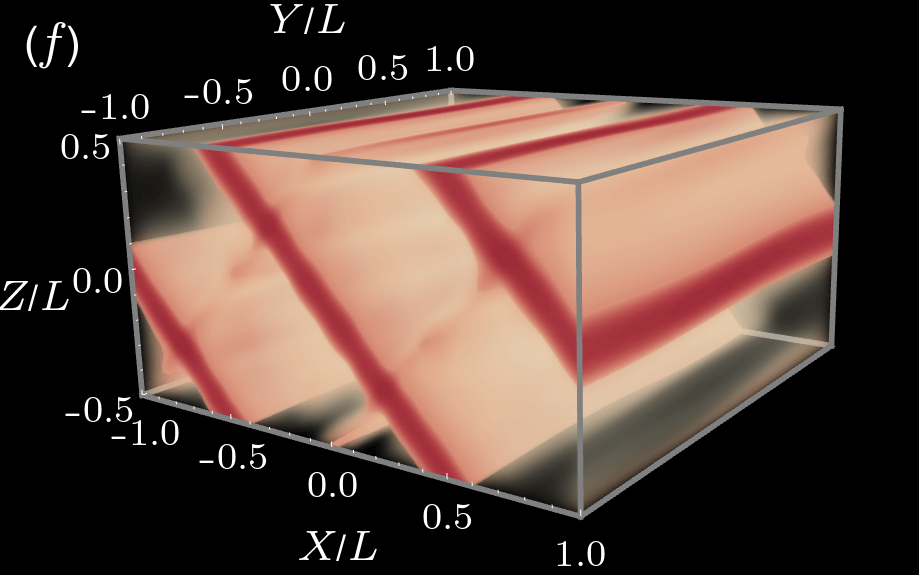}
    \end{tabular}}
    ~\vspace{5mm}\\
    \includegraphics[width=.75\textwidth]{imgs/rslt_figs/colorbar_2019}
    \caption{Snapshots of the effective temperature distribution $\chi(\bX, t)$ for a quasi-static simulation. Pure shear deformation is imposed via a domain transformation with an initial condition corresponding to a sequence of blips of elevated $\chi$ lying roughly along the superdiagonal of the simulation domain. This simulation uses periodic boundary conditions in all three directions. $\chi_{bg} = 550\text{~K}$ in the opacity function for all panels. (a) $t = 0 t_s$, $a = 0.75$, $\eta = 1.2$. (b) $t = 5\times 10^4 t_s$, $a = 0.75$, $\eta = 1.2$. (c) $t = 8\times 10^4 t_s$, $a = 0.75$, $\eta = 1.25$. (d) $t =  10^5 t_s$, $a = 0.75$, $\eta - 1.25$. (e) $t = 2 \times 10^5 t_s$, $a = 0.4$, $\eta = 1.6$. (e) $t = 4 \times 10^5 t_s$, $a = 1.1$, $\eta = 2.45$.}
    \label{fig:ps_LE}
\end{figure*}

\begin{figure*}
\fcolorbox{black}{black}{
    \begin{tabular}{cc}
        \includegraphics[width=\subpanelwid]{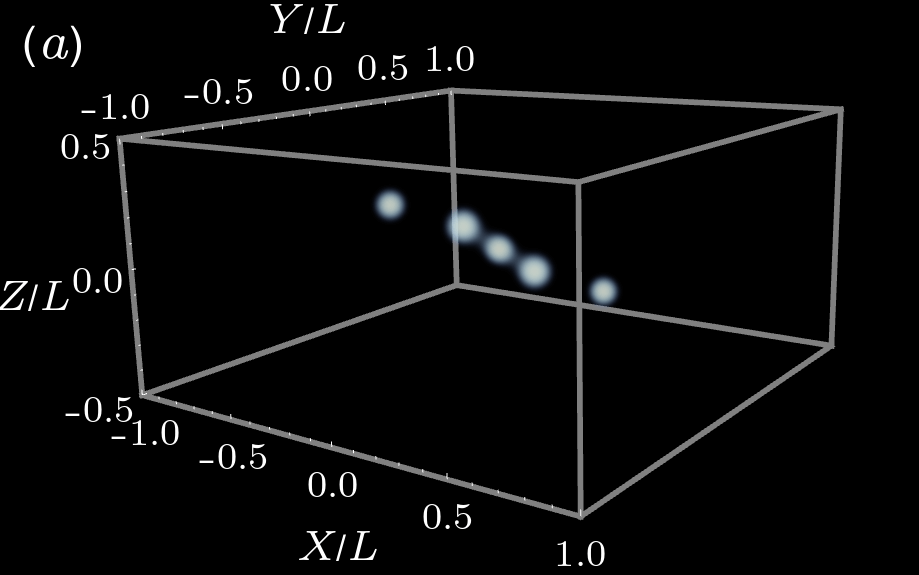} &
        \includegraphics[width=\subpanelwid]{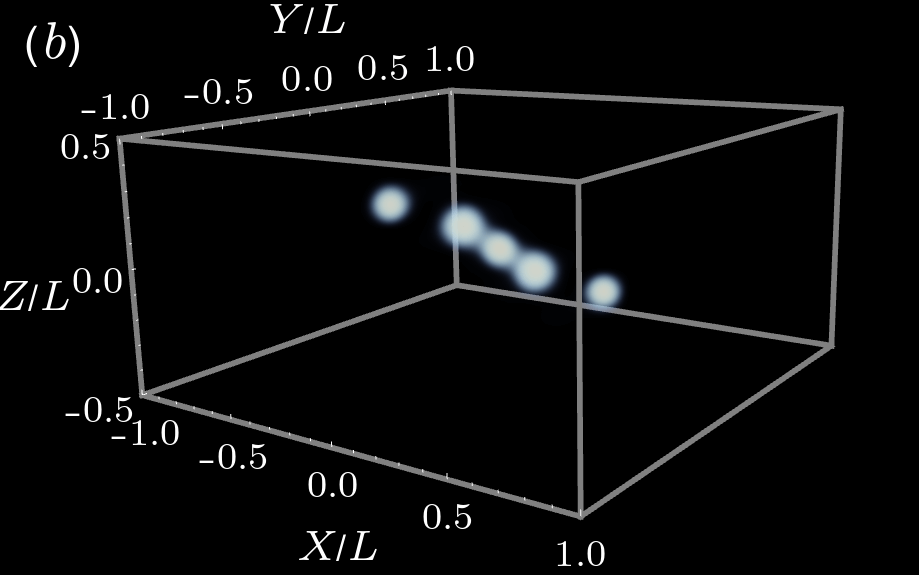} \\
        \includegraphics[width=\subpanelwid]{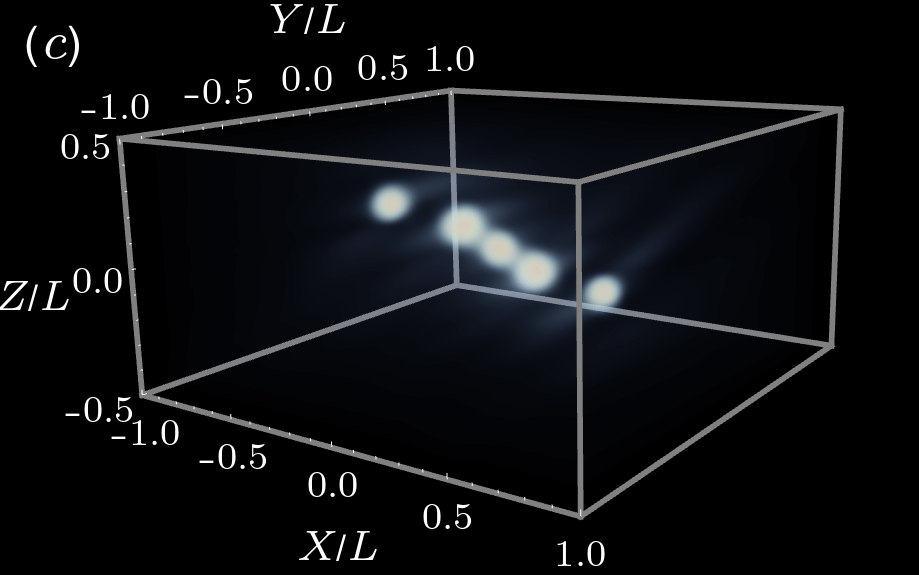} &
        \includegraphics[width=\subpanelwid]{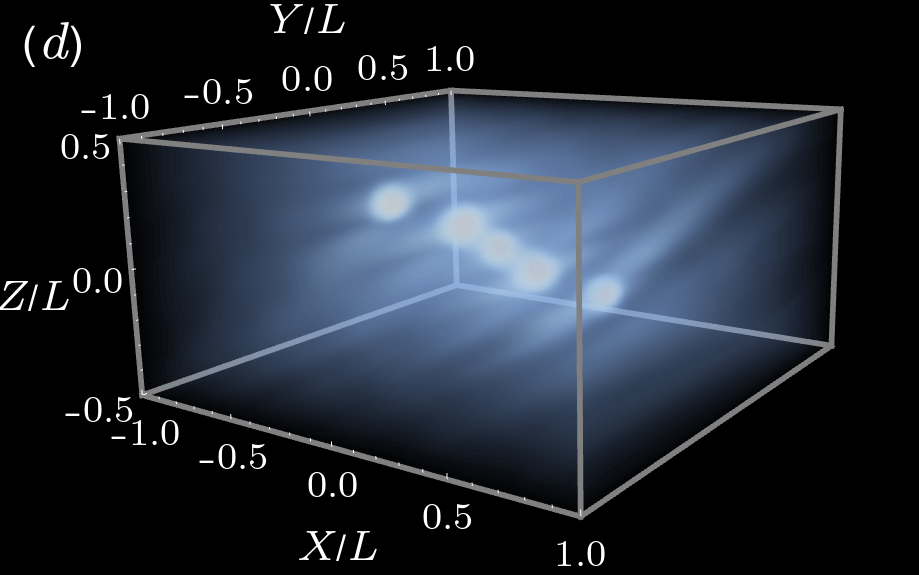} \\
        \includegraphics[width=\subpanelwid]{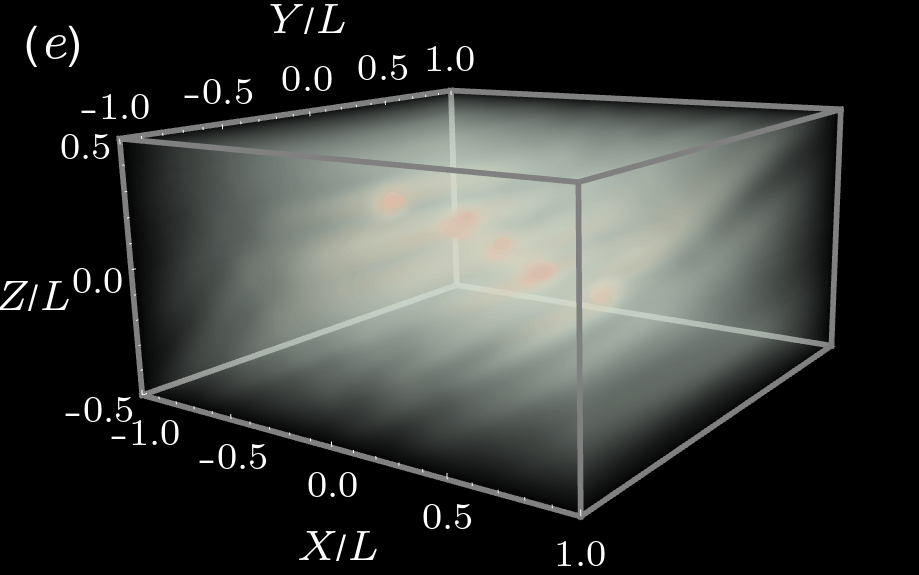} &
        \includegraphics[width=\subpanelwid]{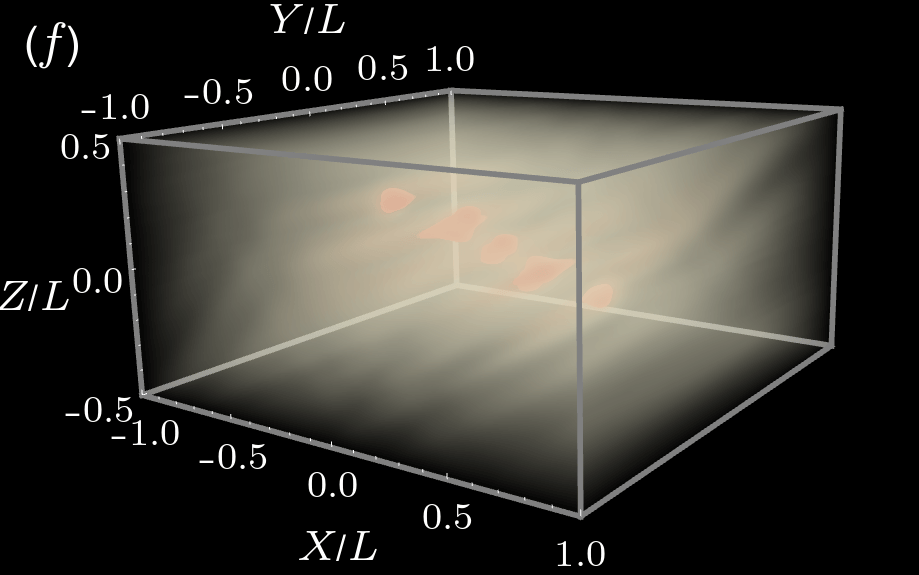}
    \end{tabular}}
    ~\vspace{5mm}\\
    \includegraphics[width=.75\textwidth]{imgs/rslt_figs/colorbar_2019}
    \caption{Snapshots of the effective temperature distribution $\chi(\bX, t)$ for a quasi-static simulation. Pure shear deformation is imposed via a domain transformation with an initial condition corresponding to a sequence of blips of elevated $\chi$ lying roughly along the superdiagonal of the simulation domain. This simulation uses non-periodic boundary conditions in $Z$ and is periodic in the $X$ and $Y$ directions. $\chi_{bg} = 550\text{~K}$ in the opacity function in all panels. (a) $t = 0t_s$, $a = 0.75$, $\eta = 1.2$. (b) $t = 5\times 10^3 t_s$, $a = 0.75$, $\eta = 1.2$. (c) $t = 10^4 t_s$, $a = 0.75$, $\eta = 1.45$. (d) $t = 1.5\times 10^4 t_s$, $a = 0.75$, $\eta = 1.45$. (e) $t = 4\times 10^5 t_s$, $a = 1.75$, $\eta = 1.75$.}
    \label{fig:ps_clamp}
\end{figure*}

Results for periodic and non-periodic boundary conditions are shown in Figs.~\ref{fig:ps_LE} and \ref{fig:ps_clamp} respectively. The initial conditions are shown in Figs.~\ref{fig:ps_LE}(a) and \ref{fig:ps_clamp}(a). In both Figs.~\ref{fig:ps_LE}(b) and \ref{fig:ps_clamp}(b) at $t = 5\times 10^4t_s$, some spreading in the $\chi$ field is seen near the defects. Shortly thereafter, the dynamics in the nonperiodic and periodic cases begin to differ dramatically.

At $t = 8\times 10^4 t_s$ in Fig.~\ref{fig:ps_LE}(c), three diagonal bands are seen connecting the defects. The bands become more pronounced at $t = 10^5 t_s$ in Fig~\ref{fig:ps_LE}(d). This continues into $t = 2\times 10^5 t_s$ in Fig.~\ref{fig:ps_LE}(e), along with the addition of diagonal bands perpendicular to the original bands. Both bands continue to grow larger and stronger by $t = 4 t\times 10^5 t_s$ in Fig.~\ref{fig:ps_LE}(f).

The deformation dynamics with non-periodic boundary conditions are significantly different. By $t = 8\times 10^4 t_s$ in Fig.~\ref{fig:ps_clamp}(c), diagonal bands have started to nucleate off of each defect in a direction roughly perpendicular to the first bands formed in the periodic simulation. By $t = 10^5 t_s$ in Fig.~\ref{fig:ps_clamp}(d), this nucleation has grown more prominent, and an increase in the background $\chi$ field is seen across the simulation. At times $t = 2\times 10^5 t_s$ and $t = 2.5\times 10^5 t_s$ in Figs.~\ref{fig:ps_clamp}(e) and \ref{fig:ps_clamp}(f) respectively, the qualitative structure remains the same, but the background $\chi$ field continues to increase. Unlike in the periodic case, true system-spanning shear bands do not fully form.

\subsubsection{A randomly fluctuating effective temperature field}
\label{sssec:rndm_ps}

In this section, we consider the same sequence of random initializations as in Sec.~\ref{sssec:rndm_ss}, but now subject to pure shear deformation. Simulations are performed across values of $\mu_\chi = 450\tK$, $500\tK$, $525\tK$, $550\tK$, $575\tK$, and $600\tK$ with a fixed value of $\sigma_\chi = 15\tK$. The diffusion length scale is set to $\frac{3}{2}h$ and the quasi-static timestep is set to $\Delta t = 200 t_s$. All simulations are conducted on a $512\times 512\times 256$ cell grid. A pure shear transformation of the form Eq.~\ref{eqn:ps_trans} is used with $A(t) = e^{\xi t}$ and a value of $\xi = \frac{1}{4t_f}$ with $t_f$ = $2\times 10^6 t_s$ so that $A(t_f) = e^{1/4}\approx 1.284$. Simulations are performed with fully periodic boundary conditions in all directions; non-periodic simulations produce qualitatively similar differences as in the case of simple shear. In all figure panels, $\chi_{bg}$ is set to be $\mu_{\chi} - 25\text{~K}$. Timing data for the simulations is reported in Table~\ref{tab:ps_times}.

\begin{table}
\centering
\resizebox{\textwidth}{!}{%
\begin{tabular}{|c|c|c|c|c|c|c|}
\hline
 & $\mu_\chi=450\text{~K}$ & $\mu_\chi=500\text{~K}$ & $\mu_\chi=525\text{~K}$ & $\mu_\chi=550\text{~K}$ & $\mu_\chi=575\text{~K}$ & $\mu_\chi=600\text{~K}$ \\ \hline
Total time (hours) & 232.7589 & 171.3655 & 159.3944 & 129.3347 & 164.1213 & 206.7765 \\ \hline
V-cycle time (hours) & 176.7667 & 125.4595 & 115.4887 & 81.0119 & 114.6882 & 143.3524 \\ \hline
\# of V-cycles & 66214 & 63871 & 62263 & 59750 & 57772 & 55149 \\ \hline
Time/V-cycle (seconds) & 9.6107 & 7.0734 & 6.6774 & 4.8811 & 7.1467 & 9.3577 \\ \hline
\multicolumn{1}{|l|}{Processor} & \multicolumn{1}{l|}{\begin{tabular}[c]{@{}l@{}}Dual 10-core \\ 2.20~GHz Intel Xeon\\ E5-2630 v4\end{tabular}} & \multicolumn{1}{l|}{\begin{tabular}[c]{@{}l@{}}Dual 10-core \\ 2.20~GHz Intel Xeon\\ E5-2630 v4\end{tabular}} & \multicolumn{1}{l|}{\begin{tabular}[c]{@{}l@{}}Dual 10-core \\ 2.20~GHz Intel Xeon\\ Silver 4114 v4\end{tabular}} & \multicolumn{1}{l|}{\begin{tabular}[c]{@{}l@{}}Dual 16-core \\ 2.10~GHz Intel Xeon\\ E5-2683 v4\end{tabular}} & \multicolumn{1}{l|}{\begin{tabular}[c]{@{}l@{}}Dual 14-core \\ 1.70~GHz Intel Xeon\\ E5-2650L v4\end{tabular}} & \begin{tabular}[c]{@{}l@{}}Dual 14-core \\ 1.70~GHz Intel Xeon\\ E5-2650L v4\end{tabular} \\ \hline
\end{tabular}
}
\caption{Data describing the total time, total time spent in multigrid V-cycles, total number of multigrid V-cycles, average time spent per multigrid V-cycle, and processor details for each simulation. This data applies to the randomly initialized simulations with periodic boundary conditions and pure shear deformation. The number of required multigrid V-cycles decreases as the background $\chi$ field increases, likely due to more homogeneous dynamics. Each simulation uses $32$ processes.}
\label{tab:ps_times}
\end{table}

\begin{figure*}
\fcolorbox{black}{black}{
    \begin{tabular}{cc}
        \includegraphics[width=\subpanelwid]{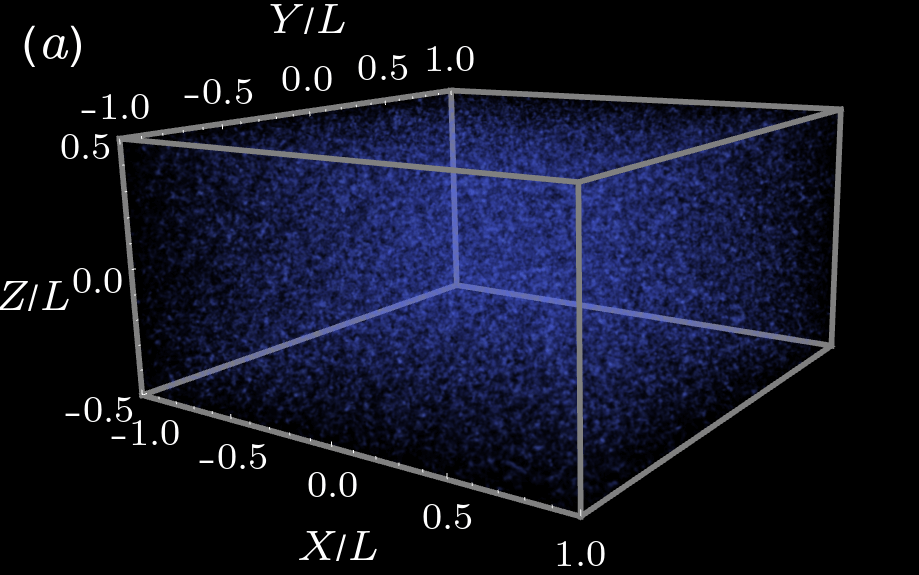} &
        \includegraphics[width=\subpanelwid]{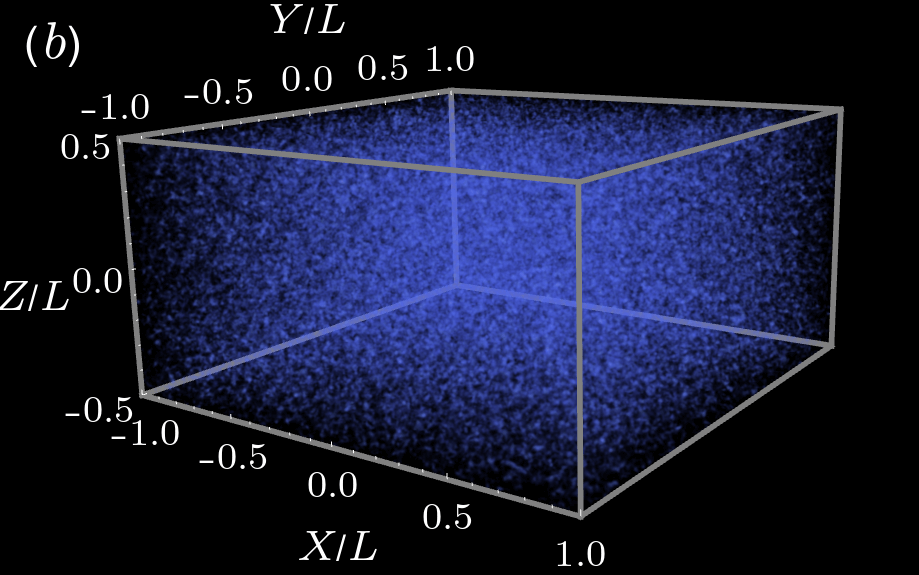} \\
        \includegraphics[width=\subpanelwid]{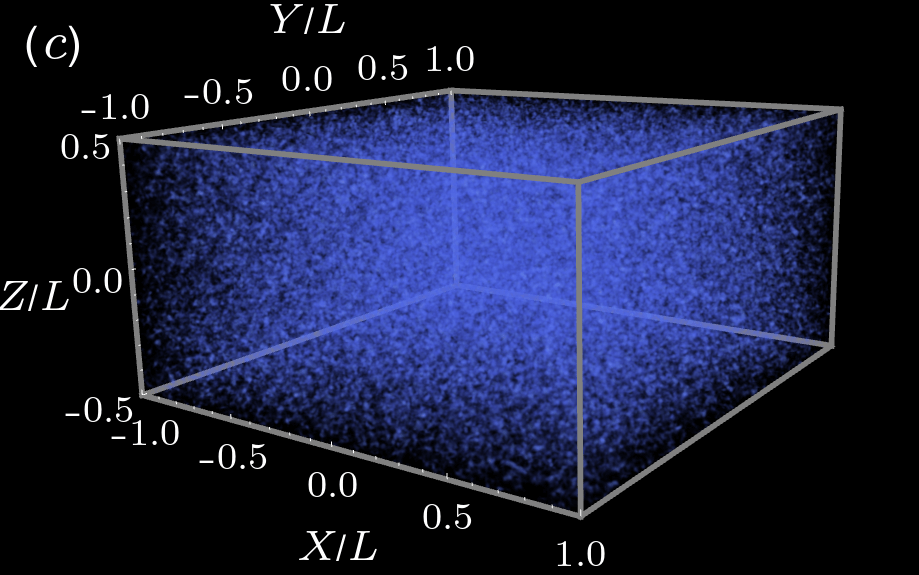} &
        \includegraphics[width=\subpanelwid]{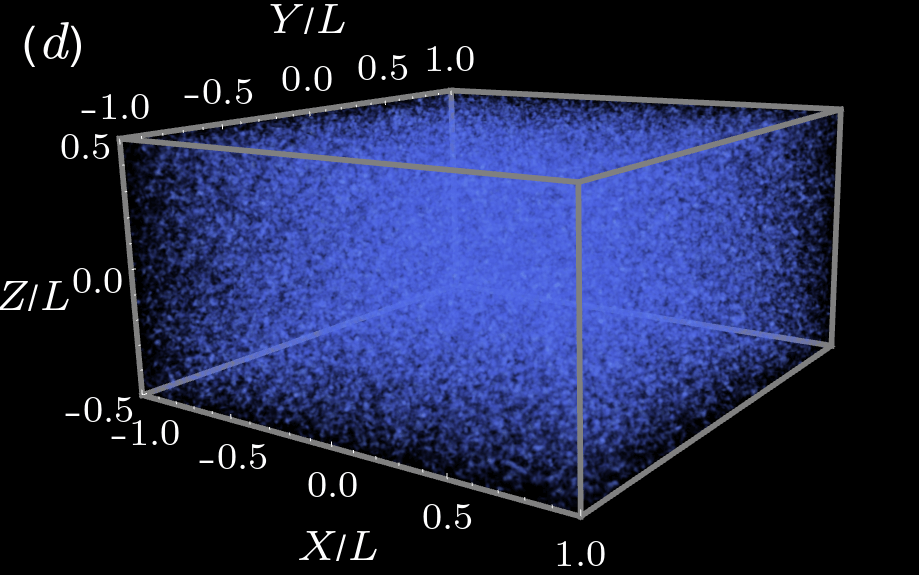} \\
        \includegraphics[width=\subpanelwid]{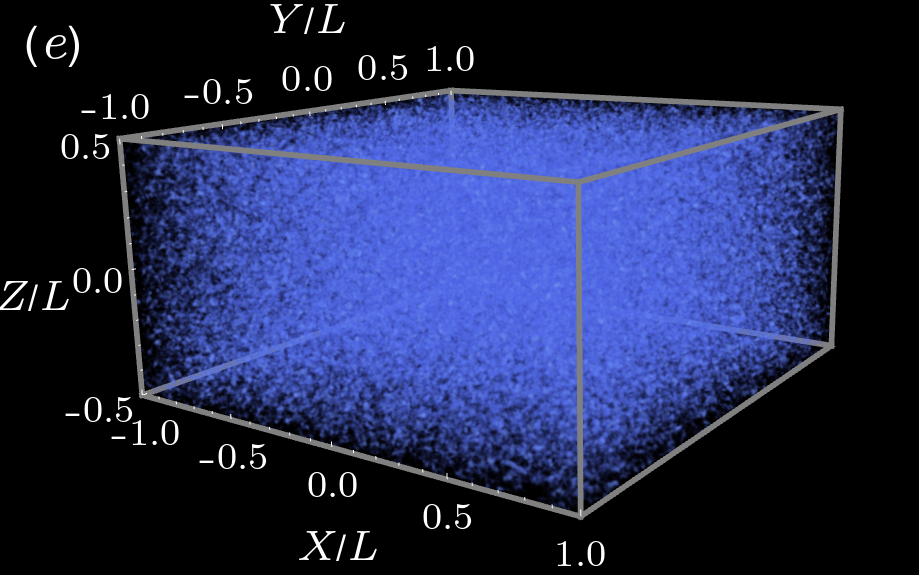} &
        \includegraphics[width=\subpanelwid]{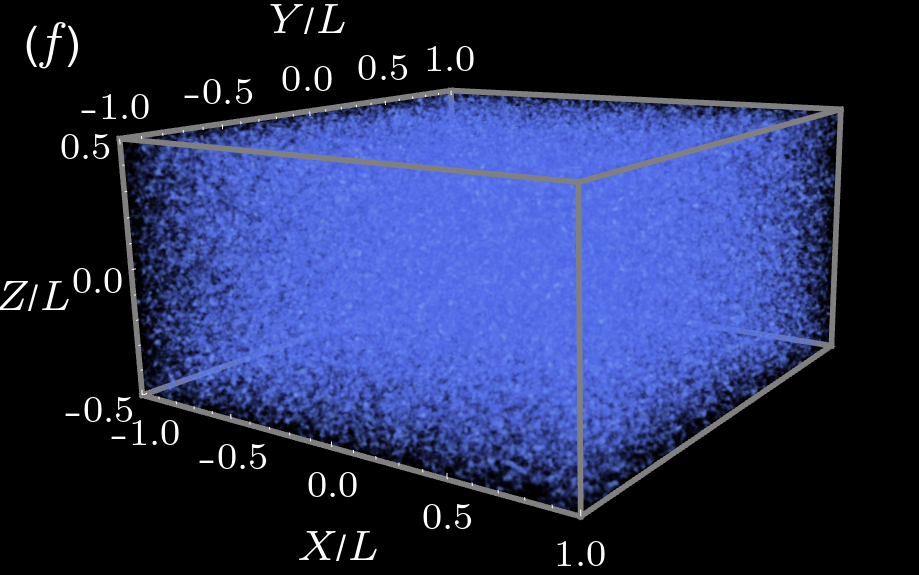}
    \end{tabular}}
    ~\vspace{5mm}\\
    \includegraphics[width=.75\textwidth]{imgs/rslt_figs/colorbar_2019}
    \caption{Snapshots of the effective temperature field at $t = 0t_s$ with pure shear transformation imposed on the domain. All simulations use periodic boundary conditions. For all plots, a value of $a = 0.25$ and $\eta = 1.3$ is used in the opacity function, and $\chi_\text{bg}$ is set to $\mu_\chi - 25\text{~K}$. Figures (a)--(f) have $\mu_\chi = 450\tK, 500\tK, 525\tK, 550\tK, 575\tK, 600\tK$ respectively.}
    \label{fig:ps_le_0}
\end{figure*}

\begin{figure*}
\fcolorbox{black}{black}{
    \begin{tabular}{cc}
        \includegraphics[width=\subpanelwid]{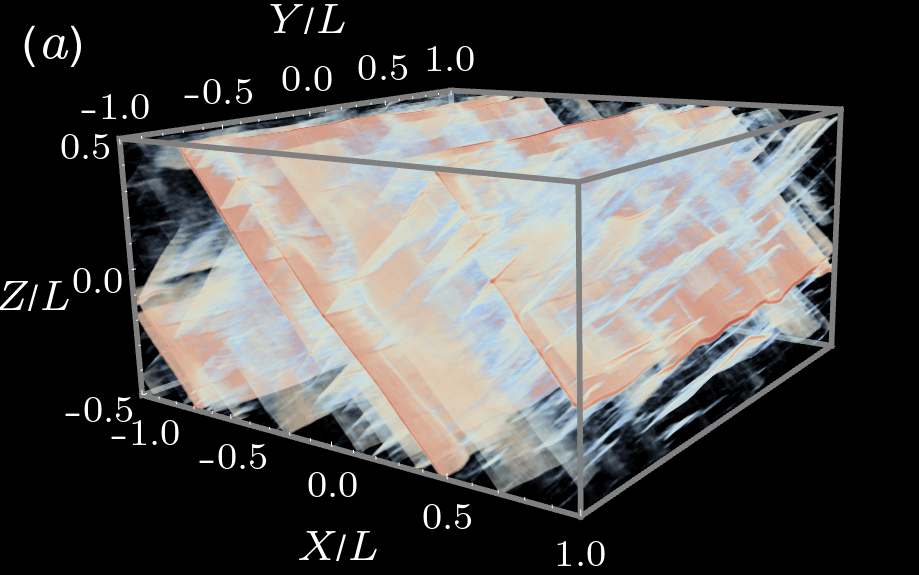} &
        \includegraphics[width=\subpanelwid]{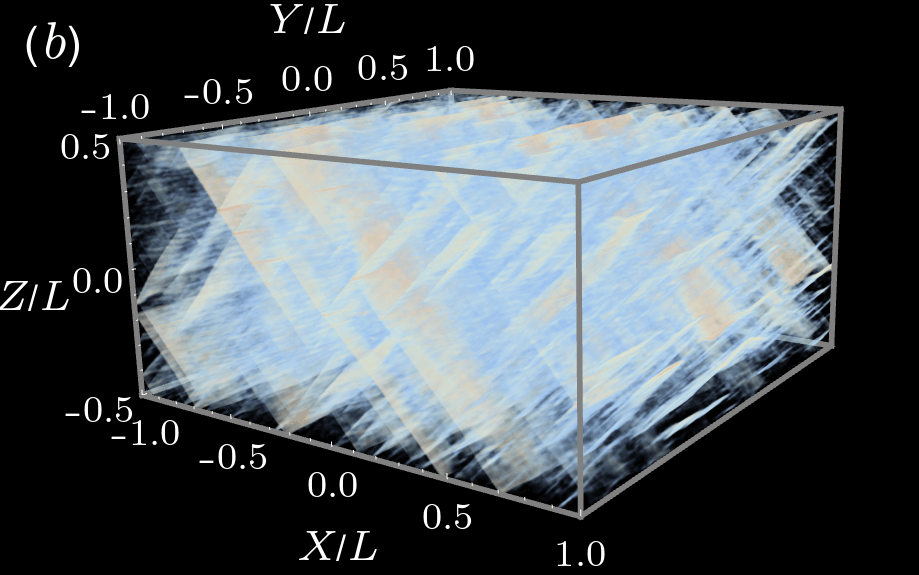} \\
        \includegraphics[width=\subpanelwid]{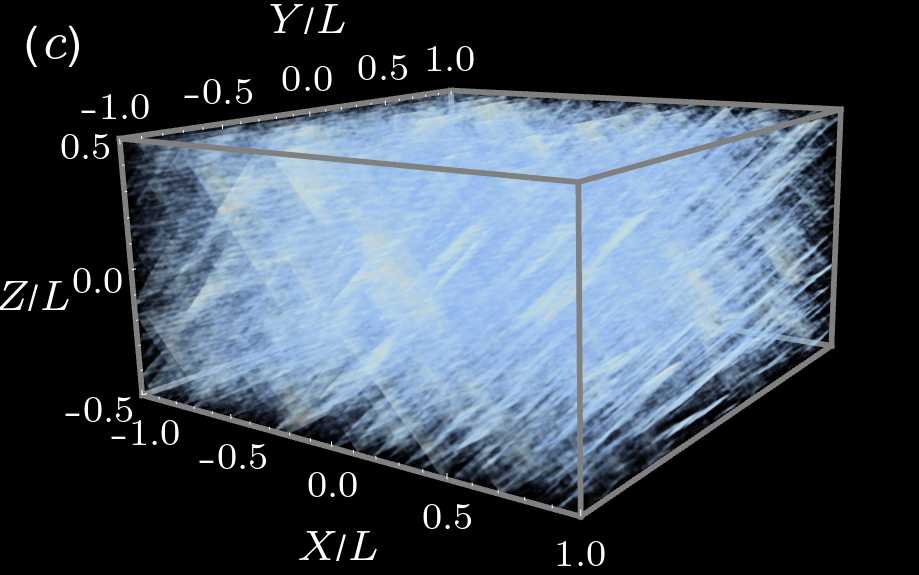} &
        \includegraphics[width=\subpanelwid]{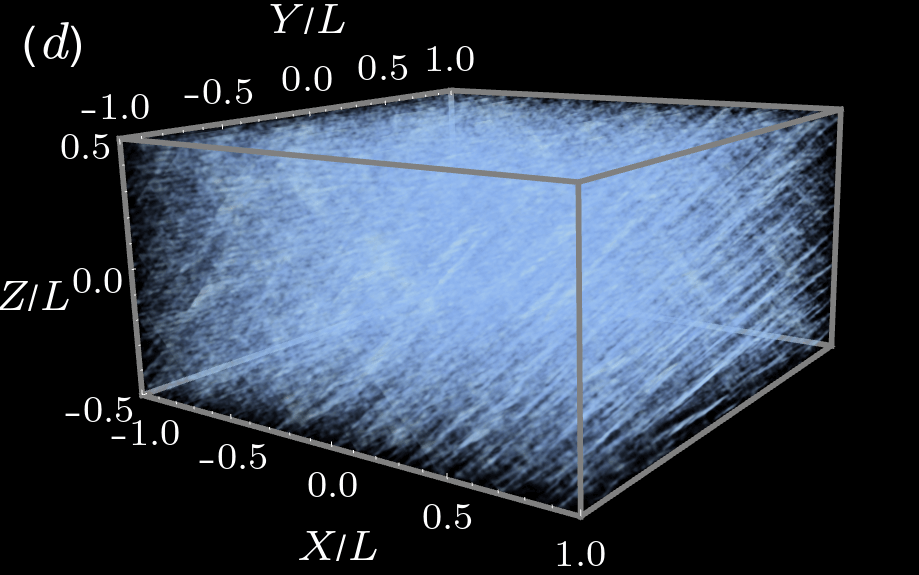} \\
        \includegraphics[width=\subpanelwid]{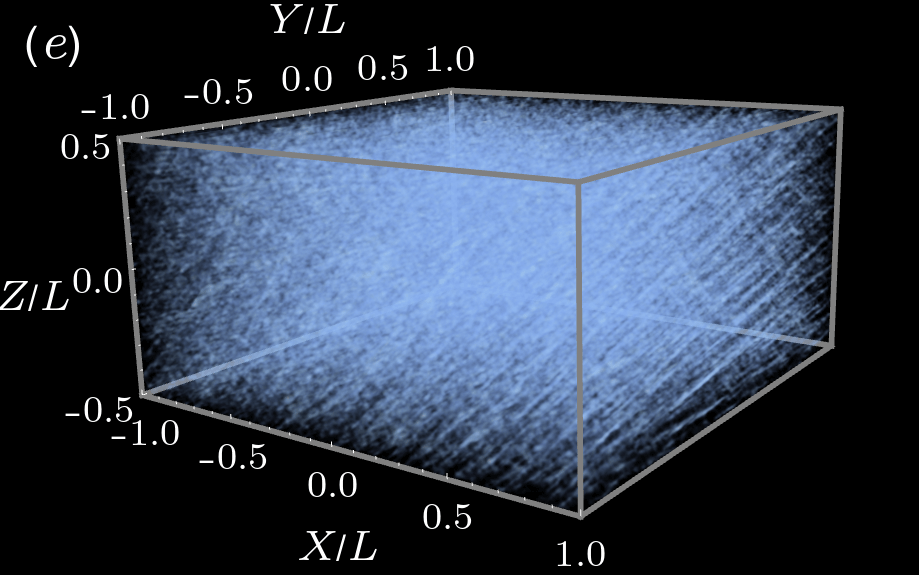} &
        \includegraphics[width=\subpanelwid]{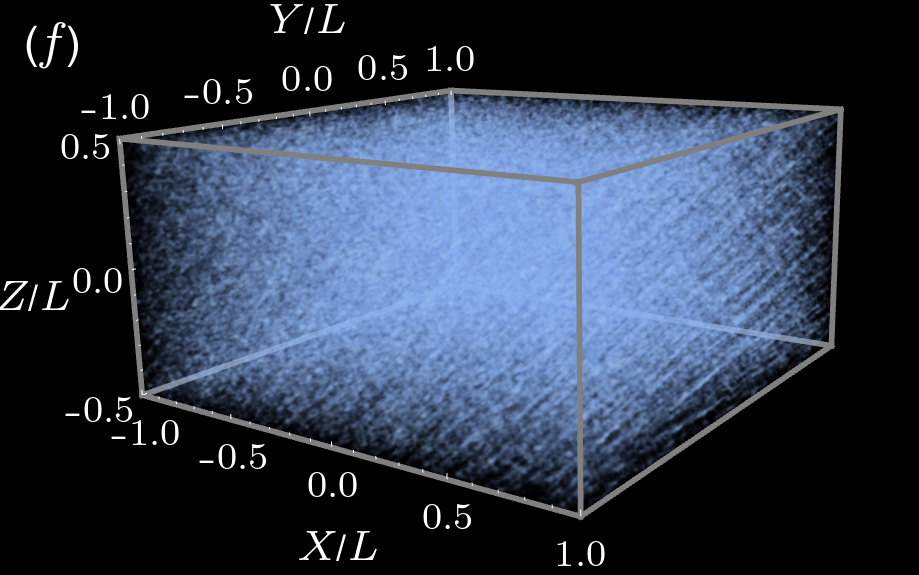}
    \end{tabular}}
    ~\vspace{5mm}\\
    \includegraphics[width=.75\textwidth]{imgs/rslt_figs/colorbar_2019}
    \caption{Snapshots of the effective temperature field at $t = 3\times 10^5 t_s$ with a pure shear transformation imposed on the domain. All simulations use periodic boundary conditions. For all plots, a value of $a = 0.55$ and $\eta = 1.5$ is used in the opacity function, and $\chi_\text{bg}$ is set to $\mu_\chi - 25\text{~K}$. Figures (a)--(f) have $\mu_\chi = 450\tK, 500\tK, 525\tK, 550\tK, 575\tK, 600\tK$ respectively.}
    \label{fig:ps_le_30}
\end{figure*}

\begin{figure*}
\fcolorbox{black}{black}{
    \begin{tabular}{cc}
        \includegraphics[width=\subpanelwid]{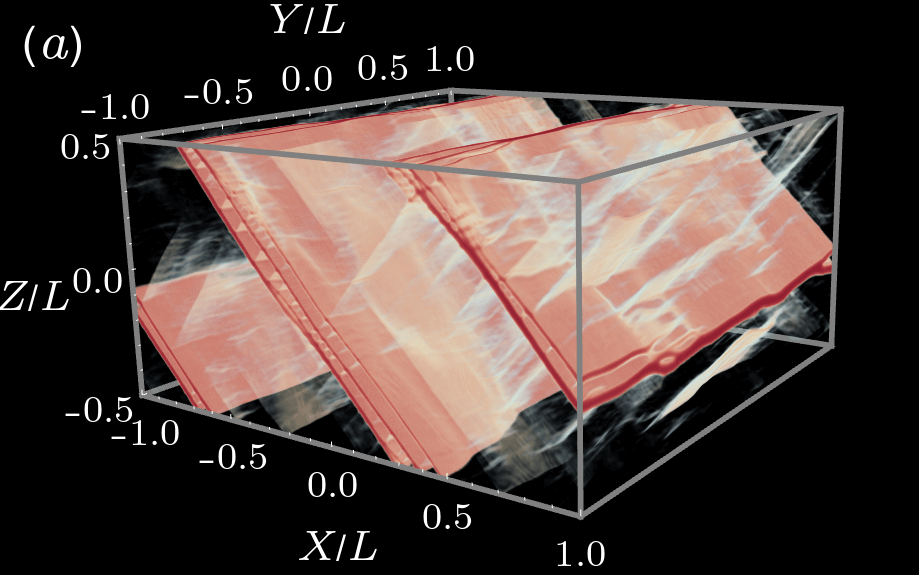} &
        \includegraphics[width=\subpanelwid]{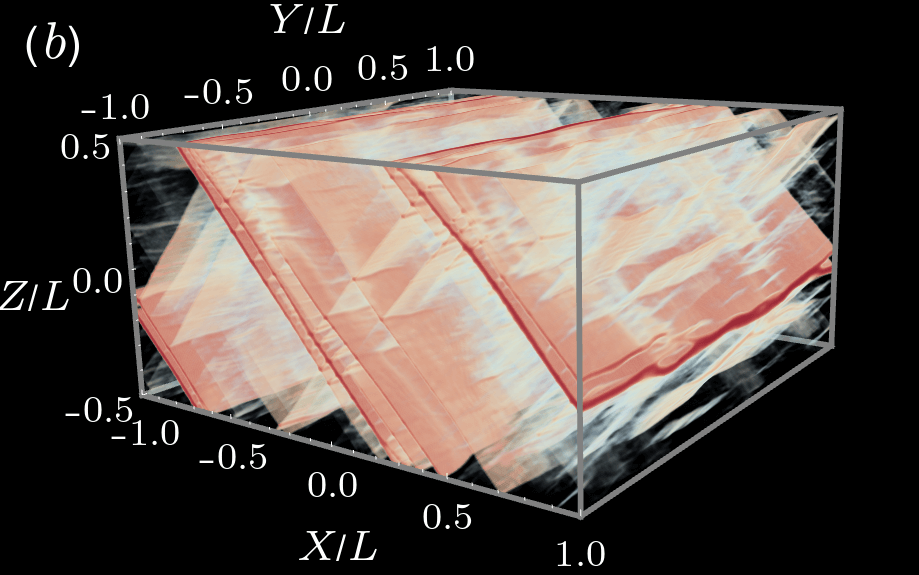} \\
        \includegraphics[width=\subpanelwid]{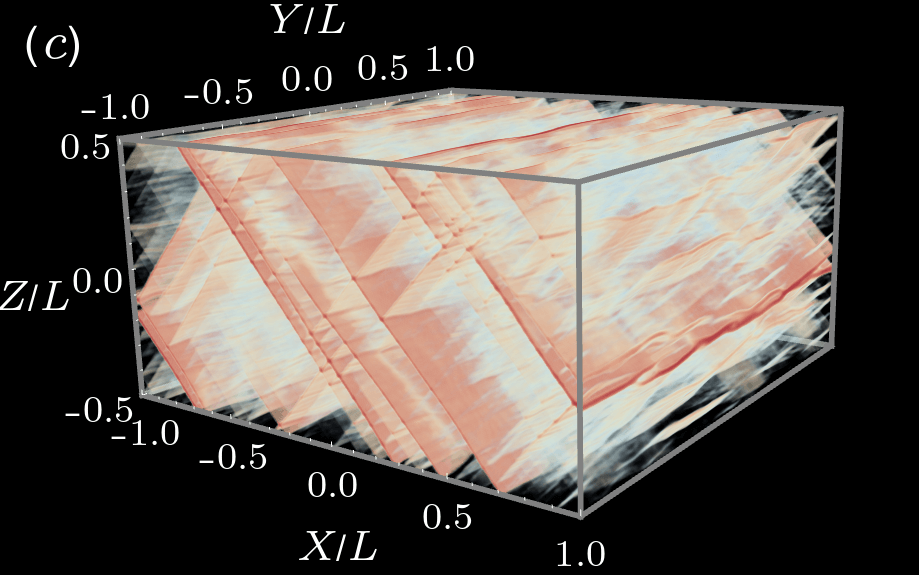} &
        \includegraphics[width=\subpanelwid]{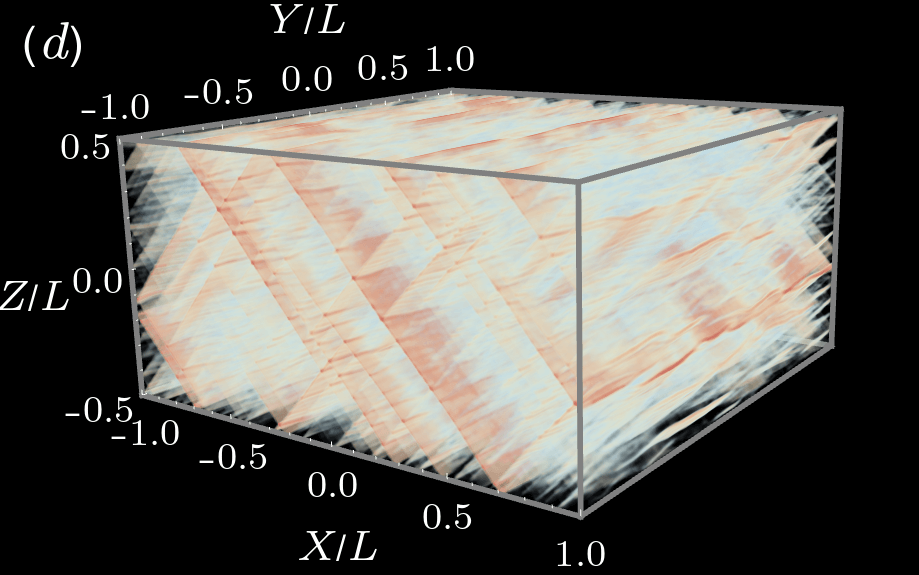} \\
        \includegraphics[width=\subpanelwid]{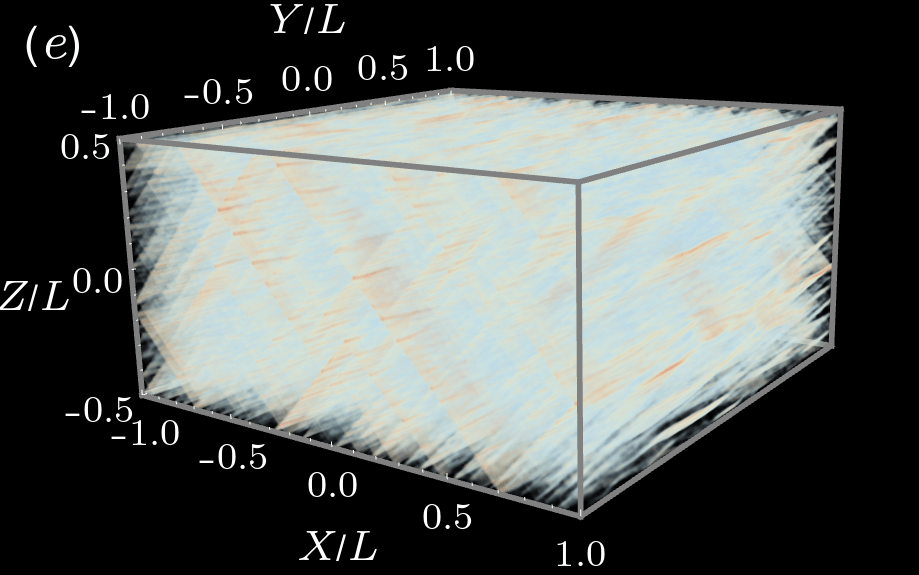} &
        \includegraphics[width=\subpanelwid]{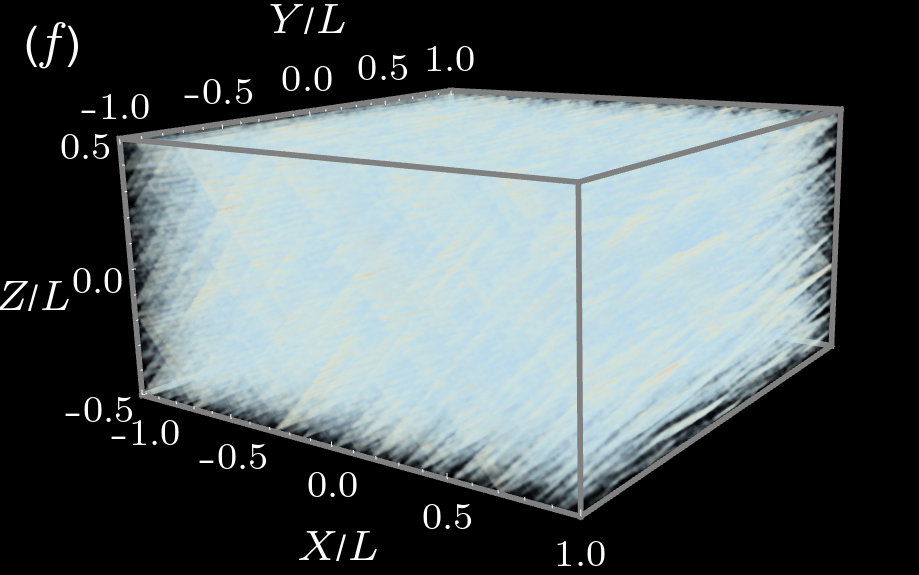}
    \end{tabular}}
    ~\vspace{5mm}\\
    \includegraphics[width=.75\textwidth]{imgs/rslt_figs/colorbar_2019}
    \caption{Snapshots of the effective temperature field at $t = 6\times 10^5 t_s$ with a pure shear transformation imposed on the domain. All simulations use periodic boundary conditions. For all plots, a value of $a = 0.75$ and $\eta = 1.6$ is used in the opacity function, and $\chi_\text{bg}$ is set to $\mu_\chi - 25\text{~K}$. Figures (a)--(f) have $\mu_\chi = 450\tK, 500\tK, 525\tK, 550\tK, 575\tK, 600\tK$ respectively.}
    \label{fig:ps_le_60}
\end{figure*}

\begin{figure*}
\fcolorbox{black}{black}{
    \begin{tabular}{cc}
        \includegraphics[width=\subpanelwid]{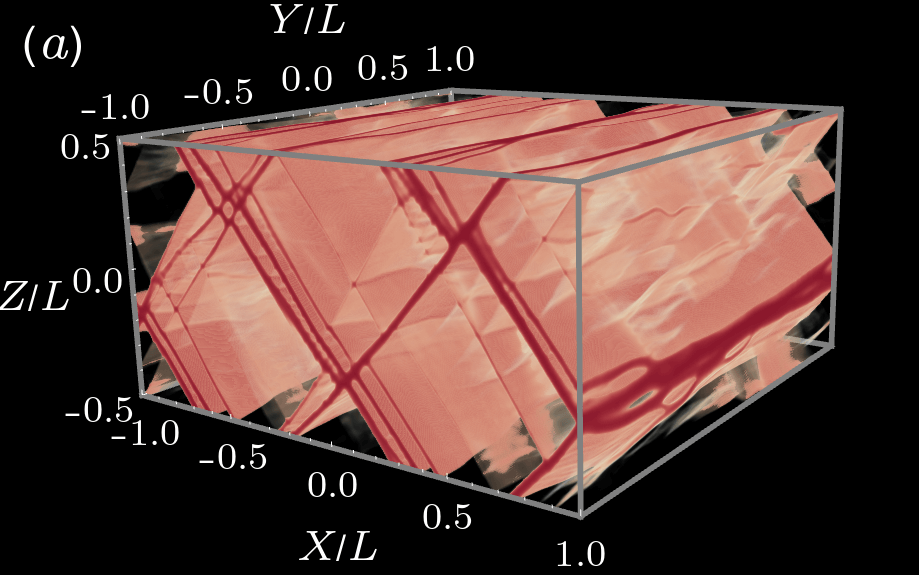} &
        \includegraphics[width=\subpanelwid]{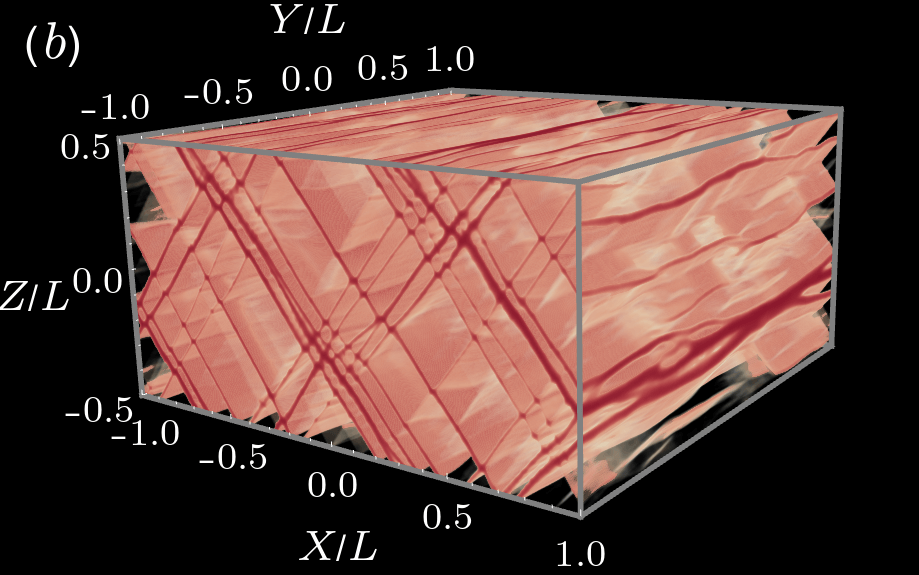} \\
        \includegraphics[width=\subpanelwid]{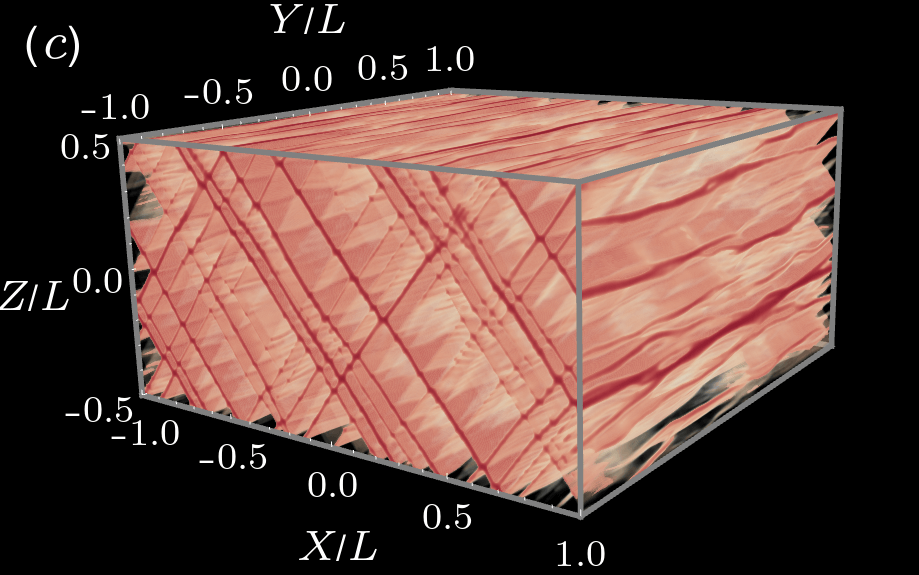} &
        \includegraphics[width=\subpanelwid]{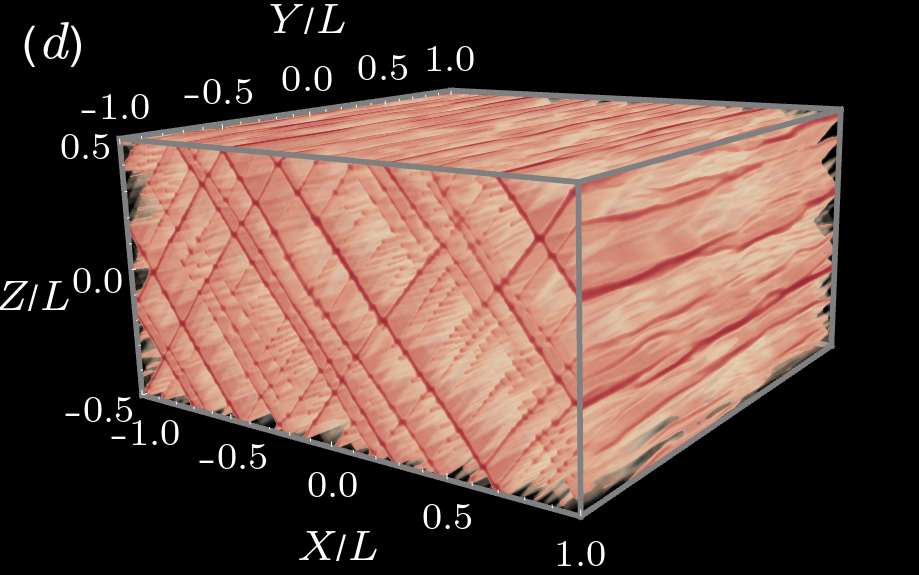} \\
        \includegraphics[width=\subpanelwid]{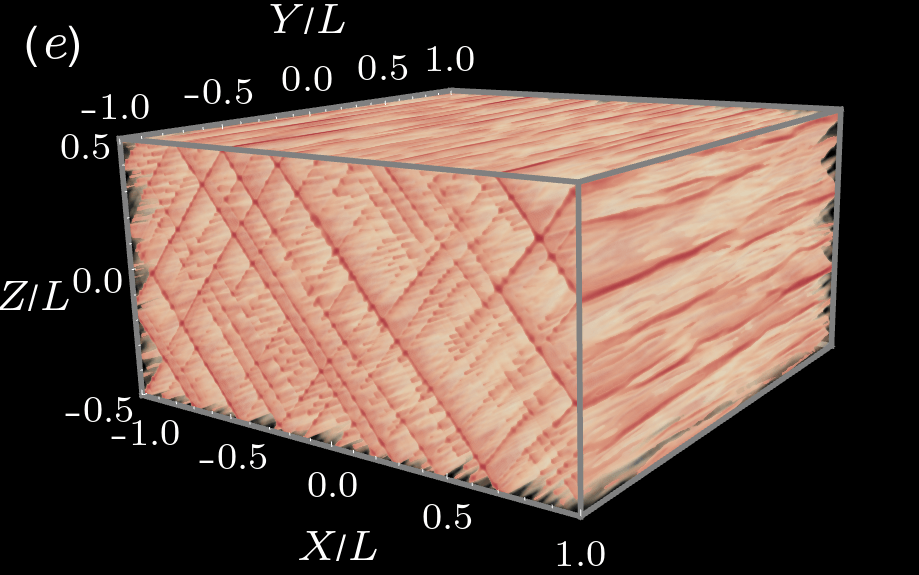} &
        \includegraphics[width=\subpanelwid]{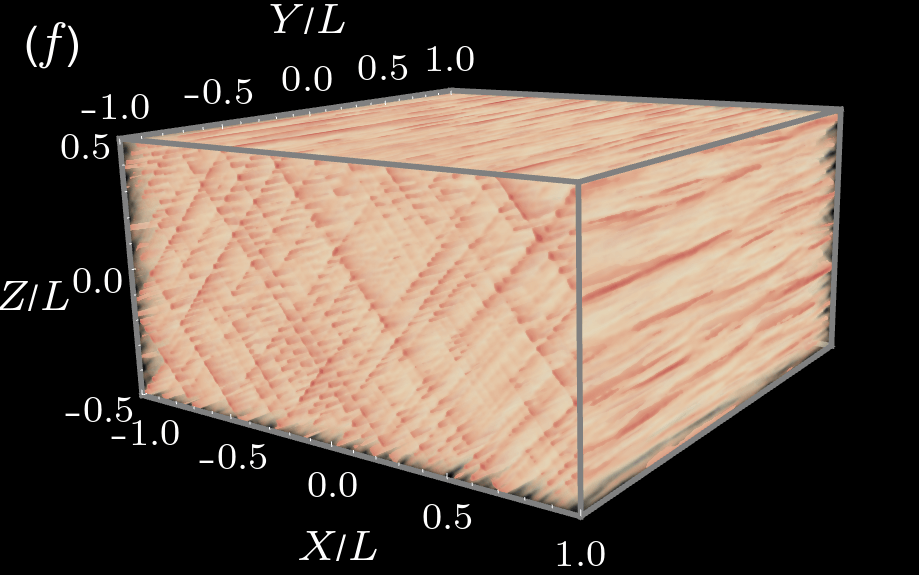}
    \end{tabular}}
    ~\vspace{5mm}\\
    \includegraphics[width=.75\textwidth]{imgs/rslt_figs/colorbar_2019}
    \caption{Snapshots of the effective temperature field at $t = 1.5\times 10^6 t_s$ with a pure shear transformation imposed on the domain. All simulations use periodic boundary conditions. For all plots, a value of $a = 1.35$ and $\eta = 1.5$ is used in the opacity function, and $\chi_\text{bg}$ is set to $\mu_\chi - 25\text{~K}$. Figures (a)--(f) have $\mu_\chi = 450\tK, 500\tK, 525\tK, 550\tK, 575\tK, 600\tK$ respectively.}
    \label{fig:ps_le_150}
\end{figure*}

The results are shown in Figs.~\ref{fig:ps_le_0}--\ref{fig:ps_le_150}, with the initial condition shown in Fig.~\ref{fig:ps_le_0}. All simulations undergo an increase in $\chi$ until the formation of diagonal shear bands begins. Much like the defect simulations seen in the previous section, shear bands nucleate diagonally at roughly $45\degree$ angles to the $X$--$Y$ plane. As in the simple shear simulations, distributions in $\chi$ with higher mean values have slower dynamics. The structural effect of varying $\mu_\chi$ is most easily seen in Fig.~\ref{fig:ps_le_150}. As $\mu_\chi$ increases, the number of shear bands vastly increases, forming a cross-hatched pattern throughout the domain. The cross-hatching becomes more regular and more finely spaced with higher values of $\mu_\chi$.

\section{Conclusion}
\label{sec:conc}
In this work, we derived the equations of hypo-elastoplasticity on a fixed reference domain which can be mapped to a physically deforming material through a time-varying linear transformation $\bT(t)$. The difference between this frame and the Lagrangian frame was shown, and the utility of this frame in implementing complex boundary conditions such as the Lees--Edwards conditions used in molecular dynamics and pure shear in a fully periodic setting was demonstrated. The quasi-static projection algorithm was derived in the reference frame and its convergence to the standard method was shown as the level of discretization increases. Several numerical examples were considered in the STZ model of amorphous plasticity. In particular, for a randomly-distributed initial condition in the effective temperature field, the dependence of shear banding dynamics on the mean of the distribution was discussed under conditions of simple shear and pure shear. Our work also highlights that the direction of shear bands (\textit{e.g.}~horizontal versus vertical in simple shear) can be strongly influenced by boundary conditions.

With the simple implementation of Lees--Edwards conditions afforded by the transformation method, boundary conditions can now be made equivalent in MD and continuum modeling. The development of a method to compute a precise matching between atomic configurations in molecular dynamics and effective temperature distributions in continuum simulations is a promising direction of future research. The ability to do so would place internal state variables in plasticity models (such as the effective temperature in the STZ model) on a firmer theoretical footing. In addition, hybrid computational approaches could be developed, where an MD simulation could first be used to compute an initial condition for a significantly larger scale continuum simulation. This type of approach would combine the physical accuracy of MD with the capability of continuum simulations to simulate large system sizes and long times. As an added benefit, our approach enables the study of the effect of periodic boundary conditions in general, independent of the relevance of these settings to MD. 

So far, our implementations are restricted to cases where the material fills the entire computational domain, and loading is applied via planar boundary conditions, or via the coordinate transformation framework. However, the methods presented here could be generalized to materials with free boundaries, using the level set method~\cite{sethian,osher} to track the material boundary. Methods to do this have already been implemented in two dimensions~\cite{rycroft12,rycroft12b,rycroft15}, and the same methods could be used in principle in three dimensions. However, it is a challenging computational task, since it requires extensive modifications to the finite-difference stencils near the material boundary. In particular, since some grid points will lie outside the material, the geometric multigrid method is no longer well-suited for solving the projection step, since it relies on a regular arrangement of grid points. It may be necessary to use algebraic multigrid approaches or Krylov-based linear solvers. Nevertheless this remains a high priority for future work, since it would open up many new directions, such as studying three-dimensional cavitation~\cite{bouchbinder08,falk99}, simulating mode III fracture~\cite{karma01}, and predicting the topography of fracture surfaces~\cite{lowhaphandu00,zhang04,raghavan09}.

\appendix
\section{Advective derivative calculation}
\label{app:adv}
Consider a scalar field $\phi(\bx, t) = \phi(\bT \bX, t)$. We can compute the advective derivative of $\phi$ as follows using the chain rule,
\begin{align}
    \frac{d}{dt}\phi(\bT\bX, t) &= \left(\frac{\p}{\p t} + \vv^\Trans\frac{\p}{\p \bx}\right)\phi(\bT\bX,t)\nonumber\\
    &= \left( \frac{\p}{\p t} + \vv^\Trans\bT^{-\Trans}\frac{\p}{\p \bX}\right)\phi(\bT\bX, t)\nonumber\\
&= \left(\frac{\p \bX}{\p t}\right)^\Trans \frac{\p}{\p \bX}\phi(\bT\bX, t) + \phi_t(\bT\bX, t) + \vv^\Trans \bT^{-\Trans}\frac{\p}{\p\bX}\phi(\bT\bX, t)\nonumber\\
    &= \phi_t(\bT\bX, t) + \left(\vv^\Trans\bT^{-\Trans} + \left(\frac{\p\bX}{\p t}\right)^\Trans\right)\frac{\p}{\p \bX}\phi(\bT\bX, t)\nonumber\\
    &= \phi_t(\bT\bX, t) + \left(\vv^\Trans\bT^{-\Trans} + \left(\frac{\p\bT^{-1}}{\p t}\bT{\bX}\right)^\Trans\right)\frac{\p}{\p \bX}\phi(\bT\bX, t)\nonumber\\
    &= \phi_t(\bT \bX, t) + \vV^\Trans \Nab \phi(\bT\bX, t).
    \label{eqn:d/dt_phi_AX}
\end{align}
In the last line, we have used Eq.~\ref{eqn:V} and the identity $\frac{\p \bT^{-1}}{\p t} = -\bT^{-1}\frac{\p \bT}{\p t}\bT^{-1}$.

\section{Linear system for simple shear}
\label{app:ss}
Let $\vV = \left(U, V, W\right)^\Trans$. For a simple shear transformation as given in Eq.~\ref{eqn:shear_trans}, $\tC : \tD$ takes the form
\begin{align}
    \left(\tC:\tD\right)_{11} &= \lambda\left(\Nab\cdot\vV\right) + 2\mu\frac{\p U}{\p X} + 2 \mu U_b t \frac{\p W}{\p X}, \\
    \left(\tC:\tD\right)_{12} &= \mu\left(\frac{\p U}{\p Y} + U_b t \frac{\p W}{\p Y} + \frac{\p V}{\p X}\right),\\
    \left(\tC:\tD\right)_{13} &= \mu\left(U_b + \frac{\p U}{\p Z} + U_b t \frac{\p W}{\p Z} - U_b t\frac{\p U}{\p X} + \left(1 - U_b^2t^2\right)\frac{\p W}{\p X}\right),\\
    \left(\tC:\tD\right)_{22} &= \lambda\left(\Nab\cdot\vV\right) + 2\mu\frac{\p V}{\p Y},\\
    \left(\tC:\tD\right)_{23} &= \mu\left(\frac{\p V}{\p Z} + \frac{\p W}{\p Y} - U_b t\frac{\p V}{\p X}\right),\\
    \left(\tC:\tD\right)_{33} &= \lambda\left(\Nab\cdot\vV\right) + 2\mu\frac{\p W}{\p Z} - 2 U_b \mu t \frac{\p W}{\p X}.
\end{align}
The above set of equations leads to the linear system (Eqs.~\ref{eqn:trans_RHS} \& \ref{eqn:trans_LHS}) for the velocity field
\begin{align}
    \label{eqn:ss_proj_1}
    \frac{-1}{\Delta t}\left(\bT^n\Nab\cdot\bSig^*\right)_1 &= \left(U_b^3t^3 \mu + U_b t\mu\right)
        \frac{\p^2 W}{\p X^2} + \left(\lambda + 2 \mu + U_b^2t^2 \mu \right) \frac{\p^2 U}{\p X^2} \nonumber\\
        &\phantom{=} + \left(\lambda + \mu -2 U_b^2t^2 \mu \right) \frac{\p^2 W}{\p X \p Z} - 2 U_b t\mu \frac{\p^2 U}{\p X \p Z} + U_b t\mu \frac{\p^2 W}{\p Z^2} + U_b t\mu \frac{\p^2 W}{\p Y^2} \nonumber\\
        &\phantom{=}+ \mu \frac{\p^2 U}{\p Z^2} + \mu\frac{\p^2 U}{\p Y^2} + (\lambda + \mu)\frac{\p^2 V}{\p X \p Y},\\
        \frac{-1}{\Delta t}\left(\bT^n\Nab\cdot\bSig^*\right)_2 &= U_b^2t^2\mu \frac{\p^2 V}{\p X^2} - 2 U_b t \mu\frac{\p^2 V}{\p X \p Z} + \lambda\frac{\p^2 U}{\p X \p Y} + \mu \frac{\p^2 U}{\p X \p Y} \nonumber\\
        &\phantom{=} + \lambda\frac{\p^2 V}{\p Y^2} + \mu \frac{\p^2 V}{\p Z^2} + 2 \mu\frac{\p^2 V}{\p Y^2} + \mu \frac{\p^2 V}{\p X^2} + \lambda \frac{\p^2 W}{\p Y \p Z} + \mu \frac{\p^2 W}{\p Y \p Z},\\
    \frac{-1}{\Delta t}\left(\bT^n\Nab\cdot\bSig^*\right)_3 &= \left(\lambda + 2\mu\right)\frac{\p^2 W}{\p Z^2} + \left(\lambda + \mu\right)\frac{\p^2 V}{\p Y \p Z} + \mu \frac{\p^2 W}{\p Y^2} \nonumber\\
    &\phantom{=} + \left(\lambda + \mu\right)\frac{\p^2 U}{\p X \p Z} - U_b t\left(\lambda + 3\mu\right)\frac{\p^2 W}{\p X \p Z} - U_b t\left(\lambda + \mu\right)\frac{\p^2 V}{\p X \p Y} \nonumber\\
    &\phantom{=} - U_b t\left(\lambda + \mu\right)\frac{\p^2 U}{\p X^2} + \left(1 + U_b^2t^2\right)\mu\frac{\p^2 W}{\p X^2}.
    \label{eqn:ss_proj_3}
\end{align}
Discretization of the second-derivative terms in Eqs.~\ref{eqn:ss_proj_1}--\ref{eqn:ss_proj_3} using the finite differences in Sec.~\ref{ssec:num} enables application of the geometric multigrid method to solve for $U$, $V$, and $W$.

\section{Linear system for pure shear}
\label{app:ps}
For a pure shear transformation of the form Eq.~\ref{eqn:ps_trans} with $A(t) = e^{\xi t}$ as in the main text, $\tC : \tD$ takes the form
\begin{align}
    \left(\tC:\tD\right)_{11} &= \lambda \Nab\cdot\vV + 2\mu\left(\xi + \frac{\p U}{\p X}\right)\\
    \left(\tC:\tD\right)_{12} &= \mu\left(e^{\xi t}\frac{\p U}{\p Y} + e^{-\xi t}\frac{\p V}{\p X}\right)\\
    \left(\tC:\tD\right)_{13} &= \mu\left(e^{2 \xi t}\frac{\p U}{\p Z} + e^{-2 \xi t}\frac{\p W}{\p X}\right)\\
    \left(\tC:\tD\right)_{22} &= \lambda\Nab\cdot\vV + 2\mu\frac{\p V}{\p Y}\\
    \left(\tC:\tD\right)_{23} &= \mu\left(e^{\xi t}\frac{\p V}{\p Z} + e^{-\xi t}\frac{\p W}{\p Y}\right)\\
    \left(\tC:\tD\right)_{33} &= \lambda \Nab\cdot\vV + 2\mu\left(\frac{\p U}{\p X} - \xi\right)
\end{align}
The above set of equations leads to the linear system (Eqs.~\ref{eqn:trans_RHS} \& \ref{eqn:trans_LHS}) for the velocity field
\begin{align}
    \frac{-1}{\Delta t}\left(\bT^n\Nab\cdot\bSig^*\right)_1 &= \mu e^{3\xi t}\frac{\p^2 U}{\p Z^2} + e^{\xi t}\mu\frac{\p^2 U}{\p Y^2} + e^{-\xi t}\left(\lambda + \mu\right)\frac{\p^2 W}{\p X \p Z} \nonumber\\
    &\phantom{=} + e^{-\xi t}\left(\lambda + \mu\right)\frac{\p^2 V}{\p X \p Y} + e^{-\xi t}\left(\lambda + \mu\right)\frac{\p^2 U}{\p X^2}\\
    \frac{-1}{\Delta t}\left(\bT^n\Nab\cdot\bSig^*\right)_2 &= e^{2\xi t}\mu\frac{\p^2 V}{\p Z^2} + \left(\lambda + \mu\right)\frac{\p^2 W}{\p Y \p Z} \nonumber\\
    &\phantom{=} + \left(\lambda + 2\mu\right)\frac{\p^2 V}{\p Y^2} + \left(\lambda + \mu\right)\frac{\p^2 U}{\p X \p Y} + e^{-2\xi t}\mu\frac{\p^2 V}{\p X^2}\\
    \frac{-1}{\Delta t}\left(\bT^n\Nab\cdot\bSig^*\right)_3 &= e^{\xi t}\left(\lambda + 2\mu\right)\frac{\p^2 W}{\p Z^2} + e^{\xi t}\left(\lambda + \mu\right)\frac{\p^2 V}{\p Y\p Z} \nonumber\\
    &\phantom{=} + e^{-\xi t}\mu\frac{\p^2 W}{\p Y^2} + e^{\xi t}\left(\lambda + \mu\right)\frac{\p^2 U}{\p X \p Z} + e^{-3\xi t}\mu\frac{\p^2 W}{\p X^2}
\end{align}

\section*{Acknowledgments}
The authors thank Eran Bouchbinder and Avraham Moriel (Weizmann Institute of
Science) for useful discussions about this work. This work was supported by the
National Science Foundation under Grant Nos.~DMR-1409560 and DMS-1753203.
N.~M.~Boffi was supported by a Department of Energy Computational Science
Graduate Fellowship. C.~H.~Rycroft was partially supported by the Applied
Mathematics Program of the U.S. DOE Office of Advanced Scientific Computing
Research under contract number DE-AC02-05CH11231.

\bibliography{elas}

\end{document}